\documentclass[11pt]{article}
\usepackage[applemac]{inputenc} 
\usepackage[T1]{fontenc} 
\usepackage{geometry} 
\usepackage{amssymb,amsmath} 
\usepackage{graphicx} 
\usepackage[english]{babel} 
\usepackage{mathrsfs} 

\newcommand{\kB}{\ensuremath{k_{\mathrm{B}}}} 
\newcommand{\ave}[1]{\ensuremath{\left \langle {#1} \right \rangle}} 
\newcommand{\abs}[1]{\ensuremath{\left\lvert{#1}\right\rvert}} 
\newcommand{\eps}{\ensuremath{\varepsilon}} 
\newcommand{\ket}[1]{\ensuremath{\left \lvert {#1} \right \rangle}} 
\newcommand{\var}{{\ensuremath{\mathrm{var}}}} 
\newcommand{\fano}{\ensuremath{\mathscr{F}}} 
\newcommand{\spectral}{\ensuremath{\mathscr{S}}} 
\newcommand{\ii}{\ensuremath{\mathfrak{i}}} 
\newcommand{\jj}{\ensuremath{\mathfrak{j}}}

\newcommand{\al}{\ensuremath{\alpha}}

\newcommand{\de}{\ensuremath{\delta}}
\newcommand{\De}{\ensuremath{\Delta}}
\newcommand{\om}{\ensuremath{\omega}}
\newcommand{\Om}{\ensuremath{\Omega}}

\newcommand{\C}{\ensuremath{\mathcal{C}}}
\newcommand{\E}{\ensuremath{\mathcal{E}}}
\newcommand{\N}{\ensuremath{\mathcal{N}}}

\newcommand{\D}{\ensuremath{\mathcal{D}}}

\newcommand{\R}{\ensuremath{\mathcal{R}}}

\newcommand{\V}{\ensuremath{\mathcal{V}}}
\newcommand{\I}{\ensuremath{\mathcal{I}}}

\newcommand{  \p   }   { \ensuremath{      \big{(}          }               }
\newcommand{  \q   }   {  \ensuremath{      \big{)}        }                }

\graphicspath{{figures/}} 
\newlength{\figwidth} 

\begin{document}
\begin{titlepage} 
\title{Semi-classical theory of quiet lasers. I: Principles}

\author{ Jacques \textsc{Arnaud}
\thanks{Mas Liron, F30440 Saint Martial, France}, Laurent \textsc{Chusseau}
\thanks{ Centre d'\'Electronique et de Micro-opto\'electronique de Montpellier, Unit\'e Mixte de
Recherche n°5507 au CNRS, Universit\'e Montpellier II, F34095 Montpellier, France}, Fabrice
\textsc{Philippe}
\thanks{D\'epartement de Math\'ematiques et Informatique Appliqu\'ees, Universit\'e Paul Val\'ery,
F34199 Montpellier, France.  Also with LIRMM, 161 rue Ada, F34392 Montpellier, France},}
\maketitle

\begin{abstract}      

When light originating from a laser diode driven by non-fluctuating electrical currents is incident on a photo-detector, the photo-current does not fluctuate much. Precisely,  this means that the variance of the number of photo-electrons counted over a large time interval is much smaller that the average number of photo-electrons. At non-zero Fourier frequency $\Om$ the photo-current  power spectrum is of the form $\Om^2/(1+\Om^2)$ and thus vanishes as $\Om\to 0$, a conclusion equivalent to the one given above. The purpose of this paper is to show that results such as the one just cited may be derived from a (semi-classical) theory in which neither the optical field nor the electron wave-function are quantized. We first observe that almost any medium may be described by a circuit and distinguish (possibly non-linear) conservative elements such as pure capacitances, and conductances that represent the atom-field coupling. Configurations involving a single electron are considered. The theory rests on the non-relativistic approximation, that is, $c$ is set as infinite. Nyquist noise sources (in which the Planck term $\hbar\om/2$ is being restored) are associated with positive or negative conductances, and the law of average-energy conservation is enforced. We consider only first and second-order correlations of photo-electric currents in stationary regimes. 

Detailed semi-classical treatments of various quiet sources were listed in the first version of this paper. Only the general principles are presently considered, detailed applications being postponed.

\end{abstract}

\end{titlepage}

\tableofcontents

\newpage

\section{Introduction}\label{introduction}

\begin{quote}
"Comprendre", c'est comprendre \emph{autrement} ("Comprehend" means comprehend \emph{differently}).
\end{quote}

In the present introduction we outline our objective, main concepts, approximations employed, key results, and describe how the paper is organized.

\paragraph{Scope of the paper.}

Laser noise impairs the operation of optical communication systems and the measurement of small displacements or small rotation rates with the help of optical interferometry. Even though laser light is far superior to thermal light, minute fluctuations restrict the ultimate performances. Signal-to-noise ratios, displacement sensitivities, and so on, depend mainly of the spectral densities, or correlations, of the photo-currents. It is therefore important to have at our disposal formulas enabling us to evaluate these quantities for configurations of practical interest, in a form as simple as possible.

We are mostly concerned with basic concepts leaving out detailed practical calculations. Non-essential noise sources such as mechanical vibrations are ignored. Real lasers involve many secondary effects that are presently neglected for the sake of clarity. For example, because of the large size of the cavity in comparison with wavelength, lasers tend to oscillate on more than one mode. Even if the side-mode powers are much reduced with the help of distributed feed-backs or secondary cavities, small-power side modes may significantly influence laser-noise properties, particularly near the shot-noise level. Side-mode powers should probably be less than 40 dB below the main mode power to be insignificant. In the case of gas lasers, multiple levels, atomic collisions, thermal motions, and so on, may strongly influence noise properties, but these effects are neglected here.

The main purpose of this paper is to show that, contrary to what most previous works imply, the properties of quiet lasers may be understood on the basis of a simple semi-classical theory, that is, a theory in which neither the optical field nor the electron wave-function are quantized. The electrons may be uncoupled to one another (dilute atom gases) or strongly coupled as is the case in semiconductors, through the Pauli exclusion principle. This theory (proposed by one of us in papers from 1986 on, and in book form in 1989 \cite{Arnaud1989}) is accurate and easy to apply, yet little known. The physical concepts are hopefully better explained in the present paper than in previous ones. Once the necessary assumptions have been agreed upon, laser noise formulas for various configurations follow from elementary Mathematics. In particular, operator algebra is not needed. Previous Semi-classical theories and Quantum Optics theories may be found in \cite{Yamamoto91}.

In this introductory part, the principles are presented but application to simple circuits, to the noise of lasers incorporating multilevel atoms or having spatially varying phase-amplitude coupling factors, the linewidth of inhomogeneously-broadened lasers, and the role of electrical feed-backs, is postponed.

Our interpretations of the basic mechanisms behind quiet-laser operation\footnote{From our view-point a constant pump current entails a constant photo-current under ideal conditions. In a recent book \cite{Kim2001} the basic mechanism behind quiet-laser operation is described as follows: "Although the noise generated in the external resistor is far below the shot-noise level, this does not mean that the carrier injection into the active region is regulated[...]. The carriers supplied by the external circuit are injected stochastically across the depletion layer before they reach the active region" (Presumably, by "stochastic" the authors means "Poisson-distributed"). The authors then introduce potential fluctuations to explain the observed quiet radiation. It may be, however, that these authors description is just another way of describing the same Physics as in the present paper. }, the rôle of the Petermann  K-factor\footnote{It was observed early by E.I. Gordon \cite{Gordon1964} that in the linear regime laser line-widths are enhanced above those given by the well-known Schawlow-Townes formula for various circuits involving lumped elements or transmission lines, see \cite[p. 120]{Thomson1980}. This linewidth-enhancement, which relates to the fact that gain and loss regions occur at different locations, may be described alternatively in terms of resonant complex potentials (or integrals of resonant complex fields). The line-width enhancement factor may be observed in strictly single-mode laser or maser oscillators. The K-factor effect discovered by Petermann \cite{Petermann1979} is of great practical importance for some laser diodes. It is not in our opinion fundamentally different from the Gordon effect just described. Further, contrary to a wide-spread belief, the K-factor cannot be applied directly to above-threshold lasers. The law of average-energy conservation tells us that quiet light should be observed with a quiet pump, leaving aside non-unity quantum efficiency, current leakage, and so on, irrespectively of the value of the K-factor value. An entirely different view point has been recently expressed \cite{Maurin2005}: "Because of the non-orthogonality of the laser eigenmodes, the noise from other modes is homodyned into the lasing mode, leading to an excess noise in the lasing mode, \emph{which is the Petermann excess noise}" (our emphasis).} and non-linear gain, on some respects differ from those found elsewhere in the literature. For simple laser models we observe however exact agreement between our results and those derived from Quantum Optics, whenever a comparison can be made.

Theories found in most of the Optical Engineering literature rest on the concept that the classical oscillating field is supplemented by a random field. They involve parameters whose values are difficult to establish before-hand, because it is difficult to evaluate quantities such as the non-linear gain coefficient separately from the complete laser operation. Accurate models of real lasers (optical frequencies) are more difficult to establish than for masers (microwave frequencies) because of the small light wavelength. As a consequence, agreements reported between theory and experiments may not be conclusive as far as fundamental concepts are concerned. Discrepancies between our semi-classical theory and previous semi-classical theories occur near the shot-noise level, and are therefore difficult to establish experimentally. Detailed comparison with observations may require specially constructed lasers.

\paragraph{Main concepts of the theory and approximations.}

The important conclusion of the present discussion is the following. When an oscillator is driven by a constant power source and dissipates energy by amounts occurring at random times, the power delivered to the outside is nearly non-fluctuating. That is, the oscillator radiation is \emph{quiet}. On many respects, lasers are akin to any oscillator, whether mechanical or electrical. We consider below the so-called "grand-mother pendulum" for the purpose of showing that the main features of quiet-radiation oscillators may be understood on a purely classical basis, and are in fact, most common. This conclusion is based on elementary mechanical considerations that have nothing to do with Quantum Optics. (Note that in order to turn a "grand-mother pendulum" into a grand-mother \emph{clock} one would need introduce a period-counting mechanism that would perturb the system as it is presently described). In similar laser oscillators the Planck constant $\hbar$ enters because the coupling of the oscillator to absorbers is effected with the help of electrons that may reside in only two states separated in energy by $\epsilon=\hbar \om$. The mechanical model discussed below differs from laser models in that dissipation events remain Poisson-distributed. The regulation then originates from the fact that the energy lost at a "dissipation event" (to be defined later on) is proportional to the time interval that occurred from the previous event to the present one (this is called a "marked" Poisson process with marks depending on the Poisson-process realization). 

\begin{figure}
\setlength{\figwidth}{0.6\textwidth}
\centering
\begin{tabular}{cc}
\includegraphics[width=\figwidth]{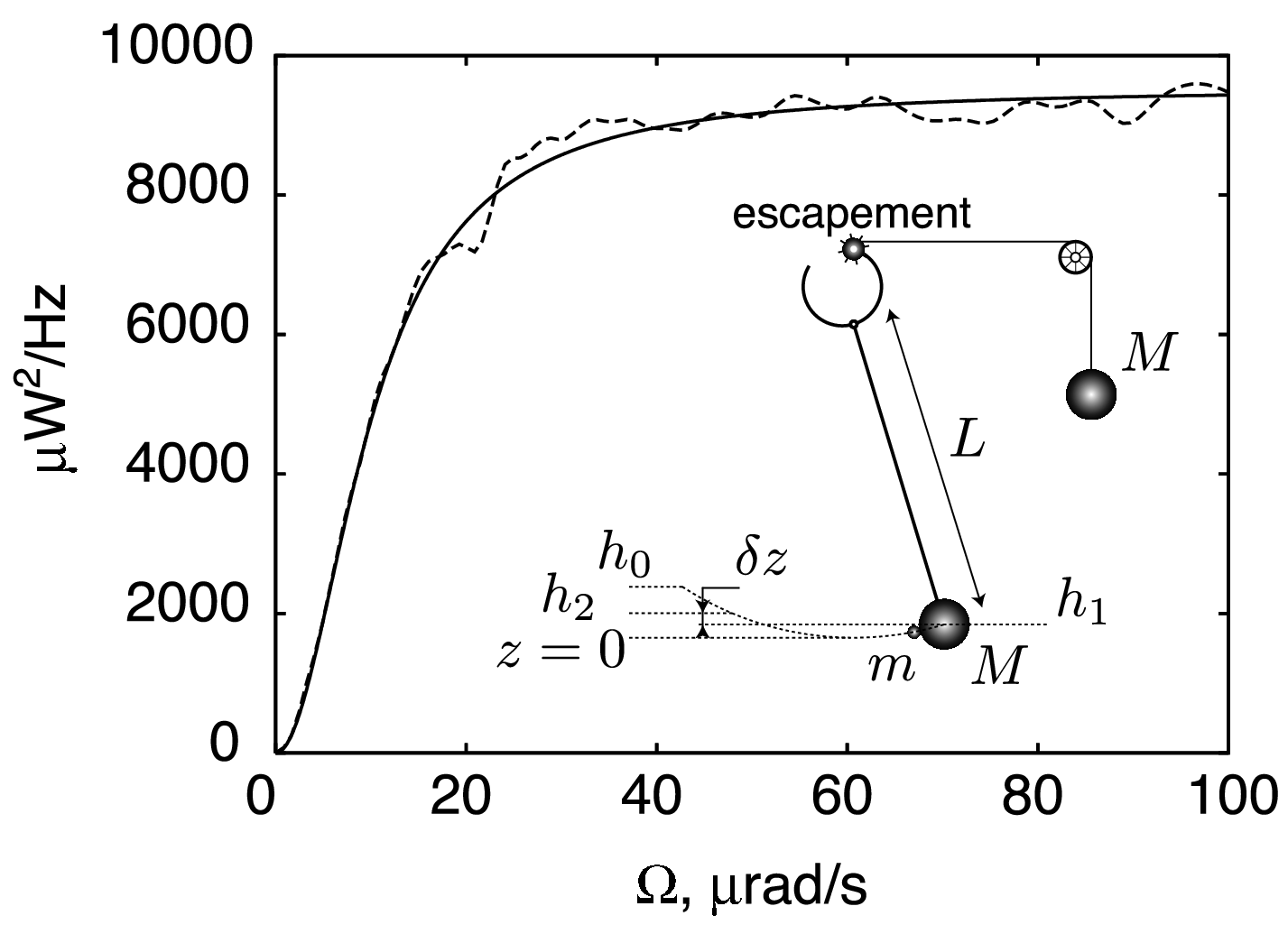}
\end{tabular}
\caption{Variation of the spectral density of the power dissipated by a "grand-mother pendulum" as a function of the Fourier frequency $\Om$. The smooth curve is theoretical, while the slightly wavy curve results from a numerical simulation of the corresponding marked Poisson process. Inset, schematics of the pendulum of mass $M$. With probability $pr\ll 1$ the pendulum picks up a molecule of mass $m\ll M$ at $z=0$ and releases it at $z=h_1$, causing damping. A suspended weight (also of mass $M$) drops by a length $\de z$ at each swing of the pendulum, thereby incrementing its mass height by $\de z$. The escapement is represented symbolically.}
\label{pendulum}
\end{figure}

The basic element of a grand-mother pendulum is a mass $M$ suspended at the end of a weightless bar of length $L$, in the earth gravitational field with acceleration $g$. As was first shown by Galileo, the oscillation period $T=2\pi \sqrt{L/g}$ does not depend on the oscillation amplitude as long as this amplitude remains small. Note also that the period does not depend either on the mass $M$. The parameters $L$, $g$, $M$ (and later on $m$ and $\de z$), and thus the period $T$, are fixed quantities in the present discussion. For simplicity, we assume that $T=1$s. If we label successive periods by $k=1,2...$ we identify time $t=kT=k$. Let us call $z(t)$ the pendulum mass height, with $z(1/4)=z(3/4)=0$, the lowest mass position. We set $z(0)\equiv h_0$, $z(1/2)\equiv h_1$ and $z(1)\equiv h_2$, see Fig. \ref{pendulum}. If the pendulum motion is not disturbed, the energy $E=Mgh_0=Mgh_1=Mgh_2$, since the pendulum mass possesses no kinetic energy when its maximum height is being reached.  

Because of dissipation, and ignoring for the moment fast fluctuations, the pendulum energy decays in the course of time according to a law of the form $E(t)=E(0)\exp(-t/\tau_p)$, where $E(0)$ denotes the energy at $t=0$ and the constant $\tau_p$ is called the pendulum life-time. The power $P_d(t)$ dissipated in the medium surrounding the pendulum is viewed as a measurable quantity, perhaps through temperature increments. The lifetime $\tau_p$ may therefore be measured, perhaps after averaging over a large number of identical pendula having the same initial heights. This life-time will be subsequently evaluated on the basis of a microscopic theory of damping.

In order to obtain stationary oscillations some energy must be supplied to the pendulum. This is achieved with the help of a mass $M$ suspended at the end of a cord (for simplicity this mass is assumed to be equal to the pendulum mass). An escapement mechanism allows the suspended mass to drop by a fixed height $\de z$ at each swing of the pendulum, say at the end of every period, and then to increment the pendulum energy by $\de E=Mg\de z$, so that the pendulum mass reaches a height exceeding the previous one by $\de z$.  The pendulum period $T=1$ being constant, this implies that the suspended mass delivers a constant power $P=\de E$, if we ignore the detailed process that may occur during a single period. Such a constant power supply may be called a "quiet pump" and is equivalent to the constant-current drive of lasers . Consider now both the power $P$ supplied by the suspended weight, as discussed above, and the dissipation mechanism characterized by the empirical constant $\tau_p$. The equation of motion of the pendulum energy reads $dE(t)/dt=P-P_d(t)=\de E-E(t)/\tau_p$. In the steady-state ($dE(t)/dt=0$) the average energy is therefore $\ave{E}=\de E~ \tau_p$.

We now model the damping mechanism at the microscopic level. We select a positive number $pr<1$, and consider a random number $x$ uniformly distributed between 0 and 1. If $x<1-pr$ no damping occurs. If $h_0$ denotes the pendulum mass height at the beginning of such an uneventful period, its height just after the period is $h_2=h_0+\de z$, as we discussed above. If, however, the random number $x>1-pr$, the pendulum picks up at $t=1/4,~z=0$, a molecule of mass $m$, initially at rest, and releases it at $t=1/2,~ z=h_1$, again with zero velocity. If $h_0$ denotes the pendulum initial height, the maximum height $h_1$ reached is given by the law of conservation of energy by $Mgh_0=(M+m)gh_1$. Note that the mass increment does not affect the period (or the quarter of the period). At the end of the period we have $h_2=h_1+\de z=h_0/(1+m/M)+\de z$. The energy delivered to the molecule collection is then $\De E=mgh_1=mgh_0/(1+m/M)$. To summarize, given the height $h_0$ at the beginning of a period, the height at the end of the period is $h_0+\de z$ with probability $1-pr$ and $h_0/(1+m/M)+\de z$ with probability $pr$.

Let the molecule-picking events occur at times $k_i$, where $i=0,1,2...$. We denote by $h_{i,0},h_{i,1},h_{i,2}$ the pendulum-mass heights as defined above, but relating to the $i$th-event. According to the previous description, molecule-picking events are Poisson distributed. At event $i$ we attach a "mark" $\mu_i$ equal to the dissipated energy $mgh_{i,1}$. The previous description shows that $\mu_i= \mu_{i-1}/(1+m/M)+(mg/(1+m/M))\de z(k_i-k_{i-1})$ is a linear deterministic function of the previous mark $\mu_{i-1}$ and of the previous time interval $k_i-k_{i-1}$. From now on we assume that $pr\ll 1, m\ll M$. The average $h$-value is then given by the average-power balance as $\ave{h}=\de z/pr$, and the average dissipated energy for a molecule-picking period is $\epsilon=mg\ave{h}=mg\de z/pr$. The pendulum lifetime may now be evaluated as  $\tau_p=M/(pr~m)$. As far as fluctuations are concerned, note that if a molecule-picking event occurs at $t=0$ and the next one at an anomalously-large time  $t\gg 1/pr$, the pendulum mass height is much incremented, and the event at time $t$ dissipates a large energy. This is how one can explain in a qualitative manner the mechanism behind dissipation regulation

The energy dissipated from $k=0$ up to $k=K$ may be written $E_d=\sum_{k=0}^K \mu_k\de(k-k_i)$, where  $\de(k-k_i)=1$ if $k=k_i$ and zero otherwise. The spectrum of $P_d(k)\equiv E_d(k)-E_d(k-1)$ may then be obtained from a Fourier transform with respect to $k$. The result is compared to the analytical result given below. The dissipated power may be measured by allowing the molecules to fall back to the $z=0$ level and letting them dissipate their kinetic energy into a bolometer, a device that measures powers through temperature changes with a time constant of a few seconds. The masses $m$ raising is analogous to the charging of an optical detector battery. Regulation is observed only when an averaging is performed over a large time duration.

We have performed a numerical simulation of the marked Poisson process outlined above for $M=1$kg, $L=g/(2\pi)^2\approx0.25$m, where $g=$9.81 ms$^{-2}$, so that $T=1$s. Further, we set $\de z=1\mu$m, so that $\de E=Mg\de z\approx10^{-5}$j, $m=0.001$kg, $pr=0.01$. We obtain for the average pendulum mass height $h= \de E/(pr~m~g)\approx10$cm, $\ave{E}=Mgh\approx 1$j and $\tau_p=M/(pr~m)\approx10^{5}$ s. The numerical result for the spectral density of the dissipated power fluctuations is shown in Fig. \ref{pendulum}, and compared to the theoretical result given below, in which we set $\tau_p=10^{5}$s, $\epsilon=mgh\approx10^{-3}$j, $\ave{P_d}=pr~\epsilon=P=\de E\approx 10\mu$W.

An analytical formula for the spectral density of the dissipated power fluctuation $\De P_d(t)$ may be obtained when the fluctuations about the average value are small. On the basis of calculations analogous to those given later on for lasers, we obtain
\begin{align}
\label{intro}
\spectral_{\De P_d}=\frac{\bigl(\Om\tau_p\big)^2}{1+\bigl(\Om\tau_p\big)^2}\epsilon \ave{P_d},
\end{align}
where $\Om$ denotes the Fourier frequency. In the present numerical application, we have $\epsilon=10^{-3}$j,  $\tau_p=10^{5}$s, and $\ave{P_d}\approx 10\mu$W. As the figure shows, there is a very good agreement between the above analytical formula and the numerical simulation of the marked Poisson process considered.

Let us consider now generators of electromagnetic waves. The only difference that exists between a microwave oscillator such as a reflex klystron and a laser relates to the different electronic responses to alternating fields. In a microwave tube the electron motion is usually not harmonic and its coupling to a single-frequency electromagnetic field may be understood accurately only through numerical calculations. In contradistinction, masers and lasers employ basically two-level molecules or atoms, and this results in simplified treatments\footnote{For two-level atoms, upward electron jumps (stimulated absorption) and downward jumps (stimulated emission) may be treated symmetrically according to the time-dependent Schrödinger equation. Strictly speaking, the two-level approximation holds rigorously only for electrons immersed in a magnetic field, the lower energy state corresponding to the case where the electron magnetic moment points in the direction of the field and the higher energy state corresponding to the electron magnetic moment pointing in the opposite direction. In the case of atoms the electron energy is bounded from below but may extend to arbitrarily large values. The symmetry between stimulated emission and stimulated absorption therefore rests on the approximation that two levels only are important. In particular, the scattering states are ignored. The two-level approximation may cause apparent violation of oscillator-strength sum rules. These difficulties are un-consequential in the present theory.}. The phenomena of stimulated emission and absorption are essentially the same for every oscillator. The noise properties are also similar. Let us quote the Nobel-prize winner W. E. Lamb, Jr. \cite[p. 208]{Lamb2001}: "Whether a charge $q$ moving with velocity $v$ in an electrical field $\E$ will gain or loose energy depends on the algebraic sign of the product $ev\E$ [...]. If the charge is loosing energy, this is equivalent to stimulated emission. [...] In the domain of electronics, a triode vacuum-tube radio-frequency oscillator was developed by L. de Forest in 1912. This was in fact the first maser oscillator made by man". One may go one step further and assert that any sustained oscillator is a laser. The act of counting oscillations, however, requires specific arrangements. 

The lasers considered oscillate in a single electromagnetic mode in the steady state. Only stationary\footnote{A fluctuation is called "stationary" when correlations of all orders are independent of the initial time. This adjective is employed differently in the expression "stationary states" where "stationary" means that the electron wave-function modulus is time independent.} fluctuations of the currents driving the active elements are allowed. The system elements are supposed not to depend explicitly on time. 

\begin{itemize}

\item Basic set-up.

An optical set up involves three basic components. First a light source driven by an electrical current (called the pump). Second, an optical circuit involving slits, lenses, beam-splitters, resonators, and so on, which we view as being conservative, that is, free of loss or gain. Third, light detectors delivering photo-currents. Light sources deliver optical power while light detectors absorb optical power. Ideally, the detector photo-currents could be employed to pump the light sources so that the complete system could operate in an autonomous manner. Such equilibrium configurations will be discussed in subsequent parts. A two-state electronic system, with energy-separation $\hbar\om_o$,  exhibits a positive conductance and absorbs power if the applied potential $U$ is slightly lower than $\hbar\om_o/e$, and exhibits a negative conductance and emits power if the applied potential slightly exceeds $\hbar\om_o/e$. In practical configurations there is therefore a slight irreversible loss of energy,  which may be carried away by acoustical waves from the electronic configuration. This energy loss, in the meV range, should not be confused with the large irreversible loss of energy that, according to the Quantum Optics view-point, occurs when a photon is radiated away into vacuum from an atom. The latter is on the order of 1 eV. 

In many experiments, we only need to know time-averaged photo-currents. This information suffices for example to verify that light passing through an opaque plate pierced with two holes exhibits interference patterns. The experiment is performed by measuring the time-averaged photo-currents issued from an array of detectors located behind the plate. Other experiments involving the transmission of information through an optical fiber require that the fluctuations of the photo-current about its mean be known. A light beam carries information if it is modulated in amplitude or phase. Small modulations may be obtained  from the present theory by ignoring the noise sources, but they are not discussed explicitly for the sake of brevity. The information to be transmitted is corrupted by natural fluctuations (sometimes referred to as "quantum noise"). 

We restrict ourselves to stationary non-relativistic configurations. That is, the free-space permeability $\mu_o$ is set equal to zero, or, equivalently, the speed of light in free space, $c$, is set at being infinite. These quantities therefore nowhere enter into the theory, and questions having to do with special relativity are irrelevant. More precisely, electron velocities $v$ are much smaller than $c$ and transition frequencies $\om$ are much smaller than $mc^2/\hbar$, where $m$ denotes the electron mass. We acknowledge that under these conditions some atomic properties are being overlooked. Relativistic effects are, for example 1) the apparent increase of the electron mass, 2) the value of the electron magnetic moment $\mu=\mu_B$ derived from the Dirac equation, and 3) the spin-orbit energy splitting. This splitting, which results from the fact that, crudely speaking, atomic electrons perform circular motions at velocity $v$ in nuclei electrical fields and thus "see" magnetic fields, is in fact small in hydrogen atoms, but becomes important for heavier atoms because $v/c$ is not negligible. Quantum Electrodynamics enables Physicists to evaluate: 4) the (Lamb) energy splitting between $2P_{1/2}$ and $2S_{1/2}$ hydrogenic states, 5) the correction $1+\alpha/2\pi...$ to the electron magnetic moment, where the fine-structure constant $\alpha\approx 1/137$ is set equal to zero in the non-relativistic approximation, 6) the radiative decay of excited-state atoms (in that case, however, indirect approximate methods based on Statistical Mechanics or the Classical Maxwell Equations with retarded potentials may be employed), and 7) the Casimir force. 

On the other hand, in previous semi-classical theories the spontaneously emitted field is considered to be the fundamental source of noise. The classical optical field is supposed to be incremented by the field spontaneously emitted by upper-state atoms with a phase uniformly distributed between 0 and $2\pi$ (hence the randomness). Instead, we view noise as basically originating from \emph{stimulated} electron jumps from one state to another, and spontaneous electronic decay is neglected for simplicity in the major part of the paper.

\item Non-fluctuating driving currents.

 We almost exclusively consider laser diodes driven by constant (non-fluctuating) electrical currents. Such currents may be obtained from a battery or a large charged capacitance and a cold series resistance. This conclusion follows from the Nyquist formula derived from Classical Statistical Mechanics that says that at material absolute temperature $T_m=0$K no fluctuations are involved in an equilibrium state. Detailed analysis shows that the Nyquist formula holds when a steady current flows through the resistance as long as the Ohm law remains applicable. Alternatively, we may generate non-fluctuating currents from space-charge-limited cathodic emission. It is now-a-day possible to inject in a device one electron at a time. A discrete realistic picture of a non-fluctuating current is accordingly the regular injection of electrons, say one every nano-second, if only small Fourier frequencies are being considered. This discrete picture is employed in numerical simulations. We may also employ a very large inductance with a current flowing through it.
 
\item Non-fluctuating radiation.

It occurred as a surprised to the physics community when Golubev and other \cite{golubev:JETP84} proved theoretically in 1984 on the basis of the Quantum Optics laws that lasers driven by a quiet pump (e.g., a non-fluctuating current) deliver sub-Poissonian (or "quiet") photon streams. From our viewpoint, this observation would be better expressed by saying that when a laser is driven by a non-fluctuating current and the output light is incident on a photo-detector, the photo-current does not fluctuate much. In the latter formulation the notion of laser-light statistics is being by-passed. The above prediction then may be viewed as a strictly classical result, resulting from the law of conservation of the average energy, as we discuss below. What is \emph{non-classical } (i.e., quantum in nature) are the shot-noise fluctuations. This so-called "Schottky  effect" has been observed long ago in vacuum tubes. This is perhaps for this historical reason that the Schottky  effect is often referred to as being a "classical effect". But because it originates from the discreteness of the electric charge, it should be viewed instead as an intrinsically quantum effect. If one considers integration times large compared with the duration between successive photo-electrons, the discrete character of the electrical charge flow tends to be washed out, the theory becomes classical in nature, and accordingly a non-fluctuating photo-current is obtained\footnote{It is interesting to note that similar concepts (relating this time to the conservation of the average angular momentum rather than to the average energy) were recently advanced by C.S. Unnikrishnan \cite{Unnikrishnan2005}. That author shows that if a pair of electrons in the singlet state is emitted and their magnetic moments are detected at separate locations at angles differing by $\theta$, the only correlation consistent with conservation of the average angular momentum is the quantum result -$\cos(\theta)/4$, if the readings are normalized to $±1/2$. What is strictly "quantum" is the discreteness of the electron spin. In the large-spin $S$ limit the correlation $-\cos(\theta)S(S+1)/3$ approaches the correlation $-\cos(\theta)S^2/3$ evaluated from classical considerations. B. d'Espagnat, though challenging Unnikrishnan's interpretation of Bell's results, seems to agree with his factual conclusions \cite{dEspagnat}. }. 

\item Law of average-energy conservation.

Let us explain in some detail how the law of average energy conservation is being employed. The electrical pump raises electrons initially in the absorbing state at rate $R_e(t)$ and thus  supplies a power $\hbar\om_e R_e(t)$, where $\hbar\om_e$ denotes the electrons transition energy. (Note that electronic states are often referred to as "atomic states". Because there exists now truly atomic lasers, the distinction is important if confusion is to be avoided). When these electrons decay back, the energy in the optical resonator is incremented. The concept of "light energy" is understood here only in a restricted sense. In order to determine the energy contained in a laser resonator at some time, say $t=0$, one may cut-off the pump and measure the number of subsequent photo-detection events. It should be noted, however, that semiconductors (incorporated in particular in laser diodes) contain some energy of their own that cannot easily be separated out from the field energy. 

Even though the optical field is not quantized here, the word "photon" is employed occasionally as another name for the energy of loss-less resonators. Precisely, the resonator energy is written as $(m+1/2)\hbar \om_o$, where $m$ is called the number of photons in the resonator, $\om_o$ the resonator (angular) frequency, and $\hbar$ the Planck constant (divided by $2\pi$). Likewise, the word "photon rate" is another name for electromagnetic power divided by $\hbar\om_o$. Conversely, the energy in the optical resonator is employed to raise detecting electrons initially in the absorbing state, with transition energy $\hbar\om_a$, to the upper state at a rate $R_a(t)$. This power is delivered to the external load, perhaps followed by an electronic amplifier. Ideally, we have $\hbar\om_a\approx\hbar\om_e\approx \hbar \om_o$, in which case the source-detector configuration may be viewed as being reversible\footnote{This situation is analogous to that of the reversible heat engines discovered by Carnot in 1824. Reversibility occurs when bodies are contacted only when their temperatures are nearly the same. As Carnot acknowledged, a non-zero power (non-zero heat flow) occurs only when there is some temperature difference between the contacted bodies. However, it is legitimate to consider the limit in which this temperature difference tends to zero. If this is the case, the mechanical energy delivered \emph{per cycle} tends to a well defined limiting value. Cycles are then very slow and the power generated is very small. }. The law of conservation of energy then says that the integral from $t=0$ to $t=T$ of the power difference $\big(R_e(t)-R_a(t)\big)\hbar \om_o$ is equal to the system energy increment from $t=0$ to $t=T$, consisting of electronic and field energy increments. These, however, are finite. It follows that in the limit $T\to \infty$, we must have $R_e(t)\approx R_a(t)$. More precisely, $\lim_{T\to\infty} \ave{R_e(t)}_T-\ave{R_a(t)}_T=0 $, where the substricts $T$ refer to average values taken over a time duration $T$. In the Fourier domain, this means that the input and output rate spectral densities must be the same in the limit $\Om\to 0$.

In the case of laser diodes, the number $n$ of electrons in the conduction band fluctuates as a consequence of the laser-diode dynamics. The Fermi-Dirac law then tells us that the potential applied to the current-driven diode fluctuates, and should be written as $U(t)$. If $J$ denotes the non-fluctuating pump current the input power $J U(t)$ is no longer a constant. However, detailed calculations show that the fluctuations of $U$ have a negligible effect on the energy balance, so that the previous argument still holds.

\item Circuit representation.

The configurations investigated in this paper are described in terms of conservative elements such as capacitances and inductances, whose values may be obtained from separate classical measurements, as is done is conventional electronics\footnote{The evaluation of a capacitance from its geometric dimensions is straightforward. If one insists in evaluating inductances from their geometric dimensions one needs suppose that they contain electrons. The latter have magnetic moments and the magnetic permeability $\mu$ may be much larger than $\mu_o$ just above the Curie temperature $T_C$, being of the form $(T-T_C)^{-1.6}$. In that case a non-zero inductance is obtained even though $\mu_o$ is set equal to zero.}, and positive and negative conductances. 

\item Average conductances.

To define conductances we will consider first a single electron treated according to the Schrödinger equation. In atoms, electrons are submitted to the static field of nuclei. We suppose instead that the electron is located between parallel perfectly-conducting plates at the same potential, pierced with holes, with plates at a negative potential outside. This configuration provides a zero field, while nuclei fields are of the form $1/r^2$ at a distance $r$ from the nucleus. The difference affects the details of the wave-functions, but not the principles.

Let us sketch the way the Schrödinger equation is being applied. The one-electron wave-function $\psi(x)$ describes an ensemble of identically-prepared systems. According to Born, $\abs{\psi(x,t)}^2$ denotes the probability density of finding the electron at $x$ if a position measurement is performed at time $t$, and $\abs{\psi(p,t)}^2$ denotes the probability density of finding the electron momentum as $p$ if a momentum measurement is performed at time $t$, where $\psi(p)$ is essentially the Fourier transform of $\psi(x)$. The electrons are submitted, besides static electrical fields, to electrical fields oscillating at some optical frequency $\om$. We only consider the resonant case, and obtain the usual Rabi oscillations. 

The Quantum Mechanical (QM)-averaged induced current $\ave{i(t)}$ is proportional to the QM electron average momentum $\ave{p(t)}$ obtained by integrating $p\abs{\psi(p,t)}^2$. The QM-average current $\ave{i(t)}$ is the product of a sinusoidal variation at the optical frequency and a much more slowly-varying envelope. Because both $v(t)$ and $\ave{i(t)}$ vary essentially at the optical frequency, the QM-average power $\ave{P(t)}=v(t)\ave{i(t)}$ may be further averaged over an optical period, introducing a factor 1/2. This power, denoted simply as $P(t)$, delivered by the alternating potential, serves to increment the electron energy and possibly the static potential source energy. Note that these energies have signs, so that the energy "delivered" may in fact be an energy received.

We first consider configurations in which the electron interacts with the field only for a finite time $\tau$, perhaps because it is flying between the plates at some fixed speed. If the electron enters in the interaction region in the lower state, it has some probability of being in the upper state when it leaves the interaction region. If this is the case, the electron energy may be delivered to some external collector. Such considerations lead to an expression for the average conductance $G$ "seen" by the optical potential applied to the plates (ratio of the induced current and applied potential), which is positive in the case just described, but may be negative if the electron enters in the interaction region in the upper state. For small values of the transit time $\tau$ compared with the Rabi period, the conductance $G$ does not depend on the strength of the optical potential. That is, the system is linear. 

The flying electron configuration just described is however not the one that we consider in the major part of this paper. We consider instead an electron present all the time in the interaction region, but ascribe to it a probability $2\gamma_2\rho_{22}$ of spontaneous transition from the upper to the lower level, and a probability $2\gamma_1\rho_{11}$ of spontaneous transition from the lower to the upper level, where $\rho_{11},\rho_{22}$ are the probabilities that the electron be in the lower or upper states, respectively, and $\gamma_1,\gamma_2$ are non-negative parameters. Supposing that $\gamma_1=0,\gamma_2\equiv \gamma$, the power $P(t)$ delivered by the optical potential serves to increment the electron energy as before, \emph{and} to supply energy to the static potential. Given the $\gamma$-parameters, we may evaluate the conductance "seen" by the optical potential in the steady-state, that is, for $t\to\infty$. This conductance is independent of the optical potential when $\gamma\gg\Om_R$, where $\Om_R$ denotes the so-called Rabi frequency. To avoid a confusion, let us note that spontaneous decay is usually viewed an irreversible loss of energy from an atom electron. In the present configuration, energy is simply transferred from the optical potential to the static potential, or the converse, the electron playing an intermediate role. To treat conveniently the situation presently described one must introduce mixed-state density matrices.

To conclude, the system considered is described by a \emph{circuit}, consisting of interconnected (conservative) capacitances, inductances, and positive and negative conductances. More generally, we may allow conductances to depend on parameters such as the optical frequency $\om$, the number $n$ of electrons in the conduction band for semiconductors or the number $n_e$ of atoms in the emitting (upper) state for atomic lasers, on the emitted photon rate $R$, or on the medium temperature $T_m$, or strain in solids.

\item Nyquist noise at optical frequencies.

As hinted above, optical set-ups are viewed as black boxes characterized by in-going and out-going photo-currents, whose statistical properties are either prescribed or sought for. Once a medium has been described by a circuit we are concerned with potentials and currents varying at, or near, some optical frequency $\om_o$. These will be called "optical potentials", $V(t)$, and "optical currents", $I(t)$, respectively, to distinguish them from static potentials, $U$, and slowly-varying currents, $j(t)$. We introduce optical potentials (or electric fields) and optical currents (or magnetic fields) for the sole purpose of ensuring that the photo-currents conserve the average energy in the sense explained above. 

One needs the quantum form given by Nyquist in his celebrated paper, supplemented by the term $\hbar\om_o/2$ previously suggested by Planck\footnote{Planck actually treated \emph{atoms} rather than field resonances as quantized oscillators, so that the nature of his contribution on that subject remains unclear.}. The complete formula will be referred to as the "Nyquist-like" formula. An experimental verification of that formula at a temperature of 1.6K is illustrated for example at the beginning of Gardiner's book on Quantum Noise \cite{Gardiner2000}. As a matter of fact, only the Planck term is important in the major part of this paper because we suppose that absorber atoms are all in the lower state (T=0K) while all the emitter atoms are in the higher state (complete population inversion). Various methods will be presented showing that, in that case, the induced-current-fluctuations spectral density is equal to $\hbar\om_o G$, where $G$ denotes the absolute value of the conductance. Current-noise sources are independent of one another. The Nyquist current noises may be supposed to be normal, i.e., jointly gaussian distributed\footnote{Within our linear or linearized approximations the noise currents are therefore also normal and cross-correlations of any order may be obtained from second-order cross-correlations. It follows that measurable noise currents are time reversible, and thus do not reflect the fact that the optical circuit elements are causal.}.

\item Dependence of $G$ on frequency $\om$.

In general, the conductance $G$ depends on frequency, as recalled above. A well-known theorem says that in the linear regime the Nyquist formula is applicable to frequency-dependent conductances $G(\om)$, as long as the temperature is uniform. 

\item Dependence of $G$ on the number $n$ of electrons.

Both the average conductance $G$ and the spectral density of the induced-current fluctuations are proportional to the number $n$ of electrons as long as these electrons are not coupled directly to one another. In semiconductors the electrons are directly coupled to one another through the Pauli exclusion principle, and the conductance $G$ depends non-linearly on the number $n$ of electrons. At an electron temperature $T_m=0$K (roughly equal to the semi-conductor temperature) a gain proportional to $n^{1/3}$ would be appropriate. However, because the fluctuations of $n$ are small, a linearized conductance of the form $G(n)=G_o+\bigl(dG/dn\bigr)\bigl(n-n_o\bigr)$ may be employed. The spectral density of the Nyquist-like noise sources are supposed to be unaffected by the small deviations of $n$ from $n_o$.

\item Dependence of $G$ on the emitted (or absorbed) rate $R$.

In semiconductors the conductance $G$ may depend significantly not only on the number $n$ of electrons in the conduction band, but also, explicitly, on the emitted power. This effect is called here "gain compression" (another name is "non-linear gain"). It plays a significant role in laser-diodes operation, increasing in particular the laser-diode relaxation-oscillation damping. The usual Nyquist-like formula remains applicable in special circumstances only.

\item Dependence of $G$ on temperature.

The Fermi-Dirac statistics tells us how the electron energy distribution broadens at the electron temperature increases, an effect that tends to lower the conductance. The temperature depends in turn on the pump power, the material heat capacity and on thermal conductances.

\item Linear and linearized regimes.

 Only two limiting cases will be considered, namely the linear regime and the linearized regime. In the linear regime optical potentials and currents are proportional to the fundamental noise sources. The response of linear systems to specified sources is straighforward, but dispersion effects need investigation. This regime is applicable to lasers below the so-called "threshold" driving current and, usually, to attenuators and amplifiers. 

The linearized regime is applicable to well-above-threshold lasers. In the linearized regime one first needs evaluate average optical potentials and currents ignoring the noise sources. This is the so-called "steady state". Next, one supposes that the \emph{deviations} of the optical potentials and currents from their average values, denoted by $\De$, are proportional to the fundamental noise sources. The latter enter again when flowing powers are being evaluated, that is, current noise sources are not given for free, so to speak, but they do enter in the power balance. This is because we take this effect into consideration that our theory differs from previous semi-classical theories. 

The intermediate situation in which the system is neither linear nor can be linearized that may occur for closed-to-threshold lasers is not considered. As said above, we treat only the stationary regime found when a laser is driven by a constant current, possibly supplemented by stationary fluctuations, and no element is explicitly time-dependent, in which case photo-detection events are stationary as well. 

We assume that the atomic polarization may be adiabatically eliminated, so that our equations involve only the optical field, proportional to the optical potential $V$, and the numbers of electrons in various levels. The latter derive from rate equations that may sometimes be simplified further by neglecting electron-population time derivatives ("slaving principle"). Because spontaneous decay plays only a secondary role in our theory it is ignored for the sake of simplicity in this introductory part.

\item Potential fluctuations and correlations.

In laser diodes employing semi-conducting materials, a constant-current drive $J$ entails a static potential $U$ across the diode that slightly exceeds $E_g/e$, where $E_g$ denotes the semiconductor energy gap, because the bottom of the conduction band is filled up with $n$ electrons, according to the Fermi-Dirac distribution. Likewise, there are $n$ holes at the top of the valence bands. The rate equations that we shall introduce later on involve random fluctuations of $n$, and thus fluctuations $\De U$ of the potential $U$. This fluctuation is very small, yet measurable. One may also measure the correlation between $\De U$ and the detected current fluctuation $\De J$. This correlation may be defined in such a way that it is independent of any linear optical loss that may occur between the laser and the detector. From our view-point, $\De U$ is a small secondary effect that may, initially, be neglected.

\item Light spectrum.

The light spectrum is a well defined quantity. To observe it, is suffices to insert between the laser and the photo-detector a narrow-band, cold and linear filter whose response is centered at some frequency $\om_m$. The average photo-current $\ave{J(t)}$ is proportional to the light spectral density $\spectral(\om_m)$. 

In the linear regime, the light spectrum may be evaluated from the modulus square of the system response to the Nyquist-like noise sources. Instead of using a narrow linewidth filter as said above, the light spectrum may be derived from the photo-current spectrum, the latter being the auto-convolution of the light spectrum.

In the \emph{linearized regime}, the light spectrum may be evaluated by first neglecting amplitude fluctuations and considering frequency fluctuation $\De\om(t)$, the latter being defined from the time derivative of the phase fluctuations of the optical wave incident on the photo-detector, which may be evaluated from the linearized-system response to the Nyquist-like noise sources. Experimentally, frequency noise may be converted to photo-current noise if the detector is preceded by frequency-selective optical circuits. A dual-detector arrangement is advisable.

\end{itemize}

\paragraph{Key results.}

When light originating from a laser diode driven by non-fluctuating electrical currents is incident on a photo-detector, the photo-current does not fluctuate much, as we emphasized earlier. Precisely, this means that the variance of the number of photo-electrons counted over a large time interval is much smaller that the average number of photo-electrons. As we shall see, this is a consequence of the law of average energy conservation. Lasers having that property are called "quiet lasers". Viewed in another way, at high power, the photo-current reduced power spectrum (the adjective "reduced" meaning that the spectrum singularity at $\Om=0$ has been removed) is of the form $\Om^2/(1+\Om^2)$, where $\Om$ denotes the Fourier frequency, and thus the spectral density vanishes at $\Om=0$. This conclusion is equivalent to the one given earlier concerning the photo-count variance.  We will say that light is sub-Poissonian when the spectral density of the photo-current is less than the average rate at small Fourier frequencies. Note that some authors say instead that light is "sub-Poissonian" when its normalized correlation (to be later defined) at zero time delay is less than unity. The two definitions are in general non-equivalent. The Quantum-Optics view-point is that quiet light is "non-classical", while, for the reasons explained earlier, we view quiet radiation as being entirely classical.

The conclusion that for quiet lasers photo-current spectral densities vanish at zero Fourier frequency holds as long as the elements involved are \emph{conservative}. Accordingly, the conclusion holds irrespectively of dispersion (related to the so-called "Petermann K-factor"), of the value of the phase-amplitude coupling factor (introduced independently in 1967 by Haken and Lax and usually denoted by $\al$), and of the amount of gain compression (introduced by Chanin and alternatively called "non-linear gain"). These effects do affect however the photo-current spectral density at non-zero Fourier frequencies, the laser linewidth, and other laser properties. Note that conventional vacuum tubes with space-charge-limited cathode emission such as reflex klystrons should also emit quiet electromagnetic radiation. We do not know whether this has actually been observed, nor whether it can be observed in consideration of the klystron modest efficiency, and of thermal or flicker noises.

\paragraph{Organization of the paper.}

Besides the introduction and the conclusion, this paper consists of six sections. The first one gives an account of the most relevant results in Physics. The second one lists mathematical results relating to deterministic or random functions. The third one is a discussion of the Circuit Theory. The fourth outlines the Classical and Quantum equations of electronic motion. The fifth presents a more general theory of electron-field interaction, based on spontaneous electronic transitions. The sixth offers alternative methods of establishing that the spectral density of Nyquist-like noise sources associated with a conductance is proportional to the absolute value of the conductance. 

Aside from historical works, many citations relate to our own work (1986-2006). It is our intention to provide in later versions a more comprehensive list. Some important references, not cited here, may be traced back, however, from the more recent papers and books cited.

\newpage

\section{Physics}\label{history}

According to the latin poet Lucretius, a follower of Democritus, there are no forbidden territories to knowledge: "...we must not only give a correct account of celestial matter, explaining in what way the wandering of the sun and moon occur and by what power things happen on earth. We must also take special care and employ keen reasoning to see where the soul and the nature of mind come from,...". And indeed, the three most fundamental questions: what is the origin of the world? what is life? what is mind? remain subjects of scientific examination. Needless to say, the present paper addresses much more restricted questions.

We will first recall how Physics evolved from the early times to present, no attempt being made to follow strictly the course of history. The theory of light or particle motion and the theory of heat followed independent paths for a long time. The Einstein contributions proved crucial to re-unite these two fields early in the 20th century. We may distinguish "pictures" based on our in-born or acquired concepts of space and time that may not answer all legitimate questions nor be accurate in every circumstances, and complete theories. Quantum theory is considered by most physicist as being accurate and complete, although some questions of interpretation remain hotly debated. We will consider in some detail the theory of waves and trajectories that are essential to understand the mechanisms behind vacuum-tube and laser operation. We also offer view-points concerning the Quantum Theory of Light.

\subsection{Early times}\label{early}

From the time of emergence of the amphibians, earth, a highly heterogeneous stuff, is our living
place. On it, we experience a variety of feelings. We feel the pull of gravity, breath air, get heat from the fire and the sun, and feed on plants growing on earth and water. Our experience, both as
human beings and as physicists, is based on these living conditions. One may presume that natural selection led human beings to an intuitive understanding of geometrical-physical-chemical quantities such as space, time, weight, warmth, flavor, and so on. At some point in the evolutionary process a degree of abstraction, made possible by an enlarged brain, facilitated our fight for survival. An example of abstract thinking is the association with space of the number 3, corresponding to the number of perceived dimensions. People "in the street" may however wish to distinguish the two horizontal-plane dimensions and the vertical dimension, considering that, for the latter, up and down are non-equivalent directions. It was not appreciated in the ancient times that the distinction between "up" and "down" is caused by the earth gravitational field, and that people living on the other-side of the earth have the same feelings as we do in their every-day life, even though, with respect to our own reference frame, they are "up-side-down". As we shall see, analogous considerations may apply to time, according to Boltzmann.

 Another naturally evolving concept is indeed the distinction between past and future and physical causality: matter acts on matter only at a later time. The so-called "arrow of time" is a much debated subject. According to Boltzmann, in an infinite universe, there may be large-scale spontaneous fluctuations of the entropy (that one may crudely describe as expressing disorder). Past $\to$ future would correspond to the direction of increasing entropy. There may be times where the entropy decreased instead of increasing. But the distinction is purely a matter of convention (in analogy with the "up and down" distinction mentioned above). This view point is consistent with the fact that the fundamental equations of Physics are (with the exception of the rarely occurring neutral-kaon decay) invariant under a change from $t$ to $-t$. There has been objection to the Boltzmann view-point, however, and most recent authors would rather ascribe the time arrow to cosmic evolution, with the universe starting at the "big-bang" time in a state of very low entropy. 
 
In contrast with the rational view concerning causality, the magic way of thinking presupposes the existence of causal relationships between our desires, fears, or incantations, and facts. Now-a-days, magic thinking co-exists with rational thinking probably because it gives people sharing similar beliefs a sense of togetherness and helps a few individuals acquire authority and power. The consequences of irrationality are often too remote to be of concern to most. 

The control of fire by man some 500 000 years ago and drastic climatic changes that occurred, mainly in Europe, some 23 000 years ago, trigerred evolutionary events. Likewise, the practice of growing crops made possible a population explosion some 10 000 years ago, particularly in Egypt, and gave an incentive for measuring geometrical figures, precisely accounting for elapsed times, and measuring weights. Let us now consider more precisely what is meant by matter, space and heat. 

Empedocle ($\sim$500 BC) viewed the world as being made up of four elements, namely earth, water, air and fire. These elements remain a source of inspiration for poets and scientists alike, but they are not considered anymore as having a fundamental nature. Democritus ($\sim$400 BC) pictured reality as a collection of interacting particles that cannot be split ("a-toms"). Aristotle wrote in his Metaphysics VIII: "Democritus apparently assumes three differences in substances; for he says that the underlying body is one and the same in material, but differ in shape, position, and inter-contact". This picture may still be viewed as being basically accurate. 

The present work is not concerned with the cosmos per se. Yet, one cannot ignore that observations of the sky have been a source of inspiration in the past and remain very much so at present. Early observers distinguished stars from planets, the latter moving apparently with respect to the former. The ancient Greeks (Ptolemeus) conceived a complicated system of rotating spheres aimed at explaining the apparent motion of these celestial objects. Aristarque (310-230 BC), however, realized that the earth was rotating about itself and about the sun, the latter being considered to be located at the center of the universe. This \emph{heliocentric} system was rediscovered by Copernic (1473-1543) and popularized by G. Bruno (burned at stake in Rome in 1600 for heresy). Next came the establishment of the three laws of planetary motion by Kepler, the dynamical explanation of these laws by Newton, which involves a single universal constant, namely $G$. The deeper theory proposed by Einstein in 1917 (General Relativity) appears to be in good agreement with observations. It involves two fundamental constants, namely $G$ and $c$.

When two bodies are in thermal contact they tend to reach the same temperature. Thus, two differently constructed thermometers may be calibrated one against the other by placing them in the same bath and comparing their readings. In the case of thermal contact the hotter body loses an amount of heat gained by the colder one but the converse never occurs. It may well be that the condition of heat-engine reversibility, discovered by Carnot in 1824, could have been made at a much earlier time and could have served as a basis for subsequent developments in Physics. The present attitude is rather that one should derive the laws of Thermodynamics from Classical or Quantum theories. It may be however that, to the contrary, the latter theories cannot be formulated unambiguously without the former. 

\subsection{How physicists see the world now-a-day}\label{how}

Beyond a qualitative understanding of the nature of heat, early observers were able to perform measurements of temperature and gas pressure with fair accuracy. Temperatures were measured through the expansion of gases at atmospheric pressure, linear interpolation being made between the freezing (0°C) and boiling (100°C) water temperatures. The concept of absolute zero of temperature emerged through the observation that extrapolated gas volumes would vanish at a negative temperature, now known to be -273.15°C=0 kelvin. The Classical Theory of Heat was established in the 18th and 19th centuries mainly by Black, Carnot and Boltzmann. The major contribution is due to Carnot (1824) who introduced the concept of heat-engine \emph{reversibility}. The fact that hot bodies radiate power was known very early (some reptiles possess highly-sensitive thermal-radiation detectors). It is however only in the 19th century that the proportionality of the total radiated power to the fourth power of the absolute temperature was established. Difficulties relating to the theory of blackbody radiation led Einstein around 1905 to the conclusion that Classical Physics ought to be replaced by a more fundamental theory, namely the Quantum Theory, even though important conclusions may be reached without it. Another motivation for studying in some detail the theory of heat is that lasers are in some sense heat engines. They may be ``pumped'' by radiations originating from a hot body such as the sun. But, just as is the case for heat engines, a cold body is also required to absorb the radiation resulting from the de-excitation of the lower atomic levels. Lasers are able to convert heat into work in the form of radiation, but their efficiency is limited by the second law of thermodynamics. Output-power average values and fluctuations may be similar for lasers and heat engines. 

The grand picture we now have is that of a world 13 billions years old and 13 billions light-years across containing about 10$^{11}$ galaxies. Apparently, 80 \% of matter is in a dark form, of unknown nature, that helped galaxy formation. Our own galaxy (milky way) contains about $10^{11}$ stars and possesses at its center a spinning black hole with a mass of 3.5 millions solar masses. Eight planets (mercury, venus, earth, mars, jupiter, saturne, uranus, neptune) are revolving around our star (sun).  Penzias and Wilson discovered in 1965 the cosmic background microwave radiation, which accurately follows the Planck law for a temperature of 2.73 kelvins. This cosmic black-body radiation is almost isotropic. Yet, minute changes of intensity according to the direction of observation have been measured, which provide precious information concerning the state of the universe some 300 000 years after the "big-bang". Numerous observations relating to ordinary stars such as the sun, neutron stars, quasars, black holes are particularly relevant to high-energy physics. It is expected that gravitational waves emitted for example by binary stars or collapsing stars will be discovered within the next ten years or so. Their detection may require sophisticated laser interferometers operating in space. In such interferometers, laser noise plays a crucial role. Reactors aim at creating on earth conditions similar to those occurring in the sun interior, i.e., temperatures of millions of kelvins, and to deliver energy, perhaps by the year 2050. An alternative technique employs powerful lasers shooting at a deuterium-tritium target. A reduction of the laser-beam wave-front fluctuations are essential in that application. For a review of the present views concerning the universe, see for example \cite{Gleiser}.

\subsection{Epistemology}\label{epis}

Epistemology is the study of the origin, nature, methods and limits of knowledge. Undoubtedly, Physics is an experimental science. Its purpose is to predict the outcome of observations, or at least average values of such observations over a large number of similar systems, from few principles using Mathematics as a language. Observations are required to set aside as much as possible human subjectivity. This is done by performing a large number of "blind" experiments, the same procedure being repeated again and again in independent laboratories. A physical theory should be "falsifiable", that is, one should be able to realize, or at least conceive, an experiment capable of disproving it. 

The average value $\ave{a}$ of a quantity $a$ is calculated by summing $aP(a)$, where $P(a)$ is the probability density of $a$. It is apparently difficult to provide an unambiguous definition of the word "probability". Let us quote Dose \cite{Dose2003} "There is a fundamental mistrust in probability theory among physicists. The need to extract as comprehensive information as possible from a given set of data is in many cases not as pressing as in other fields since active experiments can be repeated in principle until the obtained results  satisfy preset precision requirements. [In other fields] the available data should be exploited with every conceivable care and effort". As data comes in our estimate of the probability $P(a)$ improves, and eventually approaches an objective value, defined according to the frequentists view-point. 

In practice, most scientific progresses were accomplished with the help of intuitively-appealing \emph{pictures}, describing how things happen in our familiar three-dimensional space and evolve in the course of time. These pictures are supposed to tell us how things are behind the scene, or to suggest calculations whose outcome may be compared to experimental results. Let us quote Kelvin: ``I am never content until I have constructed a mechanical model of the subject I am studying. If I succeed in making one, I understand; otherwise I do not''. But many models, helpful at a time, are often discarded later on in favor of more abstract view-points. The Democritus picture of reality has been worked out in modern time by Bernouilli, Laplace and a few others. Given perfectly accurate observations made at some time, called "initial conditions", the theory is supposed to predict the outcome of future observations if the system observed is not perturbed meanwhile. (Poincaré, however, pointed out that for some systems, e.g., three or more interacting bodies in Celestial Mechanics, the error grows quickly in the course of time when the initial conditions are not known with perfect accuracy). The equations that describe ideal motions are time reversible, so that when the system is known with perfect accuracy at a time its state in the past as well as in the future is predictable. Predictions for earlier times (perhaps a misnomer) make sense if measurements were then made but not revealed to the physicist. What we have just described is sometimes referred to as the "Classical Paradigm". 

Reality is surely a concept of practical value. Anyone wishes to distinguish reality, as something having a degree of permanency, from illusions or dreams that are transitory in nature. On some matters, the opinions of a large number of people are sought, supposing that their agreement would prevent individual failures. In that sense, reality may exist independently of observers and be \emph{revealed} by observations. But according to Bohr the purpose of Physics is not to discover what nature \emph{is}, but to discover what we can say consistently about it. We stick to the Bohr view-point that observations relate only to complete set ups, including the preparation and measurement devices, the latter being considered classical. A specific measurement device is described in \cite{Allahverdyan}. In effect, the object to be measured should be able to switch another object involving a large number of degrees of freedom from one metastable state to another. 

Advanced notions are not needed in this paper. It seems nonetheless that some understanding of the Physics conceptual difficulties is useful. Seemingly reasonable pictures may fail to agree with observations in special circumstances. Indeed, consider a source and two measuring apparatuses, one located on the left of the source and the other on the right. Using Stapp \cite{Stapp1993} terminology, apparatuses may be set up to measure either size (large or small) or color (black or white) but not both at the same time. It is observed\footnote{In reality, we are referring here to Quantum Mechanical (QM) predictions rather than to real observations. There has been, however, so many experimental observations that agree with QM, that one may overlook the fact that observations have perhaps not been made for the system presently considered.  } that (l,b), (w,w) and (b,l) never occur, where large, small, white and black have been abbreviated by their first letters. The first term in these expressions correspond to the left-apparatus outcome and the second term to the right-apparatus outcome. In writing "large", for example, we of course imply that the apparatus has been set up to measure size, while in writing "black", for example, we imply that the apparatus has been set up to measure color. Let us now attempt to explain the above observations on the basis of the following picture: Assume that there exist four kind of particles, namely (lw), (lb), (sw) and (sb). The source is supposed to shoot out one of these particles to the left and one to the right according to some probability law (there are all-together 16 probabilities summing up to unity, but only 4 of them will be considered). However, the fact that (l,b) never occurs implies that pr(lw,lb)=0. Here, "pr(lw,lb)=0" means that the source is not allowed to shoot out a particle of the kind lw on the left and a particle of the kind lb on the right. Indeed, if it were allowed to do so, the left apparatus, set up to measure size, would sometimes give "l", while the right apparatus, set up to measure color, would sometimes give"b", contrary to observation. For the same reason, the source is not allowed to shoot out "lb" on the left and "lb" on the right, a condition that we write as pr(lb,lb)=0. Next we note that the observation that (w,w) never occurs implies that pr(lw,lw)=0. Finally,  the observation that (b,l) never occurs implies that pr(lb,lw)=0. Accordingly, the probability that "l" be found on both sides is, considering the four possible combinations, pr(l,l) = pr(lb,lb) + pr(lw,lb) + pr(lb,lw) + pr(lw,lw) = 0, where the probabilities obtained above have been employed. Observations reveal, however, that if "l" is found on the left side, the probability that "l" be found also on the right side is equal to 0.065, i.e., is non-zero. It follows that the picture of a source shooting out two particles disagrees with observations. Of course, the non-zero probability quoted above (6.25 per cent) applies to elementary particles having only two attributes, each of them exhibiting only two possible values, not to macroscopic objects that may have other, measurable, attributes. It is frequently the case that an effect deemed impossible according to Classical Mechanics, for example the transmission of a particle through (or above) a barrier of greater energy, is in fact observed (tunneling). This is because it is considered impossible, even in principle, to measure particle energies on top of the barrier.

We are not concerned in the present paper with Physics in general but only with stationary configurations, so that the epistemology of that part of Physics could perhaps be made more precise. The system is allowed to run in an autonomous manner, that is without any external action impressed upon it, and there is a continuous record of some quantities, particularly the times at which photo-electrons are emitted or absorbed. Systems on which we may act from the outside are not considered. Photo-electrons may be accelerated to such high energies by static fields that no ambiguity occurs concerning their occurrence times. The question asked to the physicist then resembles the one asked to people attempting to recover missing letters from impaired manuscripts: can you determine the missing letters from the known part of the text? In the present situation one would like to be able to tell whether an event occurred during some small time interval, given the rest of the record. Or at least give the \emph{probability} that such an event occurs in the specified time interval. In other words, given a large collection of similar systems, on what fraction of them does an event occur? Instead of being given impaired records, we may be given information concerning the various components that constitute the system, such as lenses, semi-conductors, and so on, characterized by earlier, independent measurements. These measurements are deterministic in nature because they are performed in the classical high-field (yet, usually, linear) regime. In view of the observed uncertainty, spontaneous noise sources must obviously be introduced somewhere in the theory. We consider that the noise sources are located solely at emitters and absorbers, viewed as being similar in nature. Pound \cite{Pound1957} described earlier lasers in terms of a Nyquist theorem extended to negative temperatures.

\subsection{Waves and trajectories}\label{waves}

Physics courses usually first describe how the motion of masses may be obtained from the Newtonian equations. But it might be preferable to let students get first  familiarity with classical waves, for example by observing capillary waves on the surface of a mercury bath. Such waves are described by a real function of space and time that one may denote $\psi(x,t)$ in one space dimension. One reason (to be explained in more detail subsequently) to consider waves as being of primary interest is that the law of refraction follows in a logical manner from the wave concept, but does not from the ray concept. Once wave concepts have been sufficiently clarified, the many-fold connections existing between waves on the one hand, and particles or light rays on the other hand, may be pointed out. Note that, historically, the motions of macroscopic bodies and light rays were established first (around 1600) and the properties of waves later on (around 1800 for light and 1900 for particles). Few precise results concerning waves seem to have been reported at the time of the ancient Greece. Yet, casual observation of the sea under gently blowing winds suffices to reveal important features. Had such observations been made, the course of discoveries in Science would perhaps have been quite different from what actually occurred. 

Waves at the surface of constant-depth seas propagate at constant speed $u$. In realistic conditions there is some dissipation and the wave amplitude may decrease but the wave speed remains essentially unchanged. This is a striking example of a physical object whose speed does not vary, no force being impressed upon it. The only condition required is that the medium parameters (the sea depth in the present situation) do not vary from one location to another. 
 
In the 1630s Galileo observed that macroscopic objects move at a constant speed when no force is exerted on them, in contradiction with the then-prevailing Aristotle teaching. A related finding by Galileo is the principle of special relativity: The laws of Physics established in some inertial laboratory are the same in another laboratory moving at a constant speed with respect to the first. In the year 1637 Descartes proposed the following interpretation for the refraction of light rays at the interface between two transparent media such as air and water. Descartes associates with a light ray a momentum that he calls "determination" having the direction of the ray and a modulus depending on the medium considered but not on direction. He observes that the $x$-component of the momentum should not vary at the interface as a consequence of the uniformity of the system in that direction, justifying this assertion by a mechanical analogy, namely a ball traversing a thin sheet. The law of refraction asserting that $\cos (\theta_1)/\cos (\theta_2)$, where the angles are defined with respect to the $x$-axis and the subscripts 1,2 refer to the two media, does not depend on the ray direction, follows from the above concepts\footnote{The law of refraction is most commonly written as $n_1\sin(i_1)=n_2\sin(i_2)$, where the $n$ are refractive indices and the $i$ angles are defined with respect to the normal to the interface.}. Note that Descartes was only concerned with trajectories in space, i.e., he was not interested in the motion of light pulses in time, so that questions sometimes raised as to whether light pulses propagate faster or slower in air or in water are not relevant to his discussion. 

No one at the time suggested that there may be a connection between particles or light rays on the one hand, and waves on the other hand. The wave properties of light were discovered by Grimaldi, reported in 1665, and explained by Huygens in 1678. The wave properties of particles were discovered much later by Davisson and Germer in 1927. In modern terms the Galileo, Descartes (and later Newton) concepts imply that particles and light rays obey ordinary differential equations. But without the wave concept the law of refraction for light or for particles relies on observation and intuition rather than logic. 

A wave packet has finite duration but includes many wave crests. A key concept is that of group velocity defined as the velocity of the peak of a wave packet, or short pulse. In particular, what is usually called the "velocity" of a (non-relativistic) body is the group velocity of its associated wave. But usually wave packets spread out in the course of time. In the non-linear regime though, wave packets, called \emph{solitons}, may exhibit particle-like behavior in the sense they do not disperse. Bore-like solitary waves created by horse-drawn barges were first reported by Russell in 1844.

Let us be more precise about waves. As said above, waves are very familiar to us, particularly gravity waves (not to be confused with the Einstein gravitational waves) on the sea generated by wind, or capillary waves generated on the surface of a lake by a falling stone. Simple reasoning and observations lead among other results to the law of refraction. Waves are defined by a real function $\psi(x,t)$ for one space coordinate $x$, and time $t$, obeying a partial differential equation. If the wave equation is unaffected by space and time translations we may set $\psi(x,t)=f(x-ut)$ for arbitrary speeds $u$. This results into an ordinary differential equation for the function $f(x)$ which in general admits solutions. Let us begin our discussion with monochromatic (single-frequency) waves propagating in the $x$ direction in a conservative linear and space-time invariant medium. The wavelength $\lambda$ is the distance between adjacent crests at a given time. We define the wave number $k=2\pi/\lambda$. The wave-period $T$ is the time it takes a crest to come back, at a given location. We define the frequency $\om=2\pi/T$. It follows from the above definitions that the velocity of a crest, called the phase velocity, is $u=\om/k$. Such waves propagate at constant speed without any action being exerted on them. For linear waves there is a definite relationship between $\om$ and $k$ independent of the wave amplitude, called the dispersion equation. For gravity waves in deep inviscid (non-viscous) waters we have, for example, $\om=\sqrt{gk}$, where $g\approx 9.81 $m/s$^2$ is the earth acceleration. When the water depth $h$ is not large compared with wavelength (shallow water), the dispersion relation involves $h$ as a parameter  \cite[see ref.~8]{Evans2003}. 

The above considerations may be related to mechanical effects. Indeed, if a wave carrying a power $P$ is fully absorbed, the absorber is submitted to a force $F$ satisfying the relation $P/\om=F/k$. This ratio, called "wave action", depends on the nature of the wave but does not vary if some parameter is changed smoothly, either in space or in time. For a wave of finite duration $\tau$, the energy collected by the absorber is $E=P\tau$ and the momentum received (product of its mass and velocity) is $p=F\tau$. 

If the water depth $h$ is changed at some time $t=0$ from, say, 1m to 2m, it is observed that $k$ is unchanged as a consequence of the wave continuity. But invariance of $k$ implies a frequency change since the  dispersion equation depends on $h$. In that case, the wave speed changes at time $t$. Conversely,  If the water depth $h$ changes at some location $x=0$ from, say, 1m to 2m, it is observed that $\om$ is unchanged as a consequence of the wave continuity. But invariance of $\om$ implies a wave number change since the  dispersion equation depends on $h$. In that case the wave speed changes at $x=0$.

Consider now a monochromatic wave (fixed frequency $\om$) propagating in two dimensions with coordinates $x$, $y$. The direction of propagation is defined as being perpendicular to the crests and the wavelength $\lambda=2\pi/k$ is defined as the distance between adjacent crests at a given time. But one may also define a wavelength $\lambda_x$ in the direction $x$ as the distance between adjacent crests \emph{in the $x$-direction} at a given time. Let the wave be incident obliquely on the interface between two media, the $x$-axis. For gravity waves the two media may correspond for example to $h(y)=1$m$, y>0$ and $h(y)=2$m$, y<0$.  Because of the continuity of the wave, $\lambda_x$ is the same in the two media. If we further assume that the propagation is \emph{isotropic}, that is, that $k$ does not depend on the direction of propagation of the wave in the $x,y$ plane, the law of refraction follows, namely that $k_x=k_1\cos (\theta_1)=k_2\cos (\theta_2)$, where the subscripts 1,2 refer to $y>0$ and $y<0$ respectively, and the angles $\theta$ are defined with respect to the interface, that is, to the $x$-axis. The law of refraction therefore follows from wave continuity and isotropy alone. 

Questions relating to the velocity of light pulses are important for the transmission of information. A wave-packet containing many wave crests moves at the so-called "group velocity" $v=d\om/dk$, which often differs much from the phase velocity $u$ defined above. Considering only two waves at frequency $\om$ and $\om+d\om$, the relation $v=d\om/dk$ may be visualized as a kind of Moiré effect. Wave crests move inside the packet, being generated at one end of the packet and dying off at the other end. For waveguides we have $uv=c^2, v<c,u>c$. For matter waves associated with a particle the group velocity $v$ coincides with the particle velocity. Since the energy $E=p^2/\p 2m\q$ and $p=mv$, a previous relation reads $p^2/\p2m\om\q=p/k$. It follows that $u=\om/k=p/2m=v/2$. For gravity waves the dispersion relation gives instead $u=2v$. A general result applicable to loss-less waves is that the group velocity $v$ is the ratio of the transmitted power $P$ and the energy stored per unit length. It never exceeds the speed of light $c$ in free space. 

Wave solutions of the form $\psi(x,t)=\psi(x-ut)$, where $\psi(x)$ is some given function and $u$ a constant, exist also for non-linear wave equations. When the $\psi(x)$ function is localized in $x$, the invariant wave-form is called a solitary wave. In some cases, solitary waves exhibit transformations akin to those of particles when two waves collide and are called "solitons" in the sense that the soliton integrity is preserved.

As said before, most continuous media may be modeled by discrete circuits. For example, a transmission line may be modeled by series inductances and parallel capacitances. Free space may be modeled by electrical rings in which electrical charges move freely and magnetic rings in which (hypothetical) magnetic charges would move freely. If each electrical ring is interlaced with four magnetic rings and conversely, the Maxwell equations in free space obtain in the small-period limit.

Under confinement along the $x$-direction, waves at some fixed frequency $\om$ may be viewed as superpositions of "transverse modes". For a transverse mode the wave-function factorizes into the product of a transverse function $\psi(x;\om)$ and a function of the form $\exp(ik(\om)z-i\om t)$. Another connection between waves and rays rests on the representation of transverse modes by ray \emph{manifolds}. These are not however independent rays. A phase condition is imposed on them that leads to approximate expressions of $\psi(x;\om)$ and $k(\om)$. Note the analogy with Quantum-Mechanics stationary states, $z$ and $t$ being interchanged.

Thus the wave-particle connection is many fold. First the medium in which the wave propagates may be approximated by a discrete sequence of elements, for example a periodic sequence of springs and masses for acoustical waves and electrical inductance-capacitance circuits for electromagnetic waves, with a period allowed to tend to zero at the end of the calculations. One motivation for introducing this discreteness is that computer simulations require it anyway. A more subtle one is that some divergences may be removed in that way. We have mentioned above capillary waves on a mercury bath. They may be treated by considering the forces binding together the mercury molecules and their inertia, ending up with equations of Fluid Mechanics. Like-wise, acoustical waves in air may be described through the collision of molecules in some limit (isothermal or adiabatic). Second, wave modes may be described approximately (WKB approximation) by ray \emph{manifolds}. Third, one may consider the behavior of wave packets in the high-frequency limit and liken these wave packets average trajectories to those of macroscopic bodies.

We have described above the motion of light and particles in terms of waves. Semi-classical theories such as the one employed in the present paper rest indeed on such wave concepts, namely Quantum Mechanics for describing electrons, and Circuit Theory for describing the relationship between potentials and currents. The speed of light in free-space is irrelevant in that theory. Quantization then only means that electrons are identical point particles. When particles such as electrons are electrically charged they may be accelerated to arbitrarily large energies by static electrical potentials. Being then in the classical domain there is no ambiguity concerning their arrival time. Uncharged point particles such as neutrons could conceivably be accelerated similarly by gravitational fields, even though this may turn out to be difficult in practice.

\subsection{Atoms and elements}\label{el}

Around 1927 it was discovered theoretically by de Broglie and subsequently verified experimentally that a wave of wave-number $k=mv/\hbar$ should be associated with electrons of velocity $v$. An approximate solution for the motion of an electron following a closed classical path in the neighborhood of a positively charged nucleus thus amounts to prescribe that an integral number $n$ of wavelengths $2\pi/k$ fits along the closed classical path.  These discrete solutions are called "stationary states" and $n$ is essentially the principal quantum number. According to the Pauli principle, at most two electrons (with spin $±\hbar/2$) may be ascribed to each of these states. At $T_m=0$K and without excitation by other particles, only the lowest-energy states are filled with electrons. Different elements (H, He, Li...) differ by the number $Z$ of protons in their nuclei.

In the next paragraph we recall how the chemical and electronic properties of the various elements found in nature follow from the above principle, and describe what happens when atoms get closer and closer to one another to form crystals. Then we recall the basic properties of semi-conductors.  

\subsection{Electron states}\label{elements}

We summarize below the most basic concepts concerning elements found in nature and their electron states. The simplest element is the hydrogen atom consisting of a proton with an electrical charge $e$ and a mass much larger than the electron mass $m$, so that for most purposes the proton may be considered as being fixed in space. This proton attracts one electron of charge $-e$ so that the assembly is neutral. According to Classical Mechanics the electron may circle around the proton at a distance $r$ with a velocity $v$ such that the centrifugal force be balanced by the attraction from the proton, namely $mv^2/r=e^2/\p 4\pi\epsilon_o r^2\q$. From this view-point any distance $r$ may occur, the velocity $v$ being appropriately chosen. According to Quantum Theory a wave-length $2\pi\hbar/mv$ is associated with electrons moving at velocity $v$. The resonance condition is that an integral number $n$ of wavelengths fits within the electron path perimeter $2\pi r$. According to this model, due to Bohr, there is only a discrete sequence of allowed electron energies, corresponding to $n=1,2...$. The more exact theory due to Schrödinger leads to symmetrical ground states, called s-states, and anti-symmetrical 3-times degenerate first-excited states, called p-states. 

The elements found in nature (roughly 100) were classified by Mendeleïev in 1869 on empirical grounds. Helium nuclei consist of two protons, lithium nuclei of three protons, and so on, with an equal number of electrons, so that atoms are electrically neutral. (There may be various numbers of neutrons bound to the protons, which depart from the number of protons by a few units, corresponding to different isotopes, some of them being unstable. Neutrons are not considered in the present discussion). Most elements have an outer layer consisting of a number of electrons going from 1 (e.g., sodium) to 8 (e.g., neon). Particularly important are 3-5 crystals, such as gallium-arsenide.

\subsection{Semi-conductors}\label{semi}

Our purpose here is to give readers unfamiliar with solid-state physics an overview of the most important phenomena. For silicon, the number of outer electrons is 4. Two silicon atoms (or more) may bind to one another by exchanging electrons of opposite spins (covalent binding). When two atoms are approaching one another, their electronic states get perturbed. As it happens, the isolated-atom electron s-state acquires an energy greater than the isolated-atom electron p-states. For a large number $N$ of atoms, the atomic separation $a$ sets up at a value that minimizes the total energy. The original s-states then split into $N$ states that are so-closely spaced in energy that they form an almost continuous band of states called the conduction band. The original 3-fold degenerate p-states split into $N$ states that are so-closely spaced in energy that they form three almost continuous band of states called the valence bands. Because the degeneracy is lifted these three bands should be distinguished. They are called respectively the heavy-hole band, the light-hole band and the split-off band. For our purposes, only the heavy-hole band needs be considered. 

The separation in energy between the bottom of the conduction band and the top of the valence band is called the band gap $E_g$, often expressed in electron-volts. At $T$=0K, the lower-energy valence band is filled with electrons while the higher-energy conduction band is empty. At that temperature the electrons are unable to respond to an external field because no state is available to them (except perhaps at extremely-high fields). If an electron is introduced in the conduction band by some means it moves in response to an electrical field with an apparent mass $m_c$ smaller than the free-space mass $m$. If, on the other hand, an electron is removed from the valence band one says that a "hole" has been introduced. This hole is ascribed a positive charge $e$ and a mass usually larger than $m$. 

When two materials having different band gaps are contacted the band gap centers align approximately, and potential steps occur both in the conduction and valence bands. In the case of a double-hetero-junction the lower-band-gap material is sandwich between two higher-band-gap materials. The potential steps tend to confine both free electrons and free holes in the central low-band-gap material (e.g., GaAs). Being confined in the same volume electrons and holes easily interact.

As the band-gap decreases electrons may undergo virtual transfers from one band to the other more easily. As a consequence the material is more easily polarized by external (static or optical) electrical fields. In other words, the material permittivity $\epsilon(\om)$ increases as the band gap decreases. This is why the permittivity (or refractive index) of the low-band-gap gallium-arsenide is significantly larger than the permittivity (or refractive index) of the large-band-gap aluminum arsenide. When a small-band-gap semi-conductor (GaAs) is sandwiched between two higher-band-gap semiconductors (AlAs), the higher-index material may guide optical waves. This fact is important for the guidance of optical waves in laser diodes employing double-hetero-junctions. An happy circumstance is therefore that electrons, holes, and light, may all get confined in the central part of the double-hetero-junctions considered. 

Gallium possesses 3 electrons in the outer shell and arsenide possesses 5 electrons. Equal numbers of these atoms may associate to form a crystal of gallium-arsenide (Ga-As), a material particularly important in Opto-Electronics. The reason for this importance is that, unlike silicon, this is a "direct band-gap" material. In direct band-gap materials the minimum of the conduction-band energy and the valence-band maximum energy correspond to the same electron momentum. Accordingly, electrons lying at the bottom of the conduction band may get easily transferred to the top of the valence band, and conversely, the law of momentum conservation being then fulfilled (photon momentum is negligible). In such a process, an energy $E_g$ is absorbed by light through stimulated or spontaneous emission processes. Unfortunately, this energy may also be absorbed by another electron (Auger effect) that subsequently cascades down, its energy being converted into heat. 

Finally, one should say a word about doping, considering as an example a silicon crystal. When a small number of silicon atoms are replaced by arsenic atoms, these atoms, referred to as "impurities", easily deliver an electron (n-doping). Conversely, when a small number of silicon atoms are replaced by gallium atoms these atoms easily capture electrons (p-doping). A p-n diode consists of two contacting semi-conductors, one with p-doping and one with n-doping. Electron currents may be injected into p-n diodes, and in particular into double-hetero-junctions. This is the current referred to in this paper as the laser-diode driving current or the photo-current of a quantum photo-detector. It is denoted by $J$.

The above discussion hopefully provides the essential concepts that one needs to get some understanding of the electrical behavior of laser diodes. Note that we denote by $z$ the coordinate along which the optical wave propagates (junction plane) and by $x$ the direction perpendicular to the semiconductor layers. Guidance along the transverse $y$ direction is also considered.

\subsection{Detectors and sources}\label{detectors-sources}

In subsequent sections we discuss sources of electromagnetic radiations and ways of detecting them. It is appropriate to consider first detectors because there exist natural sources of radiation such as the sun, and the difficulty was initially to detect such radiations rather than to generate them. Detectors convert high-frequency radiation into slowly varying currents. The mode of operation of some detectors, called "classical detectors", may be explained on the basis of the Classical Equations of Electron Motion. For others, called "quantum detectors", the Quantum Theory of Electron Motion is required. Conversely, sources convert slowly varying currents into high-frequency radiation. The mode of operation of some sources, called "classical sources", may be explained on the basis of the Classical Equations of Electron Motion. For others, called "quantum sources", the Quantum Theory of Electron Motion is required.

Historically, the first light detector was of course the human eye. Modern detectors were first vacuum tubes operating with a low-work function cathode and accelerating potentials. There exist now quantum detectors whose mode of operation is based on the phenomenon of stimulated absorption. An early man-made generator of high-frequency radiation is a vacuum-tube called the "reflex klystron". The main light sources are hot bodies and lasers.

\subsection{Classical detectors}\label{cdetectors}

Classical detectors are diodes that exhibit non-linear current-potential characteristics. If a sinusoidal potential is applied to the diode the current then exhibits a non-zero average value, which is a measure of the applied sinusoidal-potential amplitude.

\begin{figure}
\setlength{\figwidth}{0.45\textwidth}
\centering
\begin{tabular}{cc}
\includegraphics[width=\figwidth]{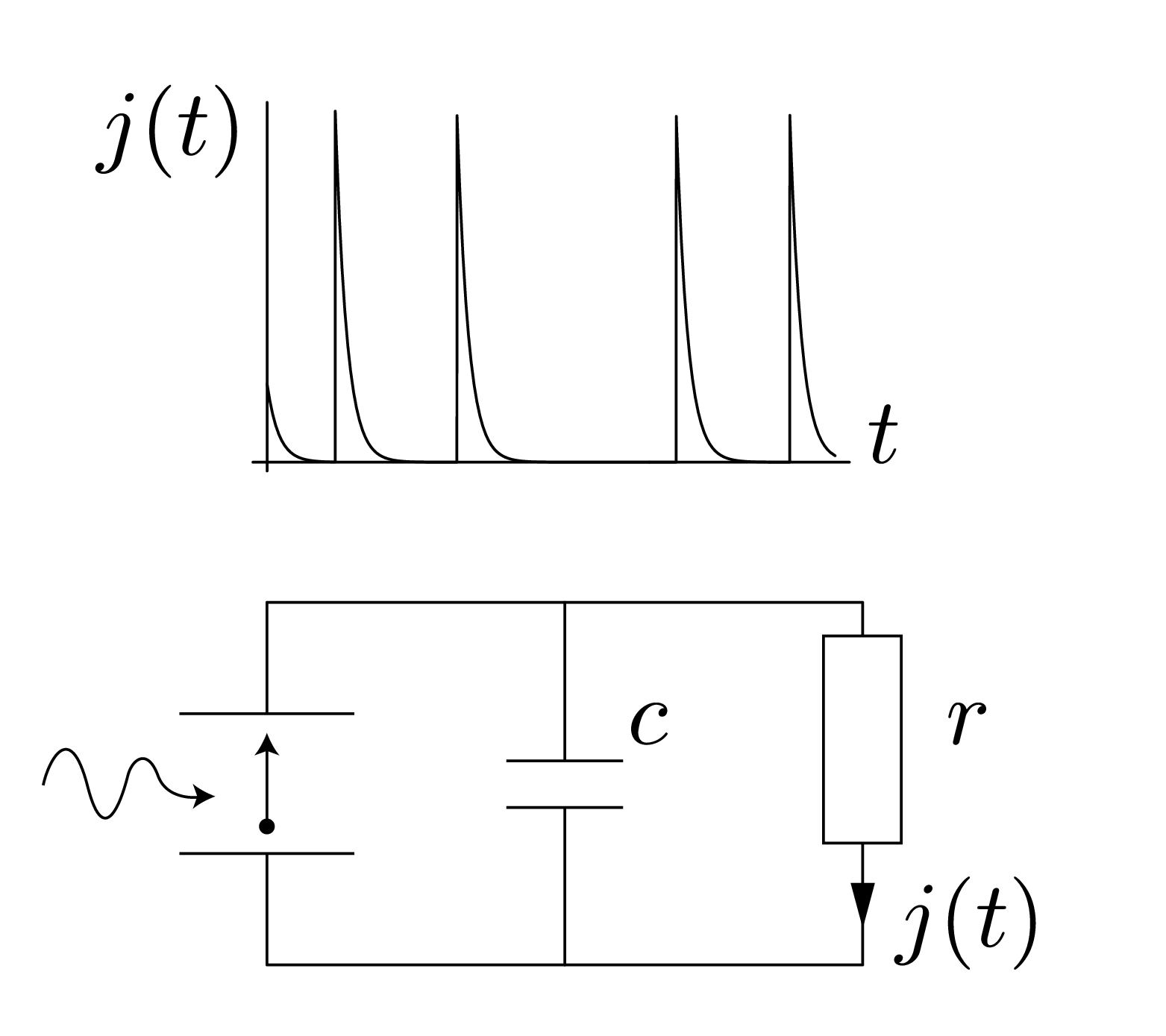} & \includegraphics[width=\figwidth]{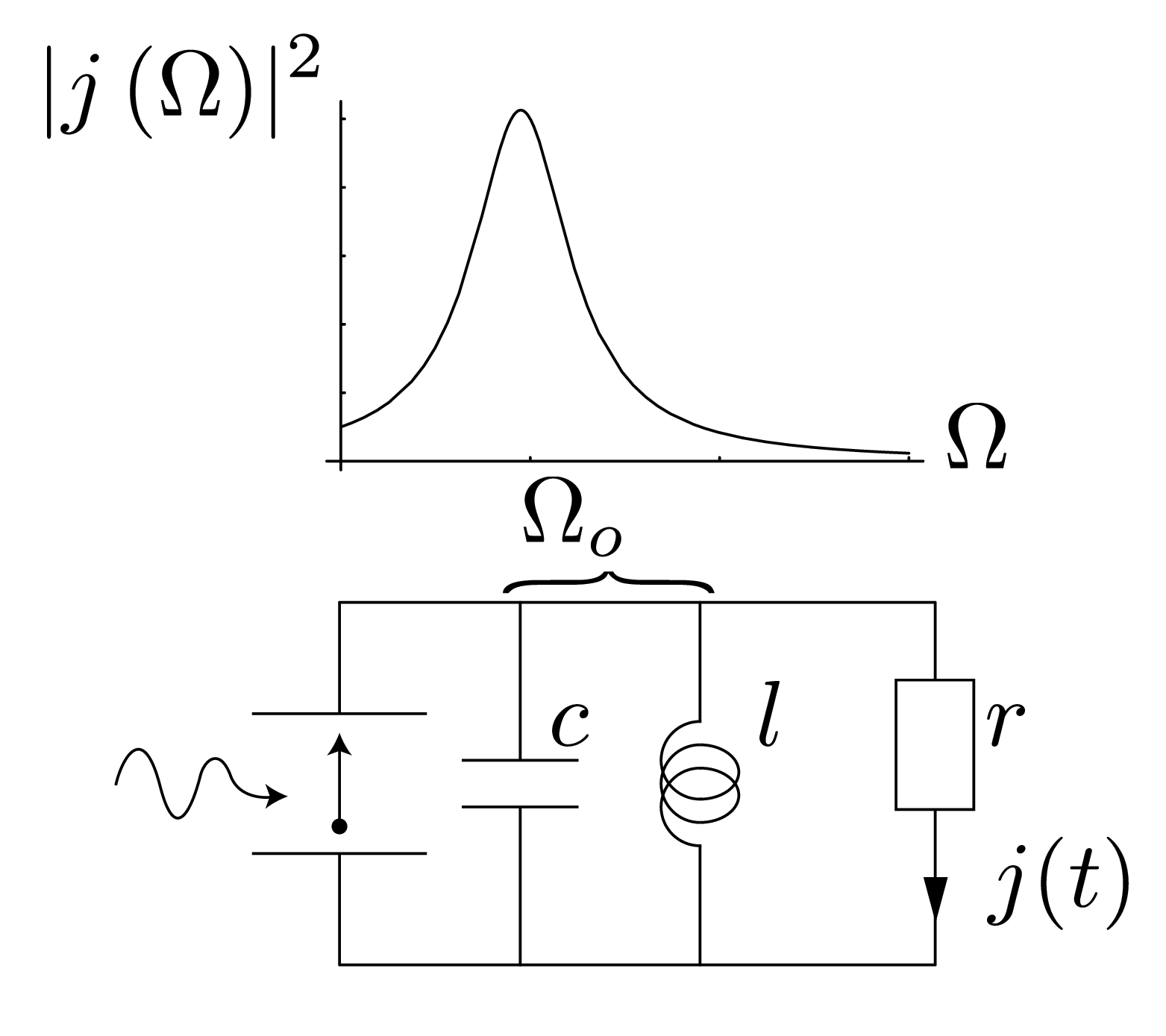} \\
(a) & (b) \\
\end{tabular}
\caption{In a), on top the photo-current is represented as a function of time, the output circuit capacitance (lower part) being taken into account. In b) the photo-current spectrum (on top) for the case where the output circuit (lower part) is a resonating circuit tuned at some Fourier frequency $\Om_o$.}
\label{photodetection}
\end{figure}

Let us recall the basic mode of operation of conventional electronic diodes, photo-detectors and photo-multipliers. Conventional electronic diodes are made up of two parallel plates (labeled in what follows the lower and upper plates) separated by a distance $d$ in vacuum. The lower plate, called "anode" is at zero potential by convention, and the upper plate, called "cathode", is raised at the potential $-U$ with $ U>0$. Suppose that at time $t=0$ an electron is freed from the upper plate and attracted by the anode\footnote{To achieve this, the cathode "work function" energy must be overcome by heat (thermo-ionic emission), high electric fields (field emission), electrons (secondary emission), or light (photo-electric emission). Electrons may be freed by thermal motion provided $\kB T_m$ be of the order of the metal work-function. If nickel is coated with barium oxide, a temperature of 1000 kelvin may suffice. Field emission occurs with kilo-volt potentials if the cathode has the shape of a needle. Electrons may be freed by light provided $\hbar\om$ exceeds the metal work-function, where $\hbar$ denotes the Planck constant and $\om$ the light frequency. Visible light for example is adequate when the cathode is coated with cesium. The non-zero initial electron velocities are presently neglected, that is, the initial electron momentum $p(0)=0$. Electrons in a metal are bound to it because they are attracted by their image charge. They may escape, though, because of a tunneling effect whose understanding requires Quantum Mechanics. But once the electron is sufficiently far away from the cathode, the Classical Equations of Electron Motion are appropriate.}. Considering only absolute values, the electron momentum increases linearly with time $t$ according to the law $p(t)=eUt/d$, where $-e$ denotes the electron charge, until it reaches the anode at time $\tau=d\sqrt{2m/eU}$, where $m$ denotes the electron mass. The electron kinetic energy is then converted into heat. In the following, $\tau$ is neglected, that is, it is set equal to zero. Fig.~\ref{photodetection} illustrates in a), the photo-current, represented as a function of time. Because the output circuit capacitance is taken into account each electron arrival corresponds to an exponentially-decaying pulse of the form $\exp(-t/rc)$. In b), photo-current spectrum for the case where the output circuit is a resonating circuit tuned at some Fourier frequency $\Om_o=1/\sqrt{\ell c}$.

In the case of photo-multipliers, the electron kinetic energy, instead of being dissipated into heat, is employed, at least in part, to free two or more electrons from the anode (secondary-emission effect). The latter are accelerated by a third plate, and so on, so that each electron freed from the cathode by light gets converted into an electron bunch containing $n$ electrons, for example, $n=10^6$. The situation is the same as if the absolute value $e$ of the electron charge had been multiplied by $n$. The purpose of photo-multipliers is not to improve the signal-to-noise ratio, which may only degrade. It is to raise the signal to such high levels that the thermal noise of subsequent electronic amplifiers is rendered negligible. 

In temperature-limited thermo-ionic diodes the electronic density is so low that the Coulomb interaction between electrons may be neglected. The electrons are emitted independently of one-another and their emission times are Poisson distributed (see the mathematical section). In that case the diode current fluctuation $\De j(t)$ obeys the so-called shot-noise formula, with a (double-sided) spectral density equal to $e\ave{j(t)}$. But when the electron flow is space-charge limited the current is sub-Poissonian, i.e., the spectral density is much smaller than the one just given.  This effect was discovered in 1940 by Thompson, North and Harris \cite{Thompson1940}.

The detectors considered above have been idealized for the sake of simplicity neglecting, e.g., dark currents and thermal noise. As said before, the current flowing out of photo-detectors may be viewed as a sum over the positive integer $k$ of delta-functions of the form $e\delta(t-t_k)$, where the $t_k$ are occurrence times. If this current is transmitted through a low-pass filter such as the one shown in Fig.~\ref{photodetection}, individual pulses may overlap, however, and not be distinguishable any more from one another. The current fluctuation then resembles gaussian noise irrespectively of the event-times statistics.

\subsection{Quantum detectors}\label{qdetectors}

Quantum photo-detectors (sometimes referred to as "narrow-band" photo-detectors), involve two electron energy levels, coupled to continua, and operate through the process of stimulated absorption. Ideally, the device is reversible in the sense that the electrical energy may be converted back into light energy through the process of stimulated emission. In contradistinction, conventional photo-diodes necessarily dissipate energy in the form of heat. Of particular interest is the visible-light-photon-counter, see \cite[p. 181]{Kim2001}, which has a quantum efficiency of 0.88, a gain of 30 000, a time response of 2ns, but unfortunately a rather large dark count of 20 000 counts per second.

\subsection{Classical sources}\label{csources}

The first high-frequency oscillator was probably a triode, with a feed-back mechanism from the anode to the grid controlling the current flow. We are particularly interested in the reflex klystron discovered by the Varian brothers in 1937, which may deliver electromagnetic radiation up to a frequency of about 10 GHz, and employs space-charge-limited cathodic emission (see e.g. \cite{Arnaud1990}). As was recalled above, Thompson and others discovered in 1940 that the current emitted by space-charge-limited cathodes is strongly sub-Poissonian\footnote{One of us (J.A.) set up in 1954 an experiment demonstrating that below threshold reflex klytrons behave as negative conductances, and that, in confirmation of the Thompson and others discovery just cited, the currrent is strongly sub-Poissonian (unpublished). The expected sub-Poissonian nature of the emitted radiation, however, was not observed.}. Reflex klystrons were mostly employed as low-noise local oscillators in radar heterodyne receivers until they were superseded by solid-state devices. 

Reflex klystrons involve two anodes made up of grids presumed to be transparent to electrons and separated by a distance $d$. Two cathodes are located just outside the anodes. The lower one emits electrons, while the upper one plays the role of a reflector\footnote{Usually the reflector is raised at a potential slightly lower than that of the emitting cathode to prevent electrons from being captured. It also helps finely tune the klystron oscillation frequency.} so that the electron motion as a function of time is a zig-zag path. The two anodes are part of a resonator. When the device oscillates an alternating potential $v(t)$ appears between the two anodes. If the electron emission time is appropriate, the electron looses its energy giving it up to the oscillating potential through an induced current. But since the electron emission times are uniformly distributed along the time axis, the net interaction with the field vanishes. Accordingly, initially, the electrons do not deliver any energy to the oscillating potential. It is as a result of the field action on the electron trajectories that non-zero energy exchanges between the field and the electrons may occur. This effect is called "bunching". Once the electron has lost most of its energy it gets captured by the anodes, and instantaneously jumps from the anode to the cathode through the static potential\footnote{In conductors the number of electrons is essentially equal to the number of atoms, a huge number. It follows that electrons are moving at very low speeds, on the order of 1$\mu$m/s, even for large currents. Accordingly, the above statement that jumps are instantaneous may seem surprising. As a matter of fact, electrons appearing on one plate are not the same as the electrons hitting the other plate. In conductors the electrical charge should be best viewed as a continuous incompressible fluid.}. The static-potential energy then gets reduced by  $eU$. In some sense, the electron plays an intermediate role. Indeed, the net effect of a complete electronic cycle is that, for each electronic event, the static potential source delivers an energy $eU$ to the oscillating potential source $v(t)$. 

A phenomenon akin to stimulated absorption may be understood similarly. This time, we suppose that the electron is emitted by one of the anodes. Without an alternating field this electron would remain permanently in the neighborhood of the anodes. However, a resonance with the alternating potential may force the electron to oscillate along the $x$ axis with increasing amplitude until its energy reaches the value $eU$, in which case it gets captured by one of the cathode. The net effect of this electronic process is that the alternating source gives energy to the static source, the opposite of what was discussed in the previous paragraph.

\subsection{Quantum sources}\label{qsources}

The first man-made quantum oscillators involving discrete matter levels were masers, operating at microwave frequencies. Subsequently maser action was discovered to occur naturally near some stars. The first laser, generating visible light, was discovered by Maiman in 1959. The fact that space-charge-limited cathodes generate light with sub-Poisson statistics was first demonstrated by Teich and Saleh in 1983 \cite{Teich1983}. 

The best-known light source is thermal radiation. A hot body like the sun radiates energy. The energy inside a closed cavity at absolute temperature $T_m$ contains an energy given by the law discovered in 1900 by Planck. An important feature of this law is that it involves a previously unknown universal constant $\hbar$ with the dimension of action or angular momentum (energy$\times$time). From the Quantum Mechanical view point, heat excites electrons to atomic levels higher in energy than the ground state energy.  These electrons then may decay spontaneously to the ground state by emitting a quantum of light. A similar mechanism is at work in the so-called "light-emitting diodes" (LED) but the spectrum of LED is narrower than that of thermal sources, though much greater than that of lasers.

A conventional neon tube generates light because the electric discharge excites neon atoms that subsequently decay to the ground state, thereby emitting ultraviolet light (subsequently converted into visible light) by the process of spontaneous emission, similar to what happens in thermal sources, but with a narrower spectrum. The so-called Helium-Neon laser\footnote{Helium plays the role of a "buffer gas", allowing the lower neon level to get depopulated.}, radiating light at a free-space wavelength of 0.63 $\mu$m, differs from conventional neon tubes in that two mirrors located at both ends, and facing each others, force the emitted light to move back and forth in the tube. Light gets amplified by the process of stimulated emission, and damped by the process of stimulated absorption. The former exceeds the latter when there are more atoms in the higher state than in the lower state (population inversion). To achieve this condition the lower-level population must be reduced through spontaneous decay to even-lower levels (3-levels lasers). Eventually a steady state of oscillation is reached. The emitted light spreads out in free space as little as is allowed by the laws of diffraction, and the laser light is nearly monochromatic (single frequency). The laser linewidth, though small, is of major importance in some applications. Laser diodes (also called injection lasers) employ  a semi-conductor with a doping that deliver electrons (n-type) and a doping that absorbs electrons (p-type)

To summarize, lasers essentially consist of single-mode resonators containing
three-level atoms or other forms of matter with a supply of energy called the pump and a sink of
energy, perhaps an optical detector. As said before, the latter converts the light
energy into a sequence of electrical pulses corresponding to
photo-detection events. When the pump is non-fluctuating the emitted light does not fluctuate
much. Precisely, this means that the variance of the number of
photo-detection events observed over a sufficiently long period of
time is much smaller than the average number of events.  Light having
that property is said to be "sub-Poissonian'', or below the "standard quantum level" (SQL).

\subsection{Quantum Electrodynamics}\label{quantum-light}

Historically, the concept that light should be quantized appeared around 1905 on the basis of thoughts expressed by Planck and Einstein. Measurements on black-body spectra were performed around 1900 with the help of gratings of appropriate periods, and described by a formula that involves the universal constant $\hbar$. On the other hand, the wave properties of electrons were discovered only decades later because the concept that electrons might possess wave-like behavior ought to wait for the observation that atoms emit light at well-defined frequencies, and because of the technical difficulty of sending electrons emitted from a small-area source on crystals (playing the role of gratings) in a very good vacuum. The interpretation of the observed diffraction patterns involves the constant $\hbar$. It is perhaps not preposterous to suggest that these two key discoveries could have occurred in the reversed order. Had this be the case, the Planck constant would have been considered as being fundamentally related to atomic behavior, and the subsequent appearance of the same constant in black-body radiation would have been viewed as a consequence of the atomic theory. 

\paragraph{The non-relativistic approximation.}

Some ancient philosophers thought that the speed of light was infinite. It would then be immaterial to say that light propagates from the sun to the eye (say) or the converse. The non-relativistic approximation is applicable to an hypothetical world in which the speed of light in free space would be arbitrarily large, the other constants ($\hbar,\epsilon_o,e,m$, defined in Section \ref{notation}) remaining as they are. Out of the latter constants we may define the electron spin along some quantization axis $±\hbar /2$, the Bohr radius $ a_o=4\pi\epsilon_o\hbar^2/me^2\approx0.53~ 10^{-10}$ meters, a speed unit $v_o=e^2/4\pi\epsilon_o\hbar=\hbar/a_om\approx 2.19~10^6 $ meters/second, a nominal metal plasma frequency $\om_p\equiv v_o/a_o\approx 4.13 ~10^{16}$rad/s, and the electron magnetic moment, equal to the Bohr magneton $e\hbar/2m$. A large part of Physics may be obtained on the basis of such a non-relativistic approximation.

\paragraph{The finite speed of light.}

Römer, however, discovered in 1676 through a kind of Doppler effect, using the motion of a Jupiter satellite as a clock, that light propagates at a finite speed $c\approx 300 ~000$ km/s. The law of causality, as it is presently understood, then implies that light propagates from the sun to the eye, for example. The Maxwell theory of electromagnetic waves suggests that radiated heat, as well as light, consists of electromagnetic waves of some sort. In 1862 Maxwell wrote "[electromagnetic waves travel] at a speed so nearly that of light that it seems we have strong reason to conclude that light itself (including radiant heat and other radiations) is an electromagnetic disturbance in the form of  waves propagated through the electromagnetic field according to electromagnetic laws.”

Note that if $v_o$, defined above, is written as $\al c$, where $\al\approx 1/137$ denotes the fine-structure constant, the non-relativistic approximation amounts to setting $\al$ as equal to zero. Because of the finite speed of light, a number of small corrections to the non-relativistic theory, on the order of $\al$, were observed. In particular, let us cite the spin-orbit splitting of electron states in atoms, the electron anomalous magnetic moment, the particle-like behavior of $\gamma$-rays\footnote{ $\gamma$-rays have been observed with an energy $\hbar \om\approx 1$ micro-joule. They appear as point particles because they are detected at rather precise times, locations, energies and momenta, independently of the nature of the absorbing material, whether it be steel or lead, say, and look like ultra-relativistic charged particles. But in the non-relativistic regime, light and electrons behaviors are quite distinct.}, the Casimir effect. A theoretical explanation of these effects is offered by Quantum Electrodynamics, in which the optical field is quantized. 

Casimir \cite{Grenet2003, Lamoreaux2006} discovered in 1948 that two conducting plates of area $A$ separated by a distance $d$ attract each other with a force
\begin{align}\label{casimir}
F=\frac{\pi^2}{240}\frac{A\hbar c}{d^4}=\frac{\pi^2}{240\al}\frac{A}{d^2}\frac{e^2}{4\pi\epsilon_o d^2}
\end{align}
The first expression was obtained theoretically by Casimir by ascribing an energy $\hbar\om/2$ to each electromagnetic mode of frequency $\om$. However, the same result was later obtained by Lifchitz from the Nyquist-like noise currents associated with small losses, letting the losses go to zero at the end. The second expression in \eqref{casimir} suggests an interpretation of the force as the product of the number of cells in the area $A$ and the force that an electron would exert on a hole separated from it by the distance $d$. The Casimir prediction agrees with experimental results with an accuracy of one per cent, with $d$ typically equal to $1\mu m$.

At face value, the Casimir force would become arbitrarily large in the non-relativistic approximation, $c\to\infty$. This is not the case, however, because the expression given above is valid only if $d\gg c/\om_p$, where $\om_p$ denotes the nominal plasma frequency in metals defined earlier. It follows that in the non-relativistic limit the Casimir force vanishes. Conductors attract each other only through the van der Waals forces that binds molecules together. From now on we only consider the non-relativistic approximation.

\paragraph{Quantum Optics.}

Most physicists opinion is that the photon concept is essential to understand "non-classical" (e.g., sub-shot-noise) states of light, even in the limit where $c\to\infty$. For Quantum Theories of Light, see e.g., \cite{Gerry2005, Loudon1983, Mandel1995, Walls1994, Meystre1991, Scully1997, Cohen-Tannoudji1988}. Let us recall some of the arguments given in favor of the photon concept, which evolved into the modern second-quantization procedures. 

Light-quanta (later on called "photons") were introduced by Einstein on the basis of the following argument.  Consider a collection of two-level atoms in a state of thermal equilibrium with the black-body radiation field. When an atom in the upper state decays to the lower state by emitting light spontaneously it recoils if the light  emission is \emph{directed} but would not if light were radiated (almost) isotropically. Einstein calculations indicate that a directed emission is required if the Maxwellian atomic velocity distribution is to be recovered. The Quantum Optics theory considers that the light emitted spontaneously by an atom may be expanded into quantized spherical waves. This concept, though accurate, does not fit well with the above Einstein \emph{picture}. But the Einstein picture fits well with the view that light is emitted only if it is directed toward some absorber\footnote{For an arbitrary distance $R$ between an emitting and an absorbing atom, see \cite{Andrews2004}. These authors employ orthodox Quantum Electrodynamics and comment that "In a sense, every photon is virtual, being emitted and then, sooner or later, absorbed", and "virtual photons are messenger particles that cannot be directly detected".}. 

The Quantum-Optics view point is well expressed by Carmichael (op.cit., p.1213) in those terms: "What role does photo-electric detection actually play in the return of the atom to its ground state [in resonance fluorescence] after each photon emission? Indeed, what does it means to speak of photon emissions as realized events separate from photo-electron detection? my viewpoint is that photo-electric detection does not cause atomic-state reduction. Projection of the atom into its ground state is caused by the dissipative nature of the atomic dynamics, and reoccurs, on average, at the mean spontaneous-emission rate, with complete indifference to the presence or absence of the observer. Photo-electric detection merely monitors emitted (realized) photons. It does not intrude into a coherent Quantum Dynamics in the manner implied by a measurement-induced wave-packet reduction; it is the irreversible decay into the vacuum that interrupts the coherence of the source dynamics. No doubt, in general outline, this view-point is already widely held". One may wonder whether the conceptual ambiguity outlined in the above quotation is not a consequence of the requirement that energy be conserved exactly (one photon being absorbed whenever a photo-detection event occurs) instead of being conserved on the average only. 

Another argument often given in favor of the concept that light consists of lumps of energy $\hbar \om$ is the observation that when a light beam of constant small intensity is incident on an ideal photo-detector (i.e., free of dark current and thermal noise) photo-current events sometimes occur long before the required optical energy $\hbar \om$ has been collected, in apparent violation of the law of conservation of energy. However, one should require that only the law of \emph{average} energy conservation be enforced. For a single system, there exists no independent way of measuring the "light intensity" as a function of time. The only information one may obtain concerning the intensity of a light beam is through the output of photo-detectors, and this brings us back simply to the observation made. A number of authors have shown that many effects that were at a time supposed to prove the reality of the photon concept such as the photo-electric effect, may in fact be interpreted in a semi-classical manner.

Most Quantum Theories of laser action begin with a discussion of the statistical operator $\rho$ of the optical field in empty loss-less resonators, which are treated in analogy with mechanical oscillators. Pumping and losses are subsequently introduced in an approximate manner by enforcing the preservation of the commutation relations. For the case presently considered (stationary linearized laser) the results of such calculations exactly coincide with ours. One must thus presume that the approximations are the same.

Other Quantum Optics treatments consider atoms in either their upper (pumping atoms) or lower (detecting atoms) states introduced at specific times into the optical cavity and spending there a fixed time $\tau$. Whether the atoms leaving the cavity are in their lower or upper state may in principle be measured, and the corresponding probabilities may be evaluated. In that way some properties of stationary lasers may be predicted. However, the flying-atoms configuration is quite different from the one discussed in the major part of this paper.

Another kind of Quantum-Optics theory considers instead continuous resonant photo-detection processes. That is, atoms in the absorbing state are coupled to the optical cavity at all times. This is the configuration considered in this paper. A recent treatment \cite{Oliveira2003} employs the operators introduced by London in 1926 having the property that $E_{\pm}\ket{m}=\ket{m{\pm}1}$, where $\ket{m}$ denotes a state with exactly $m$ photons. To our knowledge, this continuous-detection theory has not been applied to quiet lasers. This is probably the kind of Quantum Optics theory that one should employ to establish in a logical manner a link with the present Semi-Classical theory (i.e., rather than just comparing final formulas).

As far as electrons are concerned, physicists "second-quantize" the Schrödinger wave-function.  Electron second-quantization is a "book keeping" method, convenient when the optical field itself is quantized, but presently unnecessary. The statistical properties of non-interacting electrons that may exchange heat with a reservoir (this is the canonical ensemble, to be distinguished from the grand-canonical ensemble) have been obtained from second quantization of the electronic wave function. However, a much simpler direct solution has also been found, based on the partition of integers \cite{Arnaud1999}.

\paragraph{Physical paradigms.} 

A physical "paradigm" rests on a number of universal constants, on particles parameters, and on recipes to relate the theory to observations. For example Newtonian Celestial Mechanics employs a single universal constant, namely $G$, point particles have as sole parameter their mass (referred to that of a particular piece of platinum, called the kilogram). Given the position and speed of the particles at a given time, the theory provides their positions and speed at all times.

The present theory employs as universal constants $\hbar$, $\frac{1}{4\pi\epsilon_o}$ and, in some circumstances, $\kB$. However, $G$ and $c$ are not used. Particles (electrons) are characterized by their mass, electrical charge, and, in some circumstances, by spin and magnetic moment. Given the elements constitutive of a particular device, we determine the statistics of photo-electrons, which may, in principle, be measured with unlimited accuracy.

\subsection{Units and notations}\label{notation}

Our notations and conventions may differ from those employed by engineers, physicists, or experimentalists, which are not fully consistent. We attempted to follow the majority rule unless this leads to confusion. To simplify formulas we sometimes set as unity quantities such as the characteristic conductance of transmission lines. Otherwise, SI units (see below) are employed throughout.

\paragraph{Numerical values.}

Useful numerical values in the realm of Non-Relativistic Physics are
\begin{align}\label{num}
G~\textrm{(Newton gravitational constant)}&\approx 6.67~10^{-11}~\textrm{SI}\nonumber\\
g~\textrm{(earth gravitational acceleration at see level)}&\approx 9.81~\textrm{meters per second squared}\nonumber\\
e~\textrm{(absolute electron charge)}&\approx1.60~10^{-19}~\textrm{coulombs}\nonumber\\
m~\textrm{(electron mass)}&\approx9.10~10^{-31}~\textrm{kilograms}\nonumber\\
\hbar ~\textrm{(Planck constant divided by $2\pi$)}&\approx1.05~10^{-34}~\textrm{joules}\times \textrm{second}\nonumber\\
\kB~\textrm{(Boltzmann constant)}~&\approx1.38~10^{-23}~\textrm{joules/kelvin}
\end{align}
and
\begin{align}\label{eps}
\frac{1}{4\pi\epsilon_o}=10^{-7}\p 2.99792458~10^{8}\q^2~\textrm{farads/meter}
\end{align}
which is \emph{exact}, i.e., not subjected to revision as a consequence of later measurements, and involves a finite number of digits. The constants $e,m,\hbar, 4\pi\epsilon_o, \kB$ are the only ones that enter into the present theory because, in agreement with the non-relativistic approximation, we set the free-space permeability $\mu_o=0$ or equivalently $c=\infty$. We sometimes employ as energy unit the electron-volt$\approx1.60~10^{-19}$ joules. Note that capacitances may be evaluated from $\epsilon_o$ and their geometric dimensions, namely the electrodes areas and their spacings. As far as inductances are concerned we suppose that they are known from measurement.

\paragraph{Notation.} 

We list here only some of the notation employed in this paper. Different functions are distinguished by explicitly writing out their arguments. For example the Fourier transform of a function $\psi(x)$ is denoted by $\psi(k)$, i.e., with the same symbol, even though these are different functions. When the arguments are similar the same function, however, is intended. For example $w(t)$ and $w(\tau)$ represent the same function. One should not confuse a constant $U$ (no argument) with a function $U(x)$, for example. To avoid confusion between the arguments of a function and products we sometimes employ parentheses of different size, e.g., $f(x/d)$ is a function $f$ of $x/d$, while $f \p x/d\q$ represents the product of $f$ and $x/d$. As usual, $\cos^2(x)$ means $\p\cos(x)\q^2$, and likewise for other trigonometric functions. $\psi^\star(x)$ is equivalent to $\left(\psi(x)\right)^\star$.

For a two-state electron the lower and upper levels ("working levels") are denoted 1 and 2, or "a" (absorbing) and "e" ("emitting") levels, respectively. For 4-level electrons, levels of increasing energy are denoted 0,1,2,3. Pumping occurs from 0 to 3. The spontaneous-emission rate of an electron from 2 to 1 is denoted by $1/\tau_s\equiv 2\gamma$. It is often denoted in other works by $2\beta$, $\gamma$, or $\Gamma$. The decay time of an  oscillator with loss is denoted by $\tau_p$ (the subscript "p" stands for "photon", even though the word is not always appropriate).

Probabilities are denoted by $pr(a>b)$ and probability densities by $P(x)$, for example, but the letter $P$ may also refer to power. We employ double-side spectral densities, so that the usual shot-noise formula $2e\abs{J}$ is written as $e\abs{J}$, i.e., without a factor of 2 ($-e$ denotes the electron charge and $J$ the average current). We will perform three main kinds of averaging. First a Quantum Mechanical (QM) averaging over an ensemble of similarly prepared systems (if these systems exhibit macroscopic differences, these are supposed to be too small to affect the QM averages). Second, statistical averaging. We consider here an ensemble of similarly-prepared systems that differ from one another significantly. Third, an averaging over an optical period. These averagings could be denoted as $\ave{\ave{\ave{P}_{QM}}_{Stat.}}_{Opt. period}$, where, as an example, $P$ represents the power supplied by an optical potential source. This quantity may still be a slowly-varying function of time, which we denote for brevity $P(t)$. In some cases, for stationary systems, we may further introduce an averaging of $P(t)$ over an arbitrarily large period of time. Other notations will be outlined in Section \ref{math}.

\paragraph{Decibels.}

We recall here a notation commonly employed in Electrical Engineering. Usually, an amplifier is loaded with a nominal conductance such as 20 milli-siemens (resistance of 50 $\Om$) and the input impedance is equal to that of the load. If the input power of an amplifier is $P_{in}$ and the output power is $P_{out}$, the amplifier gain in decibel (abbreviation "dB") is defined as $10\log_{10}(P_{out}/P_{in})$. If the amplifier is linear the gain does not depend on the input power. In terms of the potentials $V_{in},V_{out}$ at the input and output ports, the gain reads $20\log_{10}(\abs{V_{out}/V_{in}})$ dB, because powers are proportional to the modulus-squares of the potentials in the situation considered. Likewise, an attenuation is defined as $10\log_{10}(P_{in}/P_{out})=-10\log_{10}(P_{out}/P_{in})$. A gain of 3dB means that the input power is multiplied by a factor close to 2. Note that dBm means decibels above a power of 1 mW. For example, 30dBm represents approximately a power of 1 watt.

 \paragraph{Electrical circuit schematics.}

In schematics, current sources are represented by circles with an arrow in them, while potential sources are represented by a circle with + and - signs, to define what is meant by positive current or positive potential, as shown later in Fig.~\ref{circuitfig}. By potential (or current) \emph{sources} we mean potentials (or currents) that do not depend on the current (potential) delivered. These are called in Quantum Optics "prescribed classical sources".

\newpage

\section{Mathematics}\label{math}

Ideally, the present section should derive from axioms all the mathematical results subsequently employed. This goal is not accomplished for lack of space, time (and ability). We therefore provide references to text books. A difficulty is that many text-books consider general situations from which it is not always easy to retrieve the desired information. Often, physical intuition or numerical calculations help. For example the celebrated Wiener-Khintchine theorem that relates spectrum and correlation was first obtained by Einstein in an intuitive manner.

We will clarify in the present section the notation employed, recall elementary mathematical formulas, and give the main properties of random point-processes. We then offer a simple picture of quiet light generation, and show that random deletion of photo-electrons leaves unaffected the reduced spectrum. This section is entitled "Basic Mathematics" because most results are mathematical in nature, although rigor is overlooked and the physical motivation is often pointed out.

\subsection{Complex numbers}

A complex number is denoted either as $z=\Re\{z\}+\ii \Im\{z\}$ or as $z=z'+\ii z''$, and $z^\star=z'-\ii z''$ denotes the complex conjugate of $z$. We denote $\abs{z}^2\equiv z z^\star=z'^2+z''^2$. A complex notation is often employed for describing quantities that vary sinusoidally in time. According to the complex notation, the function $i(t)=\sqrt2\abs{I}\cos(\om t+\phi)$, where the frequency $\om$ and the phase $\phi$ are real constants, is written as $i(t)=\sqrt2 \Re\{I \exp(-\ii\om t)\}$, where the complex number $I$ is defined as $I=\abs{I}\exp(-\ii \phi)$. Similar definitions apply to potentials $v(t)$ varying sinusoidally in time, that is $v(t)=\sqrt2 \Re\{V \exp(-\ii\om t)\}$. We have chosen to introduce the factor $\sqrt2$ so that the average optical power, defined as the time average of the current-potential product $v(t)i(t)$ be simply equal to the real part of the product $V I^\star$, i.e., without the factor 1/2 that would otherwise occur. $V$ and $I$ are called rms (root-mean-square) complex potentials and currents, respectively, or more briefly, optical potentials and currents. The minus sign in $ \exp(-\ii\om t) $ is employed in optics because waves propagating forward in space then involve a term of the form  $ \exp(\ii kx) $, where $k$ denotes the wave-number, that is, with a plus sign. But for slow variations the function $j(t)=\sqrt2\abs{J}\cos(\Om t+\phi)$ is denoted $j(t)=\sqrt2 \Re\{J \exp(\jj\Om t)\}$, where the complex number $J$ is defined as $J=\abs{J}\exp(\jj \phi)$, as is usually done in electrical engineering. The complex notation considerably simplifies calculations for real, causal, linear and time-invariant circuits submitted to sinusoidal potentials or currents. Even though the squares of $\ii$ and $\jj$ are both equal to -1 these two numbers should be distinguished. A "bi-complex" notation describes in an exact manner sinusoidally-modulated sinusoidal signals. It is recalled in \ref{bicomplex}. 

\paragraph{Second and third-degree equation.}\label{elem}

Elementary algebraic operations are concisely recalled, for example, in \cite{Abramowitz1965}. We give below the solution of second and third-degree equation in the form appropriate to our intended application. 

Consider the polynomial $ap^2+bp+c=0$. The solutions are
\begin{align}
\label{fou5}
p_{\pm}=\frac{-b\pm\sqrt{b^2-4ac}}{2a}.
\end{align}
For the third-degree polynomial $p^3+a_2 p^2+a_1p+a_0=0$ we evaluate sequentially 
\begin{align}
\label{fou}
q=\frac{a_1}{3}-\frac{a_2^2}{9}\qquad r=\frac{a_1a_2}{6}-\frac{a_0}{2}-\frac{a_2^3}{27}\qquad s=\sqrt{q^3+r^2}\nonumber\\
s_1=(s+r)^{1/3}\exp(\ii \pi/3)\qquad s_2=(s-r)^{1/3}\exp(\ii 2 \pi/3)\qquad s±r>0.
\end{align}
The three roots are
\begin{align}
\label{three}
p_1&=s_1+s_2-\frac{a_2}{3}\nonumber\\ 
p_2&=s_1\exp(\ii 2 \pi/3)-s_2\exp(\ii \pi/3)-\frac{a_2}{3}\nonumber\\
p_3&=-s_1\exp(\ii \pi/3)+s_2\exp(\ii2 \pi/3)-\frac{a_2}{3}.
\end{align}

\subsection{Vectors and matrices}\label{vector}

We consider a fixed coordinate system and define a vector (bold-face letter) as a collection of complex numbers denoted for example
\begin{align}\label{vec}
\boldsymbol{a}=
\left( 
\begin{array}{ccc}
a_{1}\\
a_{2}
\end{array}
\right).
\end{align}
Transposition, indicated by a "t" in upperscript, interchanges lines and columns. For example
\begin{align}\label{v1}
\boldsymbol{a}^t=
\left( 
\begin{array}{ccc}
a_{1}& a_{2}
\end{array}
\right).
\end{align}
The scalar product of two vectors $\boldsymbol{a}$ and $\boldsymbol{b}$ is defined as
\begin{align}\label{v2}
\boldsymbol{a}.\boldsymbol{b}=a_1~b_1+a_2~b_2
\end{align}
The modulus square of the length of a complex vector is 
\begin{align}\label{v3}
\abs{\boldsymbol{a}}^2\equiv\boldsymbol{a}.\boldsymbol{a}^\star =a_1~a_1^\star+a_2~a_2^\star≥0.
\end{align}
A matrix is denoted for example
\begin{align}\label{vec4}
\boldsymbol{M}=
\left( 
\begin{array}{ccc}
M_{11} & M_{12}\\
M_{21} & M_{22}
\end{array}
\right).
\end{align}
The trace of a square matrix is the sum of the diagonal elements
\begin{align}\label{vec4trace}
trace\{\boldsymbol{M}\}=M_{11} + M_{22},
\end{align}
and the determinant
\begin{align}\label{vec5}
det\{\boldsymbol{M}\}=M_{11} M_{22}-M_{12} M_{21}.
\end{align}
The sum of two matrices is obtained by summing their elements. The product $\boldsymbol{L}$ of two matrices $\boldsymbol{M}$ and $\boldsymbol{N}$ is for example
\begin{align}\label{vec6}
L_{ij}=M_{i1} N_{1j}+M_{i2} M_{2j}\qquad i,j=1,2.
\end{align}
We have$\left(\boldsymbol{M}\boldsymbol{L}    \right)^t= \boldsymbol{L}^t\boldsymbol{M}^t       $. $\boldsymbol{M}^{-1}$ denotes a matrix such that $\boldsymbol{M}^{-1} \boldsymbol{M}=\boldsymbol{1}\equiv \left(\begin{array}{ccc}1&0\\0&1\end{array}\right)$.   When $det\{\boldsymbol{M}\}=0$ the matrix is singular, and cannot be inverted. A matrix is said to be symmetrical when $\boldsymbol{M}^t=\boldsymbol{M}$, Hermitian when $\boldsymbol{M}^{t\star}=\boldsymbol{M}$, unitary when $\boldsymbol{M}\boldsymbol{M}^{t\star}=\boldsymbol{1}$. We have $trace\{\boldsymbol{A}\boldsymbol{B} \}=trace\{\boldsymbol{B}\boldsymbol{A} \}$. The trace of the product of two Hermitian matrices is real.

\paragraph{Cauchy-Schwartz inequality.}

Consider two unit vectors $\boldsymbol{a},\boldsymbol{b}$, that is, such that $\abs{\boldsymbol{a}}=\abs{\boldsymbol{b}}=1$, The Cauchy-Schwartz inequality reads 
\begin{align}\label{vec7}
\abs{\boldsymbol{a}.\boldsymbol{b}^\star}^2\equiv (\boldsymbol{a}.\boldsymbol{b}^\star)(\boldsymbol{b}.\boldsymbol{a}^\star)≤1.
\end{align}
This relation is obtained by replacing $\boldsymbol{a}$ in \eqref{v3} by $\boldsymbol{a}-(\boldsymbol{a}.\boldsymbol{b}^\star)\boldsymbol{b}$. 

\paragraph{Density matrices.}

Let $\boldsymbol{\rho}$ denote an Hermitian matrix of trace 1. 

The pure-state density matrix is constructed from the vector $\boldsymbol{a}$ with $\abs{\boldsymbol{a}} ^2=1$ as
\begin{align}\label{dens}
\boldsymbol{\rho}=
\left( \begin{array}{ccc}
a_{1}\\
a_{2}
\end{array}\right)
\left(   a_1^\star~~a_2^\star  \right) 
= \left( \begin{array}{ccc}
a_{1}a_{1}^\star&a_{1}a_{2}^\star\\
a_{2}a_{1}^\star&a_{2}a_{2}^\star
\end{array}   \right)     .
\end{align}
We readily find that $\boldsymbol{\rho}^2=\boldsymbol{\rho}$, and thus $trace\{\boldsymbol{\rho}^2\}=trace\{\boldsymbol{\rho}\}=1$. Note that $\boldsymbol{\rho}$ is unaffected by a change of the phase of $\boldsymbol{a}$. As examples, we may have
\begin{align}\label{ds1}
\boldsymbol{\rho}=\left(\begin{array}{ccc}   1&0\\0&0     \end{array}\right),\qquad \boldsymbol{\rho}=\frac{1}{2}\left(\begin{array}{ccc}   1&\ii\\-\ii&1     \end{array}\right)
\end{align}

 Let $\boldsymbol{\rho}_a,\boldsymbol{\rho}_b$ be two such matrices. By explicit calculation we find that $trace\{ \boldsymbol{\rho}_a~ \boldsymbol{\rho}_b\}=(\boldsymbol{a}.\boldsymbol{b}^\star)(\boldsymbol{b}.\boldsymbol{a}^\star)$. Thus, from \eqref{vec7} we have $trace\{ \boldsymbol{\rho}_a~ \boldsymbol{\rho}_b\}≤1$.

The mixed-state density matrix is defined as 
\begin{align}\label{dens1}
\boldsymbol{\rho}\equiv  \sum_{k}p_k\boldsymbol{\rho}_k\qquad p_k≥0\qquad \sum_k p_k=1,
\end{align}
where the $p_k$ may be called weights. Since $trace\{ \boldsymbol{\rho}_k~ \boldsymbol{\rho}_l\}≤1$ we obtain
\begin{align}\label{dens2}
trace\{ \boldsymbol{\rho}^2   \}=trace\{ \sum_{k}p_k\sum_{l}p_l\boldsymbol{\rho}_k\boldsymbol{\rho}_l  \}≤\sum_{k}p_k=1
\end{align}
If $trace\{ \boldsymbol{\rho}^2   \}=1$ we have $(\boldsymbol{a}.\boldsymbol{b}^\star)(\boldsymbol{b}.\boldsymbol{a}^\star)=1$. The density matrix $\boldsymbol{\rho}$ is then of the pure-state form in \eqref{dens}.

Let us consider a pure-state density matrix $\boldsymbol{\rho}_k$, and suppose that the QM (Quantum Mechanical) average of some quantity, such as the power $P_k$, may be obtained from the formula
\begin{align}\label{den10}
\ave{P_k}_{QM}=trace\{\boldsymbol{\rho}_k \boldsymbol{P}\}
\end{align}
where $\boldsymbol{P}$ denotes some known Hermitian $2\times 2$ matrix. Next, suppose that the pure-state density matrix $\boldsymbol{\rho}_k$ occurs with probability $pr_k$, $k=1,2...$. The statistical and QM-average of the power, denoted by a double bracket is, using the properties of the trace
\begin{align}\label{den0}
\ave{\ave{P}_{QM}}_{statistical}=\sum_k pr_k ~trace\{\boldsymbol{\rho}_k \boldsymbol{P}\}=trace\{\boldsymbol{\rho} \boldsymbol{P}\}\qquad \boldsymbol{\rho}\equiv \sum_k pr_k \boldsymbol{\rho}_k.
\end{align}
Note that the mixed-state density matrix is of the form in \eqref{dens1} with $p_k=pr_k$. From now on, the double bracket is replaced by a simple bracket. The diagonal elements $\rho_{11},\rho_{22}$, called "populations", are non-negative and sum up to 1. The off-diagonal elements $\rho_{12}=\rho_{21}^\star$ are called "coherences".

Setting $x\equiv 2\rho_{12}',  y\equiv 2\rho_{12}'', z \equiv \rho_{22}-\rho_{11} $ (unrelated to coordinates in space), the density matrix may be written as
\begin{align}\label{dns}
     \boldsymbol{\rho}=\frac{1}{2}
  \left( 
  \begin{array}{ccc}
      1-z    &   x+\ii y     \\   
      x-\ii y&   1+z             
\end{array}  
 \right)     .
\end{align}
It follows that 
\begin{align}\label{ens1}
trace\{\boldsymbol{\rho}^2\}=\frac{1}{2}(1+x^2+y^2+z^2),
\end{align}
and $trace\{\boldsymbol{\rho}^2\}≤1\Longleftrightarrow x^2+y^2+z^2≤1$.

If $\boldsymbol{\rho}$ depends on time,
$\frac{d}{dt}trace\{\boldsymbol{\rho}^2\}=x\frac{dx}{dt}+y\frac{dy}{dt}+z\frac{dz}{dt}$. It follows that if at some time we have a pure-state density matrix, $x^2+y^2+z^2=1$, we must have at that time $x\frac{dx}{dt}+y\frac{dy}{dt}+z\frac{dz}{dt}≤0$. For later use, suppose that
\begin{align}\label{ens2}
\frac{dx}{dt}&=-2a\gamma x\nonumber\\
\frac{dy}{dt}&=-z-2a\gamma y\nonumber\\
\frac{dz}{dt}&=y-2\gamma z+2b\gamma,
\end{align}
where $a,\gamma$ are non-negative parameters and $-1≤b≤1$. The above condition holds provided that
\begin{align}\label{ens5}
(a-1)z^2+bz-a≤0\qquad \abs{z}≤1,
\end{align}
that is, $2a≥1-\sqrt{1-b^2}$. When $b=±1$ we must have $2a≥1$. Physical arguments given later on show that for a one-electron model $2a$ is in fact unity.

\subsection{Fourier transforms}\label{fouriertitle}

The Fourier transform $\psi(k)$ of the  function $\psi(x)$ and the reciprocal relation are
\begin{align}
\label{fourier}
\psi(k)&=\int_{-\infty}^{\infty}{dx\exp(-\ii kx)\psi(x)}\\
\label{fourier'}
\psi(x)&=\int _{-\infty}^{\infty}{\frac{dk}{2\pi}\exp(\ii kx)\psi(k)},
\end{align}
where $\ii^2=-1$, and $k$ is called the wave-number. Note that the position of the $2\pi$ factor varies from one author and another, without of course affecting the end results. Obviously, the Fourier transform of $\p\ii k\q^n \psi(k)$ is equal to the $n$th derivative of $\psi(x)$ with respect to $x$. If $\psi(x)$ is real, we have $\psi^\star(k)=\psi(-k)$. 

Relations similar to \eqref{fourier} and \eqref{fourier'} hold with $x$ changed to $t$, $k$ to $\om$ and (to be consistent with our conventions for optical signals) $\ii$ changed to $-\ii$. Then the element of integration in \eqref{fourier'} is $d\nu\equiv d\om/2\pi$, where $\nu$ denotes as usual the optical frequency
\begin{align}
\label{fourier1}
\psi(\om)&=\int_{-\infty}^{\infty}{dt\exp(\ii \om t)\psi(t)}\\
\label{fourier1'}
\psi(t)&=\int _{-\infty}^{\infty}{\frac{d\om}{2\pi}\exp(-\ii \om t)\psi(\om)},
\end{align} 
In the Fourier-frequency domain, changing $\om\to\Om,\ii\to-\jj$, we write
\begin{align}
\label{fourier2}
\psi(\Om)&=\int_{-\infty}^{\infty}{dt\exp(-\jj\Om t)\psi(t)}\\
\label{fourier2'}
\psi(t)&=\int _{-\infty}^{\infty}{\frac{d\Om}{2\pi}\exp(\jj \Om t)\psi(\Om)},
\end{align}

Note the following physical application. For particles moving in time-independent potentials $V(x)$, stationary states $\psi(x)$ are real functions of $x$ (to within an arbitrary over-all phase factor that we set equal to 1). If furthermore $V(x)$ is an even function of $x$, $\psi(x)$ is either an even or odd function of $x$. It follows from the above considerations that the $\psi(k)$-functions are, respectively, real even or imaginary odd. In the present mathematical section we set $\hbar=1$ and do not distinguish the electron momentum $p$ from the wave-number $k$.

We will need the following expression of the Dirac $\de$-distribution\footnote{The $\de(t)$-distribution may be viewed alternatively as a function equal to $1/h$ for $-h/2<t<h/2$ and 0 otherwise, so that the area under the function is unity, letting $h$ go to zero at the end of the calculations. Many other forms of the $\de$-function may be used, with less-singular derivatives than for the one just given.}
\begin{align}
\label{delt}
\de(x)=\int_{-\infty}^{\infty}\frac{dk}{2\pi}\exp(\ii k x),
\end{align}
implying that its Fourier transform is unity.
Using this expression one may prove that 
\begin{align}
\label{planche}
\int_{-\infty}^{+\infty}\frac{dk }{2\pi}\abs{\psi(k)}^2=\int_{-\infty}^{+\infty}dx \abs{\psi(x)}^2.
\end{align}

Let $\abs{\psi(x)}^2 dx$ be interpreted as the probability of finding the position of a particle between $x$ and $x+dx$ if a measurement is performed, and $P(p)dp\equiv\abs{\psi(p)}^2dp$ be the probability of finding the electron momentum between $p$ and $p+dp$ if a measurement is performed (this latter measurement may be accomplished by letting the particle free at some time $t$ and observing its position on some far-away screen). We are led to define the wave function in momentum space as
\begin{align}
\label{pa}
\psi(p)=\frac{1}{\sqrt{2\pi}}\int_{-\infty}^{+\infty}dx\exp(-\ii px)\psi(x),
\end{align}
according to \eqref{planche}. In general the wave-function depends on time, and the average value of $p$, evaluated at some time $t$, depends on time and is denoted $\ave{p(t)}$.

For two functions $\psi(x)$ and $\phi(x)$ and their respective Fourier transforms $\psi(k)$ and $\phi(k)$ we obtain from the expression in \eqref{delt} of the $\de(.)$ distribution the identity
\begin{align}
\label{par}
\int_{-\infty}^{+\infty}\frac{dk}{2\pi} \psi(k) \phi^\star(k)=\int_{-\infty}^{+\infty}dx \psi(x) \phi^\star(x).
\end{align}

\paragraph{Transformations.}

If we set in \eqref{par} $\psi(k)\equiv \psi_1(k)$ and $\phi(k)=k^n \psi_2(k)$, we have, using the observation following \eqref{fourier'} that the Fourier transform of $\p\ii k\q^n \psi(k)$ is equal to the $n$th derivative of $\psi(x)$ with respect to $x$
\begin{align}
\label{pars}
\ave{k^n }_{12}\equiv \int_{-\infty}^{+\infty}\frac{dk}{2\pi}k^n \psi_1(k) \psi_2^\star(k)=
\int_{-\infty}^{+\infty}dx~ \psi_1(x) \p \frac{d}{\ii dx}  \q^n \psi_2^\star(x),
\end{align}
provided the integrals exist.

Through two integrations by parts we obtain that
\begin{align}
\label{bypart}
\int_{-\infty}^{+\infty}dx~\psi(x)\p\frac{d}{dx}\q^2\phi(x)=\int_{-\infty}^{+\infty}dx~\phi(x)\p\frac{d}{dx}\q^2\psi(x) .
\end{align}
On the other hand, for any derivable function $\psi(x)$,
\begin{align}
\label{ff}
2\frac{d}{dx} \psi(x)=\left(\p\frac{d}{dx}\q^2 x-x \p\frac{d}{dx}\q^2\right)  \psi(x) .
\end{align}

\paragraph{Average wave-number.}

Setting  $n=1$ in \eqref{pars} we find that
\begin{align}
\label{km}
\ave{k }_{12}&\equiv \int_{-\infty}^{+\infty}\frac{dk}{2\pi}k \psi_1(k) \psi_2^\star(k) =-\ii \int_{-\infty}^{+\infty}dx~\psi_1(x)\frac{d\psi_2^\star(x)}{dx}.
\end{align}
Using  \eqref{ff} with $\psi(x)=\psi_2(x)$, the above expression may be written as
\begin{align}
\label{km6}
\ave{k }_{12}=-\frac{\ii}{2} \int_{-\infty}^{+\infty}dx~\psi_1(x)\left(\p\frac{d}{dx}\q^2 x-x \p\frac{d}{dx}\q^2\right)  \psi_2^\star(x) .
\end{align}
Finally, employing \eqref{bypart} with $\psi(x)=\psi_1(x)$ and $\phi(x)\equiv x\psi_2^\star(x)$ we obtain
\begin{align}
\label{km2}
\ave{k }_{12}=-\frac{\ii}{2} \int_{-\infty}^{+\infty}dx~x\left(\psi_2^\star(x)\frac{d^2\psi_1(x)}{dx^2}-\psi_1(x)\frac{d^2\psi_2^\star(x)}{dx^2}\right).
\end{align}
In particular, we have from \eqref{km2}, suppressing the subscripts, the relation
\begin{align}
\label{kkk}
\ave{k }&\equiv \int_{-\infty}^{+\infty}\frac{dk}{2\pi}k \psi(k) \psi^\star(k) 
=\frac{\ii}{2} \int_{-\infty}^{+\infty}dx~x\left(\psi(x)\frac{d^2\psi^\star(x)}{dx^2}-\psi^\star(x)\frac{d^2\psi(x)}{dx^2}\right),
\end{align}
which will prove useful in establishing the first Ehrenfest equation.

\paragraph{Eigen-functions.}

If $\psi_1(x)$ and $\psi_2(x)$ are solutions of the eigen-equations $d^2\psi_{1,2}(x)/dx^2+e_{1,2}\psi_{1,2}=0$, we obtain from \eqref{km2} that
\begin{align}
\label{kmm}
\ave{k }_{12}=\frac{\ii \p e_1-e_2^\star\q}{2} \int_{-\infty}^{+\infty}dx~x~ \psi_1(x)\psi_2^\star(x)\equiv\frac{\ii \p e_1-e_2^\star\q}{2}x_{12} .
\end{align}
On the other hand, setting $n=2$ in \eqref{pars} we obtain that ($i,j=1,2$)
\begin{align}
\label{quick}
\ave{k^2 }_{ij}&\equiv \int_{-\infty}^{+\infty}\frac{dk}{2\pi} \psi_i(k) k^2\psi_j^\star(k)=-\int_{-\infty}^{+\infty}dx \psi_i(x) \frac{d^2\psi_j^\star(x)}{dx^2}=e_j^\star\int_{-\infty}^{+\infty}dx \psi_i(x) \psi_j^\star(x).
\end{align}
The above expressions will be employed in Section \ref{quantum} in relation with the time-dependent Schrödinger equation.

\subsection{Convolution and Laplace transforms }\label{convolution}

Let us consider a real, causal, linear and time-invariant system. These conditions imply that for a potential source $v(t)$ the current $i(t)$ (or more generally the response to a source of any kind) is given by
\begin{align}
\label{conv}
i(t)=\int_{-\infty}^{+\infty}du~ h(u) v(t-u)\equiv h*v=h*y,
\end{align}
where the kernel $h(u)$ is real, equal to 0 for $u<0$, and middle stars denote convolution products. Convolutions are associative, so that parentheses in convolution products are unnecessary, and commutative. For example, for a conductance $G$, we have $i(t)=Gv(t)$, and thus $h(t)=G\de(t)$, where $\de(.)$ denotes the Dirac distribution. If $v(t)=V(p)\exp(pt)$ where $p$ denotes a complex number (not to be confused with particle momenta), $i(t)=I(p)\exp(pt)$, where $I(p)=H(p)V(p)$ and
\begin{align}
\label{convbis}
H(p)=\int_{0}^{+\infty}dt~\exp(-pt)h(t),
\end{align}
a Laplace transform, defines $H(p)$ for complex $p$. In most of this work we set $p=-\ii\om$, and $H(-\ii\om)$ is denoted $Y(\om)$ and called the admittance.  
 
In particular the Laplace transform of $\exp(\lambda t)$ is $1/(p-\lambda)$. It follows that if the reciprocal of a polynomial in $p$ may be written as a sum of terms of the form $1/(p-p_k)$ the inverse Laplace transform is easily obtained. More generally, the Heaviside theorem says that if $f(p)$ is a polynomial with distinct roots (not to be confused with probabilities) $p_k$, $k=1,2...n$ ($f(p_k)=0$), the inverse Laplace transform of $1/f(p)$ is
\begin{align}
\label{conv4}
L^{-1}\{\frac{1}{f(p)}\}=\sum_{k=1}^n \frac{\exp(p_k t)}{\bigl(  df(p)/dp  \bigr)_{p=p_k}}.
\end{align}

The Laplace transform of the convolution of any number of functions is the product of their Laplace transforms. For example, the Laplace transform of $G_k(t)\equiv w(t)*w(t)...*w(t)$ ($k$-times) is the $k$th power  $w(p)^k$ of $w(p)$, where $w(p)$ denotes the Laplace transform of $w(t)$. Thus, the Laplace transform of $G(t)\equiv w(t)+w(t)*w(t)+....$ is the sum of an infinite geometric series \cite[p.~53]{Cox1980} 
\begin{align}
\label{convter}
G(p)=\frac{w(p)}{1-w(p)}.
\end{align}
Let us define an average waiting time
\begin{align}
\label{tmean}
\ave{t}\equiv\int_0^\infty dt~ t~w(t)=-\bigl( \frac{dw(p)}{dp}     \bigr)_{p=0}.
\end{align}
If \eqref{convter} holds, $G(t\to\infty)$ is finite and the other terms are decaying exponentials, so that $G(p\to 0)\approx G(t= \infty)/p$, we obtain
\begin{align}
\label{conv9}
\frac{1}{\ave{t}}=G(t=\infty).
\end{align}

\subsection{Relative noise, normalized correlation and variance.}

In the present section we clarify the notation employed, and explain why it differs from the one used in other works. More precise definitions and result are given in a subsequent section.

We are mostly concerned with photo-currents $j(t)\equiv -e\D(t)$. Here, $\D(t)$ is the sum over $k$ of $\de(t-t_k)$-distributions where the $t_k$ form a stationary point (or "event") process of average rate $D$. We call "reduced spectrum" $\spectral_{\De D}(\Om)$ the spectral density $\spectral_{\D} (\Om)$ of that process with the singularity at $\Om=0$ removed. The "relative spectrum" $\spectral_{\De D/D}(\Om)$ is obtained by dividing $\spectral_{\De D}(\Om)$ by $D^2$. Finally, the "relative noise" $\N(\Om)\equiv\spectral_{\De D/D}(\Om)-1/D$ vanishes for shot-noise, also referred to as the "standard quantum limit". We are mostly interested in circumstances where the relative noise is negative. In the engineering literature most authors call "relative-intensity noise" the quantity $2\spectral_{\De D/D}(\Om)$, and express it in decibels/hertz. Our main objection to using it is that, unlike the relative noise employed in this paper, the relative-intensity noise does not enjoy the property of being independent of (cold, linear) attenuations. Further, the expression "decibel per hertz" is difficult to comprehend.

We call "normalized correlation" $g(\tau)$ the auto-correlation of the $\D(t)$ process with the singularity at $\tau=0$ being removed, divided by the square of the average rate $D$. It may take any non-negative value. The reason why we do not use the Quantum Optics notation $g^{(2)}(\tau)$ is two-fold. One is that correlations of order other than the second are not employed, so that no confusion may arise. More importantly, $g^{(2)}(\tau)$ is usually defined in terms of the so-called "light intensity" $I(t)$ according to $g^{(2)}(\tau)\equiv \ave{I(0)I(\tau)}/I^2$. For mathematical reasons a quantity so-defined cannot be less than unity. Since $g^{(2)}(\tau)$-values less than unity are apparently measured, the Quantum-Optics view point is that $I(t)$ should be considered as an operator instead of an ordinary function of time. In this paper the concept of "light intensity" does not enter and $g(\tau)$ refers to photo-currents exclusively.

We denote by $\V(T)$ the variance of the number of photo-detection events occurring within the time-interval $T$, divided by $D$, minus 1. In Quantum Optics, the Mandel $Q$-parameter is defined by a similarly looking expression, namely $Q\equiv \mathrm{variance}(m)/\ave{m}-1$, but the operator $m$ sometimes refers to the number of photons in the cavity, rather than to the number of photo-electrons. Thus our $\V(T)$ and the $Q$-factor may have different physical meanings. Some authors employ the Fano factor $\fano\equiv Q+1$. Sometimes in the literature, however, the Fano factor refers, not to the number of photons in a cavity, but to the normalized spectrum of electrical-current fluctuations. Again, because of this ambiguity we set the (double-sided) spectral density of the current driving a laser diode as $\xi \ave{J}$, where $\xi=0$ for a quiet pump and $\xi=1$ for a Poissonian pump, and consider that the Fano factor $\fano$ relates to the number of photons (ratio of energy and $\hbar\om_0$) in an optical resonator.

\subsection{Random processes}\label{random}

Instead of attempting to clarify the meaning of the word "probability", let us give the axioms. Given two events $A$ and $B$, $A+B$ denotes an event that occurs when $A$ or $B$, or both, occur. A probability is denoted by "pr(.)". The theory rests on the two following axioms:
\begin{itemize}
\item The probability $pr(A)$ of an event $A$ is a non-negative number, unity for a sure event.
\item If $A$ and $B$ are mutually exclusive (i.e., the occurrence of one at a given trial precludes the occurrence of the other) $pr(A+B)=pr(A)+pr(B)$.
\end{itemize}
The product $AB$ of two sets of events $A$ and $B$ is the set of the events that are common to $A$ and $B$. Two events are called independent if $pr(AB)=pr(A)pr(B)$.

\paragraph{Random variable}

To every outcome $\zeta$ of an experiment (such as rolling a dice) we associate a number $x(\zeta)$, called a random variable. $\{x≤X\}$ denotes the set of all outcomes $\zeta$ such that $x(\zeta)≤X$. The distribution function $F(X)$ is the probability $pr\{x≤X\}$ that $x(\zeta)$ be less than or equal to $X$. The probability density is defined as $P(X)=dF(X)/dX$. Since, in Physics, $X$ has usually a dimension (e.g., time) the dimension of $P(X)$ is the reciprocal of that of $X$.

\paragraph{Stochastic process}

We are given an experiment such as rolling a dice, specified by its outcome $\zeta$. To every outcome we assign a time function $x(t;\zeta)$. This is called a stochastic process.  

\paragraph{Wide-sense stationarity}

A real process $x(t)$ is said to be "wide-sense stationary" when $\ave{x(t)}$ and the correlation $R(\tau,t)\equiv\ave{x(t)x(t+\tau)}$ do not depend on $t$. It follows that for real wide-sense stationary random processes $R(\tau)=R(-\tau)$. To prove it, set $t=-\tau$ and remember that $x(t)$ is an ordinary function of time so that $x(t)$ and $x(t+\tau)$ commute. 

\paragraph{Ergodicity}

Usually one observes the photo-current $j(t)$ from a photo-detector in a single set-up. The mathematical treatment of noise, on the other hand, rests on the consideration of an arbitrarily large number of macroscopically-identical set ups, averaging referring to these many set-ups. The question thus arises as to what practical conclusions may be drown from formulas derived from the formalism. The answer is that, provided a stationary system is \emph{ergodic}, the statistics may be obtained from a single set-up. If this is the case, statistical averages are equivalent to time averages. A system may be ergodic only if the correlation $\ave{j(0)j(\tau)}$ tends to  $\ave{j(0)}^2$ as $\tau$ tends to infinity. More stringent conditions must be fulfilled, however, that we suppose met. In principle, once the statistical calculations have been performed, one should verify that the system considered is ergodic in order to be able to apply the results to a single system. In practice, this step is omitted.

\subsection{Spectrum}

The spectrum $\spectral_t(\Om)$ of a process is a real non-negative even function of $\Om$. It may be obtained by considering a finite duration $T$, evaluating of average of the modulus square of the Fourier transform of $x(t)$, dividing by $T$, and letting $T$ go to infinity. The motivation for introducing a subscript "$t$" is that we intend to introduce later on a spectrum denoted $\spectral(\Om)$ (without a subscript) obtained from $\spectral_t(\Om)$ by removing the singularity at $\Om=0$. Alternatively, the spectrum may be expressed as the Fourier transform of $R(\tau)$ (Wiener-Khintchine theorem)
\begin{align}\label{def}
\spectral_t(\Om)&=\lim_{T\to \infty}\frac{1}{T}\ave{\abs{\int_{0}^{T}{dt~ x(t)\exp(\jj\Om t)}}^2}\nonumber\\
&=\int_{-\infty}^{+\infty}{d\tau~ R(\tau)\exp(\jj\Om \tau)}=\int_{-\infty}^{+\infty}{d\tau~ R(\tau)\cos( \Om \tau)}.
\end{align}
The two above expressions agree in the mean if the integral from 0 to $\infty$ of $\tau R(\tau)$ is finite \cite[p. 336]{Papoulis1965}. We have employed above the electrical-engineering $ \exp(\jj\Om t)$ notation. To compare with the previous notation in \eqref{fourier} change $\Om$ to $k$, $\tau$ to $x$, and $\jj$ to $-\ii$. Evaluating $\ave{\p x(\tau)±x(0)\q^2}$ we notice that $-R(0)≤R(\tau)≤R(0)$. This condition does not suffice however to make $R(\tau)$ positive definite, that is, its Fourier transform could still be negative. 

In the special case where $x(t)$ does not depend on time we have $R(\tau)=\ave{x^2}$= constant. Substituting in \eqref{def} we find that $\spectral_t (\Om)=2\pi \ave{x}^2\de(\Om)$, where $\de(.)$ denotes the Dirac $\de$-distribution. We are thus led to define a reduced spectrum $\spectral(\Om)\equiv\spectral_t(\Om)-2\pi \ave{x}^2\de(\Om)$.

Conversely, the correlation may be expressed in terms of the spectrum through the inverse Fourier transform according to
\begin{align}\label{convterbis}
R(\tau)=\int_{-\infty}^{+\infty}\frac{d\Om}{2\pi}    S_t(\Om)   \cos( \Om \tau).
\end{align}
If we define $y(t)\equiv x(t)-\ave{x(t)}$ we have $\ave{y(t)}=0$. The function $C(\tau)=\ave{y(0)y(\tau)}$ is called the (auto) covariance of the process $x(t)$.

We are mostly interested in the case where $x(t)=\sum_{k}\de(t-t_k)$ with $0<t_k<T$, where the $t_k$ are referred to as  "points" with a constant density $D$ (point processes are discussed in more detail in section \ref{stationary}). We have $\ave{x(t)}=D$. The spectrum of $x(t)$ exhibits a singularity $2\pi D^2\de(\Om)$ at $\Om=0$. The first expression in \eqref{def} give the reduced spectrum
\begin{align}\label{d}
\spectral(\Om)=\lim_{T\to \infty}\frac{1}{T}\ave{\abs{\sum_{allowed~k}^{}{\exp(\jj\Om t_k)}}^2},
\end{align}
where $\Om=2\pi n/T$, $n=1,2...$. Note that $n=0$ is not allowed, but $\Om$ can be made as small as one wishes by setting $n=1$ and letting $T$ go to infinity. This expression is useful to evaluate spectra through numerical calculations that generate runs, each with a different $t_k$ sequence. Because of the assumed ergodicity, a single run suffices. Total energies are obtained by integrating over frequency from $-\infty$ to $+\infty$ the quantity $\p d \Om/2\pi\q\spectral (\Om)$. For two independent processes $x(t)$ and $y(t)$ of spectral densities $\spectral_x$ and $\spectral_y$, respectively, the spectral density of  $z(t)=ax(t)+by(t)$ is $\spectral_z=\abs{a}^2\spectral_x+\abs{b}^2\spectral_y$. 

The moments of a quantity such as $x$, denoted $\ave{x^n}$, are defined as the integrals over $x$ from $-\infty$ to $+\infty$ of $x^nP(x)$, where $n=1,2...$ and $P(x)$ denotes the probability density of $x$, or the sum from $k=1$ to $\infty$ of $k^npr(k)$, where $pr(k)$ denotes the probability of having the outcome $k$. Note that the above integrals or sums may not exist, even for well-behaved probability laws. It follows from \eqref{def} and \eqref{fourier} that
\begin{align}
\label{x2}
\ave{x^2}=C(0)=\int_{-\infty}^{\infty} \frac{d\Om}{2\pi} \spectral(\Om)
\end{align} 
if $\ave{x}=0$. Thus, if the spectrum shape is known, the variance of $x$ determines the spectrum.

The sources of noise in our theory are narrow-band current sources written as $c(t)=\sqrt 2 \p C'(t) \cos(\om_o t)+C''(t) \sin(\om_o t) \q$, where $\om_o$ denotes the average laser frequency, and $C'(t),~C''(t)$ are jointly-stationary slowly-varying real functions of time. It can be shown that $c(t)$ is wide-sense stationary if and only if $\ave{C'(t)}=\ave{C''(t)}=0$ and the auto and cross correlations fulfill the conditions $R_{C'C'}(\tau)=R_{C''C''}(\tau),R_{C'C''}(\tau)=-R_{C''C'}(\tau)$. Furthermore, we assume that the statistics is independent of a phase change, and this entails that $R_{C'C''}(\tau)=0$. Let us recall the following result. If $C'(t),~C''(t)$ are uncorrelated and their spectra $\spectral_{C'}(\om)=\spectral_{C''}(\om)$ vanish for $\abs{\om}>\om_c$, then $\spectral_{C}(\om)=\spectral_{C'}(\om-\om_o)+\spectral_{C'}(\om+\om_o)$ \cite[p. 380]{Papoulis1965}.

Consider a constant-amplitude frequency-modulated signal $x(t)=\cos(\om_ot +\phi(t))$. If $\De\om(t)=d\phi(t)/dt$ is a stationary low-frequency gaussian process of (double-sided) spectral density $\spectral_{\De\om}$, the spectral density $\spectral_x(\om)$ of $x(t)$ is Lorentzian with a full-width at half power (FWHP) $\de\om=\spectral_{\De\om}$ \cite[p. 140]{Rowe1965}, that is, is of the form $\spectral_x(\om)\propto 1/[1+\p2(\om-\om_o)/\de \om\q ^2]$. Note that here $\De \om(t)$ denotes a process, while $\de \om$ is a real positive number, the spectral width.

\subsection{Point processes}\label{stationary}

For physical motivation note that, ideally, photo-currents in an experiment lasting from $t=0$ to $t=T$ are of the form $j(t)=-e\D(t),~\D(t)\equiv D+\De D(t)=\sum_{k}^{}\de(t-t_k)$ with $0<t_k<T$, where $-e$ denotes the electron charge and $\delta (.)$ the Dirac distribution. Experimentally, we may measure a number of quantities relating to $j(t)$ with the help of integrators, narrow-band filters or electron counters.

Point processes are sequences of increasing positive real numbers $t_{k}, k=1,2…$. Each $k$ value corresponds to a "point" occurring at time $t_k$. We consider $M$ such sequences, labeled by $m=1,2...M$, called runs. As said above, averages denoted by the sign \ave{.} refer to sums from $m=1$ to $m=M$ of some quantity defined for each run divided by $M$, letting $M$ go to infinity. Clearly, averaging is a linear operation, that is \ave{a+b}=\ave{a}+\ave{b}. Point processes are fully defined either by the so-called "non-exclusive multi-coincidence rates", or from the statistics of durations between successive points (related to the "exclusive multi-coincidence rates"). Stationary point processes have a time-independent density $D$. Let $d(t)$ denote the number of points occurring up to time $t$, that is the number of $k$ values such that $t_k<t$. The (rather intuitive) result that $\ave{d(T)}=DT$ is demonstrated in Section \ref{point}. 

\paragraph{Poisson processes}

The most important process is the Poisson process. Consider a Poisson process of density 1. The density that the first point occurs at time $\tau>0$ is $\exp(-\tau)$, whether or not there is a point at $t=0$. If there is a point at $t=0$, $w(\tau)=\exp(-\tau)$ is the waiting-time density.
Consider next an inhomogeneous Poisson process of density $\lambda(t)$ with a point at $t=0$. It may be reduced to a Poisson process through a transformation of the time scale $d\tau=\lambda(t)dt$. Thus the waiting-time density, that is, the probability that the next point occurs in the interval $(\tau,\tau+d\tau)$, divided by $d\tau$, reads, 
 \begin{align}
\label{ju}
w(\tau)=\lambda(\tau)\exp(-\int_0^\tau dt \lambda(t))=-\frac{d}{d\tau}\exp(-\int_0^\tau dt \lambda(t)).
\end{align}
It follows from the second form above that the integral of $w(\tau)$ from 0 to $\infty$ is unity, provided $\lambda(t)$ does not tend to 0 as $t\to\infty$. This means that the point eventually occurs. The average duration between adjacent points is, after an integration by parts
 \begin{align}
\label{juk}
\ave{\tau}=\int_0^\infty d\tau\exp(-\int_0^\tau dt \lambda(t)).
\end{align}
If, for example, $\lambda(t)=1$, we obtain $\ave{\tau}=1$ as expected. Conversely,
 \begin{align}
\label{ju5}
\lambda(t)=\frac{w(t)}{\int_t^\infty d\tau w(\tau)}=-\frac{d}{dt}\log \left( \int_t^\infty d\tau w(\tau)    \right).
\end{align}

\paragraph{Ordinary renewal process}

With $t_0=0, t_1=\tau_1, t_2=\tau_1+\tau_2,...$ and the $\tau_i, i=1,2...$ independent and distributed according to the same density $w(\tau)$, one generates an ordinary renewal process $t_k, k=1,2...$. Such a process is non-stationary, but it tends to be stationary for large times.

Let us denote
 \begin{align}
\label{juy}
w(p)=\int_0^{\infty}dt \exp(-pt)w(t)
\end{align}
the Laplace transform of the waiting-time density. Given that there is a point at $t=0$, the probability $G(t)dt$ that there is a point between $t$ and $t+dt$, is the sum of the probabilities that this occurs through one jump, two jumps,...Because the jumps are independent and have the same densities, we obtain the Laplace transform of $G(t)$ 
 \begin{align}
\label{juyi}
G(p)=\int_0^{\infty}dt \exp(-pt)G(t)
\end{align}
in the form, see Section\eqref{stationary},
 \begin{align}
\label{jyi}
G(p)=\frac{w(p)}{1-w(p)}.
\end{align}

It follows that, given the density $\lambda(t)$ of an inhomogeneous Poisson process, viewed as a renewal process, we may in principle obtain the Laplace transform of the auto-correlation function $G(t)$. As an example, suppose that $\lambda(t)=1\Longleftrightarrow w(t)=\exp(-t)$ whose Laplace transform is $w(p)=1/(1+p) $. Thus, from \eqref{jyi}, $G(p)=1/p$ and $G(t)=\lambda(t)=1$ as expected.

\subsection{Event-rate spectrum}\label{event}

The photo-current spectrum $e^2\spectral_\D(\Om)$, where $\Omega$ denotes the Fourier frequency, may be measured by letting the photo-current $j(t)$ flow through a narrow-band filter with center frequency $\Om$. The function $\spectral_{\D} (\Om)$ exhibits a $2\pi D^2\de(\Om)$ singularity, but the reduced spectrum $\spectral_{\De D} (\Om)$ is non-singular, where $\De D\equiv \D-D$. In the theory to be subsequently presented, $\spectral_{\De D} (\Om)$ is obtained by setting in rate equations $d/dt\to \jj \Om$. 

The relative noise $\N(\Omega)$ is then defined as
\begin{align}
\label{defn}
\N(\Om)\equiv \spectral_{\De D/D}(\Om)-\frac{1}{D},
\end{align}
where $\spectral_{\De D/D}(\Om)=\spectral_{\De D}(\Om)/D^2$.

One may be interested instead in the normalized second-order correlation function $g(\tau)$, a non-negative even function of the delay time $\tau$. Aside from  normalization, $g(\tau)$ is the correlation of  $\D(t)$ with the singularity at $\tau=0$ being removed. Alternatively, $Dg(\tau)d\tau$ may be defined as the probability that an event occurs between $\tau$ and $\tau+d\tau$, given that an event occurred at $t=\tau$. As shown in Section \ref{point}, $g(\tau)$ is related to the relative noise defined above by the integral relations
\begin{align}
\label{g}
\N(\Om)&= \int _{-\infty }^{\infty }  d\tau\bigl(g(\tau)-1\bigr)\exp(-\jj\Om \tau)\\
\label{n}
g(\tau)-1&=\int _{-\infty }^{\infty }  \frac{d\Om}{2\pi} \N(\Om)\exp(\jj\Om \tau).
\end{align}
The motivation for introducing $g(\tau)-1$ in \eqref{g} is that this quantity tends to 0 as $\tau$ tends to infinity because widely separated events are in that limit independent for stationary processes. The above relations are closely related to the Wiener-Khintchine relations. They are established in Section \ref{point} directly for point processes. Note that our definition of "sub-Poissonian" photo-currents is that $\N(0)<0$. This does not necessarily imply that $g(0)<1$.

\subsection{Photo-count variance}\label{variance}

As shown in Section \ref{point}, the normalized variance $\V(T)$ of the number of events occuring during some time $T$ and $g(\tau)$ are related  
\begin{align}
\label{v}
\V (T)\equiv \frac{\ave{d(T)^{2}}-\ave{d(T)}^2}{\ave{d(T)}}-1&=D \int _{-T}^{T} d\tau(1-\frac{\abs{\tau}}{T})\bigl(g(\tau)-1\bigr)\\
\label{v'} 2D\p g(T)-1\q&=\frac{d^2\p T\V (T)\q}{dT^2} 
\end{align}
In the special case of a Poisson process we have $g(\tau)=1$, $\V (T)=0$ and $\N(\Om)=0$, that is, $\spectral_{\De D}=D$.

As an example let us consider a high-power laser driven by a non-fluctuating current.  The relative noise will be obtained in the form
\begin{align}
\label{qq}
\N(\Om)=-\frac{1}{D\p 1+\p \Om\tau_p \q^2\q }
\end{align}
where $\tau_p$ is the so-called "photon life time" of the resonator. From this expression we obtain, setting $D=1$ for simplicity, that 
\begin{align}
\label{gt}
g(\tau)&=1- \int _{-\infty }^{+\infty }  \frac{d\Om}{2\pi} \frac{1}{1+\p  \Om\tau_p \q^2}\exp(\jj\Om \tau)=1-\frac{1}{2\tau_p}\exp(-\frac{\tau}{\tau_p})\\ 
\label{o}
g(0)&=1-\frac{1}{2\tau_p}
\end{align}
Of course $g(\tau)\to 1$ if $ \tau\to\infty$. In the present situation $g(0)<1$.

\subsection{Dark-room picture}\label{dark}

For the sake of illustration let us present a simple picture of regular point processes. The initial point process considered is periodic and consists of events occurring at $t=1,2...$ time units, that is $t_k=k$. Under circumstances to be defined later on (delay times much larger than unity) this process may be viewed as being almost stationary. The density is clearly unity. 

In our picture, one person (representing an electron) enters into a dark room every time unit and wanders randomly in the room until he finds the exit.  This picture may describe regularly-pumped lasers at high power because electrons entering the cavity then are quickly converted into photons. Photons wander in the optical resonator for some time and then get instantly converted into photo-electrons. The point process is written as $t_k=k+\xi_k$, where the $\xi_k$ are independent of one-another and distributed according to the same density $P(\xi_k)\equiv P(\xi)$. An appropriate distribution would be the exponential one.

Let us treat a special case that may be solved almost by inspection, namely the case where  $P(\xi)=1/\tau_r$ if $0≤\xi<\tau_r$ and 0 otherwise, and $\tau,\tau_r$ are large integers. Consider a pair $i,j≠i$ of $k$ values such that $i+\xi_i$ may be in the first time slot $(0,dt)$ and $j+\xi_j$  may be in the second time slot $(\tau, \tau+d\tau)$. Inspection shows that this is possible only if $-\tau_r<i≤0,\tau-\tau_r<j≤\tau$. Ignoring first the restriction $j≠i$, we find that the probability we are looking for is the number of allowed $i,j$ values, that is, the product of the $i,j$ ranges, times $1/\tau_r^2$, namely $\tau_r^2/\tau_r^2=1$. This result is accurate if $\tau≥\tau_r$. But if $\tau<\tau_r$ one must subtract from the numerator of the previous expression the number of $i,j$-values that are equal, namely $\tau_r-\tau$, so that the normalized correlation reads
\begin{align}
\label{corr}
g(\tau)&=1,\quad\qquad &\tau≥\tau_r\nonumber\\
g(\tau)&=\frac{\tau_r^2-\p \tau_r-\tau\q}{\tau_r^2}=1- \frac{\tau_r-\tau}{\tau_r^2},\quad &\tau<\tau_r.
\end{align}
In particular, $g(0)=1-1/\tau_r$, indicating a modest amount of anti-bunching, remembering that  $\tau_r\gg1$. The same result is obtained for the laser model in \eqref{o} if we set $\tau_r=2\tau_p$ to make the average life-times the same in the two models. 

The reduced photo-events spectrum is obtained from $g(\tau)$ through a Fourier transform according to \eqref{g}
as
\begin{align}
\label{sp}
\N(\Om)&\equiv 2\int_{0}^{1} dx \p x-1\q \cos(\Om\tau_r x)=2\frac{\cos(\Om\tau_r)-1}{\p\Om\tau_r\q^2}\nonumber\\
\spectral(\Om)&=1+\frac{\cos(\Om\tau_r)-1}{(\Om\tau_r)^2/2},
\end{align}
where we have set $x\equiv \tau/\tau_r$, remembering that the density (average rate) $D=1$. We note that $\spectral(0)=0$, as one expects from the fact that the primary process is regular and that no event has been lost or created. The spectral density of the process considered, given in \eqref{sp}, is illustrated in Fig.~\ref{darkroom}. 

\setlength{\figwidth}{0.6\textwidth}
\begin{figure}
\centering
\includegraphics[width=\figwidth]{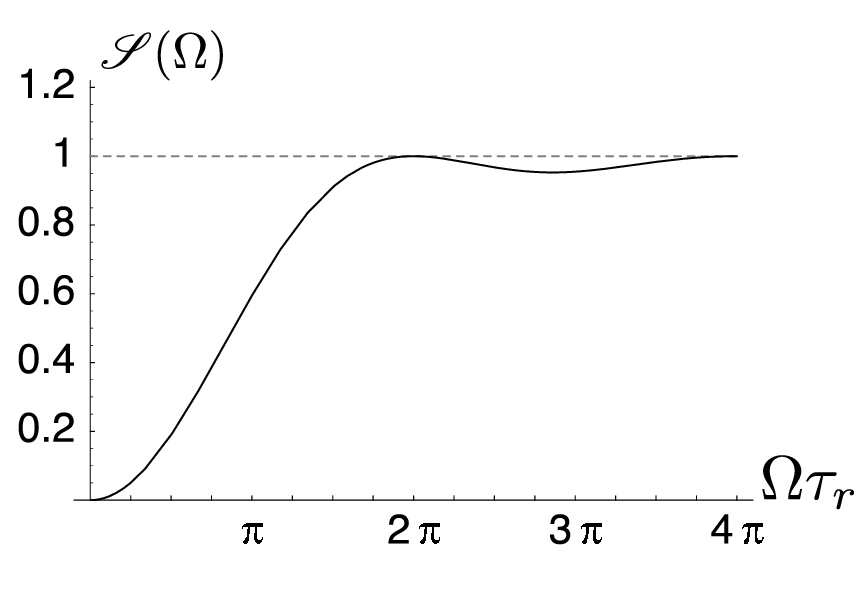}
\caption{Plain line: spectral density corresponding to the dark room picture, see \eqref{sp}. The dotted line corresponds to the shot-noise level.}
\label{darkroom} 
\end{figure}

Using \eqref{v} we obtain in the present model
\begin{align}
\label{nn}
\V(T)&=-1+\frac{\tau_r}{3T},\qquad T≥\tau_r \nonumber\\
\V(T)&=-\frac{T}{\tau_r}+\frac{T^2}{3\tau_r^2},\qquad T<\tau_r. 
\end{align}
It is easy to see that the expression of $g(\tau)$ in \eqref{v'} is verified in that example.

\subsection{Random deletion}\label{deletion}

Random deletion of events (also called "thinning") means that each event is ascribed a probability $1-pr$ of being deleted. For example, considering the first event of a given run, we flip a coin. If head, that event is preserved (probability 1/2). If tail, it is deleted. The same procedure is applied to the other events of the run and to the events of other runs, each time with a new coin flipping. Obviously the average rate $D$ of the process is multiplied by $pr$. An important result is that the function $g(\tau)$ and thus the other two functions defined above, and in particular the relative noise $\N(\Om)$, are not affected. Indeed consider the case where there is one event in the time slot $[0,dt]$ and one event in the time slot $[\tau, \tau+d\tau]$, corresponding to a product of 1. In any other circumstances the product is 0. After thinning the probability of having again (1,1) is multiplied by $pr^2$. But the denominator in the normalized correlation $g(\tau)$ is also multiplied by $pr^2$, so that the result is unchanged. The average rate may be restored by an appropriate scaling of the time axis. But since in general $g(\tau/pr)≠g(\tau)$, rescaled thinning affects the statistics with the sole exception of Poissonian processes, in which case $g(\tau)=1$.

\subsection{Point processes. Mathematical details}\label{point}

Point processes are possibly unlimited sequences of increasing positive real numbers $t_{k}, k=...-1,0,1,2…$, collectively denoted as $\{t_k\}$. Each $k$ value corresponds to an "event", or "point", occurring at time $t_k$. We consider $M$ such sequences, labeled by $m=1,2...M$.., called runs. Averages denoted by the sign \ave{.} refer to sums from $m=1$ to $m=M$ of some quantity defined for each run divided by $M$, letting $M$ go to infinity. The full specification of a point process requires correlations of all order. However we limit ourselves to first and second-order quantities. Note that the full time axis is considered, with past being distinguished from future\footnote{For random points on a line it is often required to treat the two directions "left to right" and "right to left" symmetrically. In that case the complete intensity function has to be abandoned.}.

Point processes are defined by a complete intensity  function
\begin{align}
\label{nnx}
\rho(t;H_t)=\lim_{\de\to 0+} \de^{-1}pr\{N(t,t+\de)>0\vert H_t\}, 
\end{align}
where $pr\{N(t,t+\de)>0\vert H_t\}$ denotes the probability that at least one event occurs within the interval $(t,t+\de]$, given the previous history $H_t$, that is the point process up to and including $t$ (signs "(,)" indicate that the end point is not included while signs "[,]" indicate that the end point is included). The condition that the $t_k$ are increasing may be written as $pr\{N(t,t+\de)>1=o(\de)\}$ for all $t$. This ensures that there are essentially no multiple occurrences at a given time. Such processes are called "orderly" for an obvious reason. In the special case of Poisson processes the previous history does not matter, and we only need to know the density of the process. Other important special cases are processes with independent (perhaps equally distributed) increments).

Given a point process, new point processes may be obtained through the four following operations, whose use in the present paper will be indicated.
\begin{itemize}
\item A change of time scale. In that way an inhomogeneous Poisson process may be converted into a (homogeneous) Poisson process.
\item Thinning, in which some of the points in the original process are deleted with a constant probability, independently of all the other points. This happens when the light incident on a photo-detector is attenuated.
\item Translation of individual points. See the section \ref{dark}.
\item Superposition, in which a number of separate processes are merged. Given the point process relating to a single electron submitted to a field, we shall consider the process obtained for independent electrons. Another example is that of a radio-active source involving many independent atoms. We assume that superposition of an arbitrary large number of orderly processes is a Poisson process \cite{Khintchine1960}. \footnote{As a proof of this conclusion up to second-order, consider the superposition of $M$ independent processes of density 1 (for simplicity) and the same second-order correlation $g(\tau)$. These processes, of duration $\tau$, may be cut out from a single sample provided they are sufficiently far apart to be almost independent. As before, for the original process, $g(\tau)d\tau$ denotes the probability that, given that an event occurred at $t=0$, another event occurs in the interval $(\tau,\tau+d\tau)$. Consider now the superposed process consisting of the $k=1,2...M$ independent processes. The density of the superposed process is obviously $M$. Suppose that for $k=1$ an event occurs at $t=0$. The probability that an event of that same sample occurs in the $(\tau,\tau+d\tau)$ interval is, as said above, $g(\tau)d\tau$. But events in the $(\tau,\tau+d\tau)$ interval may originate from the $M-1$ other processes, with probability $(M-1)d\tau$. It follows that, given that an event occurred at $t=0$ for the sample $k=1$, the probability that another event occurs in the $(\tau,\tau+d\tau)$ interval is $(g(\tau)+M-1)d\tau$. What has been just said for $k=1$ applies to all the $M$ samples, so that the probability that, given that an event occurred at $t=0$, another event occurs in the $(\tau,\tau+d\tau)$ interval is $M(g(\tau)+M-1)d\tau$. Dividing by $M^2$ for normalization, we obtain that $g^{(M)}(\tau)=1+\frac{g(\tau)-1}{M}$, which tends to unity in the limit that $M$ goes to infinity. Remember that the result $g(\tau)=1$ for all $\tau$ values characterizes Poisson processes up to the second order. It follows that if we consider a large collection of electrons submitted to the same optical field, the detection events tend to be Poisson distributed independently of the single-electron response.}
\end{itemize}

Let $d(t)$ be the number of events occurring up to time $t$, that is the number of $k$ values such that $t_k<t$. Obviously $d(0)=0$ since the $t_k$ are positive numbers. Let us prove that, for some measurement time $T$, $\ave{d(T)}=DT$, where $D$ is a constant called the intensity of the process.
We introduce the (positive) number $D_h(t):=d(t+h)-d(t)$ of events occurring between $t$ and $t+h$. Because the process considered is stationary $\ave{D_h(t)}$ does not depend on $t$. It is convenient to split the measurement time $T$ into time slots of duration $h=T/n$, labeled by $i=1,2,...n$. Eventually, we let $n$ go to infinity, so that it is unlikely that more than one event occur within any time slot. Thus, if $D_{i}\equiv D_{h}(\p i-1\q h)$ denotes the number of events occurring during slot $i=1,2...n$, we have either $D_{i}=1$ or $D_{i}=0$ and $\ave{D_i}$ does not depend on $i$. For later use note that ${D_i}^2=D_i$. The number $d(T)$ of events occuring during the measurement time $T$ is the sum of the $D_i$ with $i$ running from 1 to $n$, so that its average reads 
\begin{align}
\label{93}
\ave{d(T)}=\ave{\sum_{i=1}^{n} D_{i}}=n\ave{D_i}=\frac{T}{h}\ave{D_i}\equiv TD,
\end{align}
where we have set $D\equiv \ave{D_i}/h$.

Because the process is stationary its auto-correlation  $\ave{D_h(t+\tau)D_h(t)}$ does not depend on $t$ for every $h>0$ and every $\tau >0$. The degree of second order coherence $g(\tau)$ is the limit of $\ave{D_h(t+\tau)D_{h}(t)}/\ave{D_h(t)}^2$ as $h$ goes to 0. Let us set for $j>i$ 
\begin{align}
\label{94}
\ave{D_{i}D_{j}}\equiv\ave{D_{i}}^2g_{n}\bigl((j-i)\frac{T}{n}\bigr),
\end{align} 
and evaluate
\begin{align}
\label{95}
\ave{d(T)^{2}}&=\ave{\sum_{i=1}^{n}D_{i}\sum_{j=1}^{n}D_{j}}\nonumber\\
&=n\ave{D_{i}}+2\ave{D_{i}}^2\sum_{i=1}^{n} \sum_{j=i+1}^{n}g_{n}\bigl((j-i)\frac{T}{n}\bigr)\nonumber\\
&=\ave{d(T)}+2\ave{D_{i}}^2\sum_{i=1}^{n} (n-i)g_{n}(\frac{iT}{n})\nonumber\\
&=\ave{d(T)}+2D^2\frac{T}{n}\sum_{i=1}^{n} (T-\frac{iT}{n})g_{n}(\frac{iT}{n}). 
\end{align}
In the limit $n\to \infty$ the sum may be replaced by an integral and $g_{n}$ by $g$, thus 
\begin{align}
\label{99}
\ave{d(T)^{2}}=\ave{d(T)}+2D^2 \int _{0}^{T} d\tau(T-\tau)g(\tau).
\end{align}
After slight rearranging the variance of $d(T)$ may be written in the form
\begin{align}
\label{100}
\V (T)&\equiv\frac{\var(d(T))}{\ave{d(T)}}-1= \frac{\ave{d(T)^{2}}-\ave{d(T)}^2}{\ave{d(T)}}-1\nonumber\\
&=2D \int _{0}^{T} d\tau(1-\frac{\tau}{T})\bigl(g(\tau)-1\bigr) 
\end{align}
since $\int_{0}^{T} d\tau (1-\tau /T)=T/2$.
The motivation for introducing $g(\tau)-1$ in the integral is that this quantity usually tends to 0 quickly as $\tau$ tends to infinity. Intuitively, this is because widely separated events tend to be independent and consequently in that limit $\ave{D_{i}D_{j}}\approx \ave{D_{i}}\ave{D_{j}}=\ave{D_{i}}^2$. Setting $D=1$ for brevity, relation \eqref{100} may be written as,
\begin{align}
\label{xx}
P_c(\tau)=\sum_{k=0}^{\infty}k^2\frac{d^2P(k,\tau)}{d\tau^2},
\end{align} 
where $P_c(\tau)dt d\tau$ denotes the probability density of having an event between 0 and $dt$ and an event between $\tau$ and $\tau+d\tau$ or, equivalently, $P_c(\tau)d\tau$ is the probability density of another event being registered during the time interval $\tau$ and $\tau+ d\tau$, given that an event occurred at $t=0$. In \eqref{xx} $P(k,\tau)$ denotes the probability of $k$ events being registered between $t=0$ and $t=\tau$.

We define the event rate $\D(t)=D+\De D(t)$ as the sum over $k$ of $\delta (t-t_{k})$, where $\delta (.)$ denotes the Dirac distribution. Going back to time slots of small duration $h$, the event rate is the sum over $i$ of $D_i/h$ where, as before, $D_i=0$ or 1. The denominator $h$ may be omitted because of the subsequent normalization. The calculations given below parallel the ones given above in relation with the photo-count variance and some details will therefore be omitted. The spectral density is defined in terms of the event times $t_{k}$ occurring during runs of duration $T$, to be later tend to infinity, see \eqref{def}.
\begin{align}
\label{102}
T\spectral_{\Delta D}(\Omega)&=
\ave{\sum_{i=1}^{n}D_{i}\exp(-\mathfrak{j}\Omega i)
\sum_{j=1}^{n} D_{j}\exp(\mathfrak{j}\Omega j)}\\
&=n\ave{D_{i}^2}+D^2(\frac{T}{n})^2
\sum_{i=1}^{n} \sum_{j=i+1}^{n} g_{n}\bigl((j-i)\frac{T}{n}\bigr)\cos( (j-i)\Omega)\\
&=DT+D^2\frac{T}{n}\sum_{i=1}^{n} (T-\frac{iT}{n})g_{n}(\frac{iT}{n})\cos(i\Omega).
\end{align} 
The above expression may be transformed as was done earlier for evaluating the 
variance of $d$. On account of the fact that
\begin{align}
\label{107}
2\sum _{i=1}^{T}(1-\frac{i }{T}) \cos (\frac{2\pi i n }{T})+1=0,
\end{align}
where $n$ denotes any non-zero integer, converting the sum into an integral, and in the large $T$ limit, we obtain
\begin{align}
\label{108}
\N(\Om)\equiv \spectral_{\De D/D}(\Om)-\frac{1}{D}=2 \int 
_{0 }^{\infty }  d\tau\bigl(g(\tau)-1\bigr)\cos(\Om \tau).
\end{align}

This relation between the relative noise $\N(\Om)$ and the normalized correlation function $g(\tau)$ has been established directly for point processes. It is however instructive to show how this relation may be alternatively derived from the Wiener-Khintchine (WK) theorem in \eqref{def}. We consider the event-rate process $\D(t)=\sum_k \de(t-t_k)$ which, in the limit $h\to 0$ corresponds to $D_h(t)/h$, where $D_h(t)$ is the number of events between $t$ and $t+h$, as defined earlier. For simplicity we suppose, without loss of generality, that the average rate $D$ is unity. If $R(\tau)$ denotes the correlation of $\D(t)$, the WK theorem tells us that
\begin{align}
\label{wk}
\spectral_\D(\Om)=\int_{-\infty}^{+\infty}{d\tau~ R(\tau)\exp(\jj\Om \tau)}.
\end{align}
But $R(\tau)$ presents a singularity at $\tau=0$. We are thus led to define $g(\tau)=R(\tau)-\de(\tau)$. On the other hand, $\spectral_\D(\Om)$ presents a singularity $2\pi \de(\Om)$ but $\spectral_{\De D}(\Om)=\spectral_\D(\Om)-2\pi \de(\Om)$ is free from singularity. Substituting these expressions in the above WK equation and using some form of the $\de$-distribution we recover \eqref{108}.
Since presently the correlation $R(0)=\infty$, the restriction $-R(0)≤R(\tau)≤R(0)$ established in Section \ref{notation} does not entail any restriction on $g(\tau)$.

\subsection{Useful integrals}\label{integrals}

A number of integrals from $x=-\infty$ to $x=\infty$ will be needed 
in subsequent papers, which may be evaluated by contour 
integration. The method is as follows.

Complex numbers are denoted by $z\equiv z'+\ii z''$, where 
$\ii^2=-1$. The complex conjugate of $z$ is denoted $z^\star\equiv 
z'-\ii z''$. Let $f(z)$ be a function of $z$ whose only 
singularities are simple poles at $z_{1},~z_{2}\ldots$. One calls 
\emph{residue} at $z_{k}$ the coefficient of $(z-z_{k})^{-1}$ in the 
(Laurent) series expansion of $f(z)$ near $z_{k}$. The integral of 
$f(z)$ along a closed counterclockwise contour is equal to $2\pi $\ii  ~
times the sum of the enclosed pole residues.

For example, closing the real axis by an upper half-circle of infinite 
radius we obtain
\begin{align}\label{residue}
\int _{-\infty}^{\infty}\frac{\textrm{d}x}{1+x^2}=\int 
_{-\infty}^{\infty}\frac{\textrm{d}x}{(x-\ii)(x+\ii)}=2\pi 
\ii~ \frac{1}{\ii+\ii}=\pi
  \end{align}
   Here we have a single enclosed pole at $x=\ii$. The 
  coefficient of $1/(x-\ii)$ in the integrand is 
  $1/(2\ii)$ when $x=\ii$. We obtain similarly
 \begin{align}\label{first int}
\frac{1}{\pi }\int _{-\infty}^{\infty}\frac{\textrm{d}x}{(1-ax^2)^2 +x^2}=
\frac{1}{\pi }\int _{-\infty}^{\infty}\frac{\textrm{d}x~ ax^2}{(1-ax^2)^2 +x^2}=1
  \end{align}
where $a$ denotes a non-zero constant.

Further, for application to inhomogeneously-broadened lasers, let us define a weight function
 \begin{align}\label{weight}
w(x)\equiv \frac{(g-1)/\pi }{(g-1)^2 +x^2}
  \end{align}
  that reduces to the Dirac $\delta$-distribution when $g$ tends to 
  1, and 
   \begin{align}\label{Imn}
I_{mn}\equiv 8(g^2+y^2)^m \int 
_{-\infty}^{\infty}\frac{\textrm{d}x~ w(x-y)x^n}{(1+x^2)^m} 
  \end{align}
  
  We obtain
 \begin{align}\label{Imn bis}
I_{10}&=8g\\
I_{12}&=8y^2+8g(g-1)\\
I_{20}&=4y^2(g-1)+4g^2(g+1)\\
I_{21}&=8gy\\
I_{30}&=3(g-1)y^4+6g(g^2-1)y^2+g^3(3g^2+3g+2)\\
I_{31}&=2y[y^2(g-1)+g^2(g+3)]\\
I_{32}&=(g-1)y^4+2g(g^2+3)y^2+g^3(g-1)(g+2)\\
I_{33}&=2y[(3g+1)y^2+3g^2(g-1)]\\
I_{34}&=(3g+5)y^4+6g(g^2-1)y^2+g^3(3g-2)(g-1) 
\end{align}
We also need for evaluating the gain of semiconductors
   \begin{align}\label{Imnb}
\frac{1}{\pi} \int 
_{0}^{\infty}\frac{\textrm{d}x~ \sqrt 
{x}}{(x+1)(x-a)}=\frac{1}{1+\sqrt{-a}} 
  \end{align}
if $a$ is negative, and $1/(1+a)$ if $a$ is positive. In the latter 
case, the integral is understood in principal value.

\subsection{Bi-complex representation of signals}\label{bicomplex}

As recalled in Section \ref{notation} electrical engineers usually factor out a term exp(\jj$\Omega t$) to represent time-harmonic sources. $\Omega\equiv 2\pi f$ is 
the angular \emph{baseband} (or "Fourier") angular frequency. The real signal is obtained by taking the real part of the product $V$exp(\jj$\Omega t$), where $V$ denotes some 
complex number. Only time-invariant linear causal systems are presently
considered. It follows that the system response has 
the same form as the applied source. In Physics, it is usual to factor out a term of 
the form exp(-\ii$\omega t$) where $\omega\equiv 2\pi \nu $ denotes the \emph{carrier} angular frequency. 

When a source at frequency $\omega$ is modulated at frequency 
$\Omega$, the bi-complex representation described below proves useful. 
To avoid bothering with minus signs, it is convenient to set 
$i_{1}\equiv$-\ii~ and $i_{2}\equiv$\jj. Further, we set 
 $p_{1}\equiv i_{1} \omega$ and $p_{2}\equiv i_{2} 
\Omega $. The algebra of bi-complex numbers is associative and commutative. A
bi-complex number is written as
   \begin{align}\label{bi}
\mathcal {V} = a+b i_{1}+ci_{2}+di_{1}i_{2},
  \end{align} 
where $i_{1}^2=i_{2}^2=-1$, $i_{1}i_{2}=i_{2}i_{1}\equiv j,~j^2=1$, and $a,b,c,d$ 
are real numbers. The algebra of bi-complex numbers was discovered by Segre in 1892. A modern account of the bi-complex numbers algebra may be found, for example, in \cite{Rochon2004}. A bi-complex number is invertible if $a^2+b^2+c^2+d^2 ≠ ± 2\p ad-bc\q$. To prove it, multiply $a+b i_{1}+ci_{2}+dj$ by $a-b i_{1}-ci_{2}+dj$ and obtain $A+Bj$, where $A=a^2+b^2+c^2+d^2$ and $B=2\p ad-bc\q$ are real numbers. Next, note that $\p A+Bj\q \p A-Bj\q=A^2-B^2$ is real. The condition for a bi-complex number to be invertible is therefore that $A≠±B$, which is the above condition.

 The real signal $v(t)$ is equal to $\mathcal {V}(p_{1}, 
 p_{2})\exp \left(\p p_{1}+ p_{2}\q t\right)$ +ccc, where "ccc" 
 means that one must add 3 terms to the one written out, one with $p_{1}$ changed to 
 $-p_{1}$, the second with $p_{2}$ changed to $-p_{2}$, and the third 
 with both $p_{1}$ and $p_{2}$ changed to $-p_{1}$ and $-p_{2}$. Let us now state the basic theorem. If $v(t)$ denotes a modulated 
voltage represented by $\V (p_{1}, p_{2})$ as said above, $Y(p)$ denotes the usual complex circuit admittance (the 
ratio of two real polynomials in $p$, and  
$i(t)$ the real electrical current flowing through the circuit, we have  
   \begin{align}\label{theorem}
i(t)=Y(p_{1}+ p_{2})\mathcal {V}(p_{1}, p_{2}) \textrm{exp}[(p_{1}+ 
p_{2})t]+\textrm{ccc}. 
  \end{align} 

\newpage

\section{Classical Circuit Theory}\label{circuittheory}

We recall in the present section the basic concepts employed in Circuit Theory, and describe devices useful in radio, microwave, and optical frequency ranges, considering both resonators and transmission lines. The circuit equations provide currents $i(t)$, viewed as (real, linear, causal) responses to specified potential sources $v(t)$. 

We first consider conservative elements (that is, elements that conserve energy) such as capacitances, $C$, and inductances, $L$. We next consider non-conservative elements such as positive or negative conductances. In most practical cases the conductances are obtained from separate measurements. We postpone to Section \ref{qp} a microscopic conductance model, namely a single electron located between two parallel conducting plates. A potential $v(t)$ oscillating at an optical frequency $\om$ is applied to the plates. We suppose exact resonance between $v(t)$ an the electron natural oscillatory motion at frequency $\om_o$.  The (Quantum-Mechanical average) induced current $i(t)$ is proportional to the average electron momentum, and, under circumstances to be discussed, the ratio $\ave{i(t)}/v(t)$ may be a real constant $G$. When $eU$ is slightly smaller than $\hbar\om_o$, the conductance is positive and the optical potential delivers energy to the static potential. On the other hand, when $eU$ slightly exceeds $\hbar\om_o$ the conductance is negative and the optical field receives energy from the static potential. When $eU$ is precisely equal to $\hbar\om_o$ the conductance vanishes but fluctuations remain. To summarize, the complete system including static and optical potentials conserve energy aside from an irreversible loss of energy $\abs{eU-\hbar\om_o}$ that can be made as small as one wishes.

Accordingly, the circuit theory may be applied to non-conservative elements involving, besides $C$ and $L$, conductances $G$. We mainly consider sources (and responses) that vary sinusoidally in the course of time at frequency $\om=\om_o$. For a closed linear system, the equations have solutions only for discrete complex values $\om_n$ of $\om$. We will be particularly interested in circuits that have only one nearly-real frequency, the other ones having large negative imaginary parts, corresponding to strongly damped modes. 

We first evaluate the current $i(t)$ induced in a potential source $v(t)$ using the Classical Theory of Electron Motion. This is a deterministic problem and no fluctuations are involved if we ignore the random electron motion due to heat and the discreteness of the electrons. In the realm of Quantum Mechanics we may evaluate the average value and the higher moments of the current, but the function $i(t)$ itself, for a single sample of the ensemble, is undefined.

\subsection{Classical devices}\label{circuit}

Let us recall basic results. The complex notation often employed for describing quantities that vary sinusoidally in time is recalled in Section \ref {notation}. For strictly sinusoidal potentials and currents represented by the complex numbers $V$ and $I$, respectively, and linear circuits, we have the generalized Ohm law $I=Y(\om)V$, where the complex constant of proportionality $Y(\om)$, called the admittance, and its inverse the impedance $Z(\om)$, in general depend on the frequency $\om$, which may vary from minus to plus infinity.

If a potential $v(t)$ is applied to a conductance $G$ (a real number), we have by definition $i(t)=Gv(t)$, or, using the complex notation, $I=GV$, where $V$, and thus $I$, are in general complex numbers. We consider in the major part of this paper \emph{ideal} conductances defined as follows: They are supposed to be independent of the driving potential $V$ and to be independent of frequency. Furthermore, they are supposed to have a fixed energy content that may be set equal to zero since only energy differences are relevant. A physical model for ideal conductances is a piece of metal having a large number of inelastic scattering centers. Electrons accelerated by the applied field quickly loose their energy, which is converted into heat. Under such circumstances the electron kinetic energy remains negligible, and thus the total energy is fixed. In contradistinction, the input conductance of a loss-less transmission line of characteristic conductance $G_c$ terminated by an ideal conductance $G=G_c$ (matched load) is equal to $G$ at any frequency. But there is in that case a stored energy equal to $G\abs{V}^2\tau$, where $\tau$ denotes the transit time of a pulse along the transmission line (this is power divided by the group velocity times the line length). Thus, a matched transmission line does \emph{not} constitute an ideal conductance in the sense defined above, even though the input conductance $G$ is a real constant. 

\begin{figure}
\setlength{\figwidth}{0.18\textwidth}
\centering
\begin{tabular}{cccc}
\includegraphics[width=\figwidth]{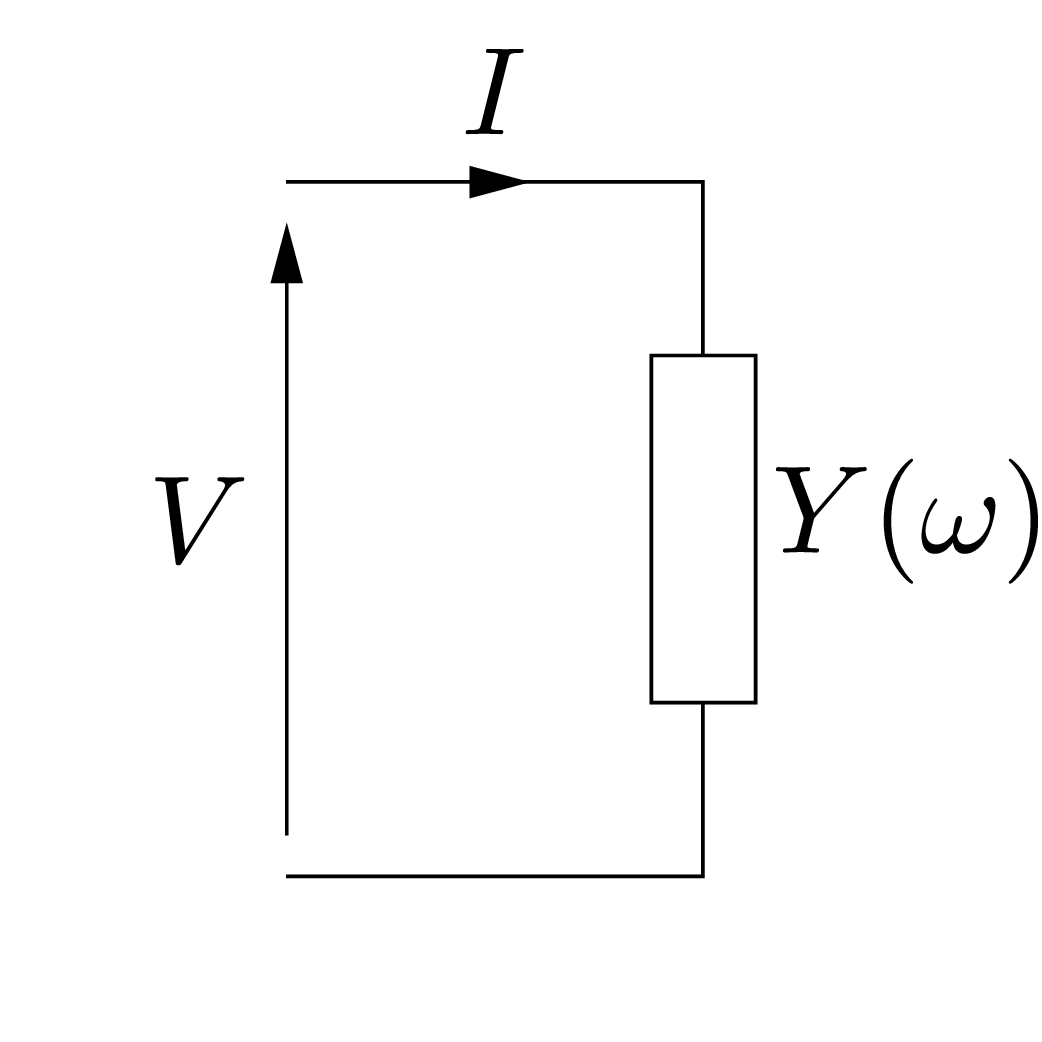} & \includegraphics[width=\figwidth]{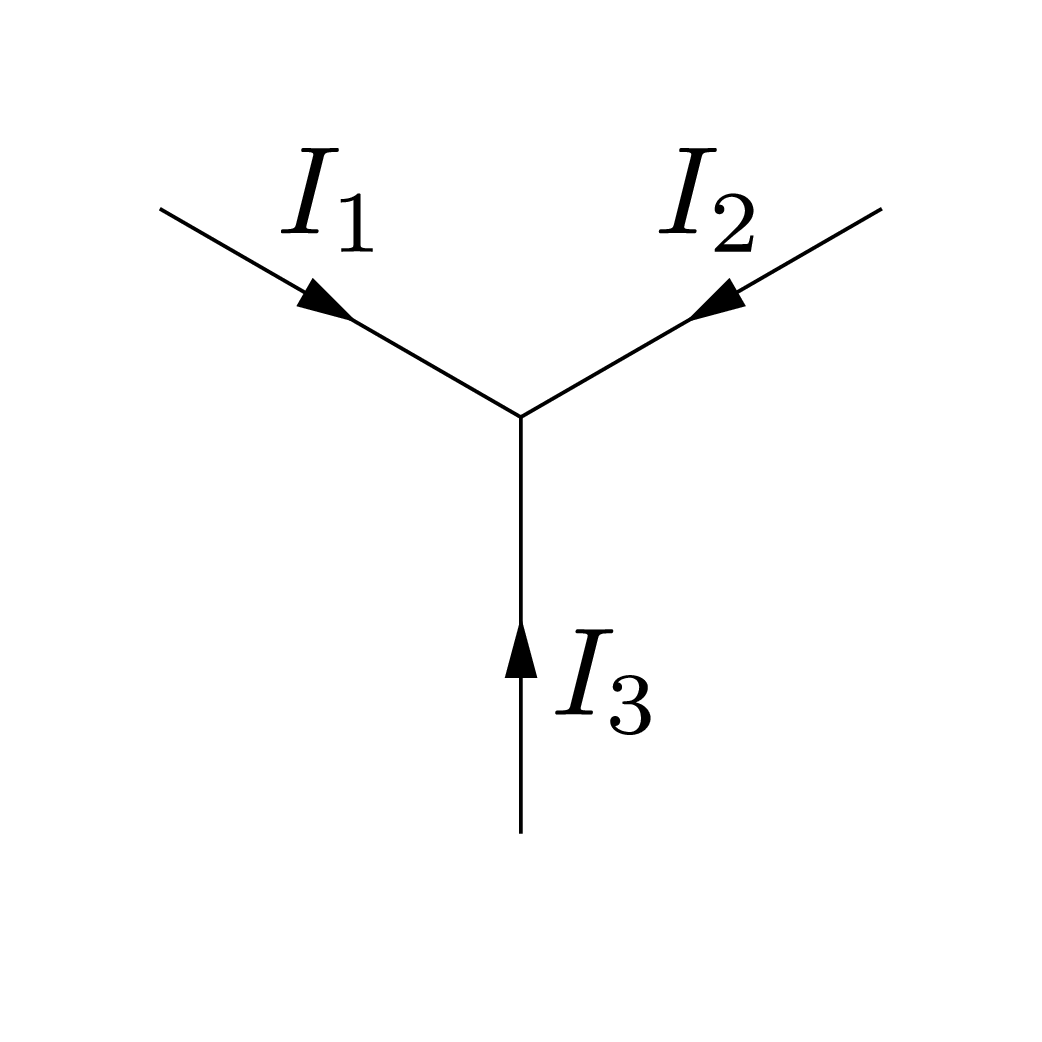} &  \includegraphics[width=\figwidth]{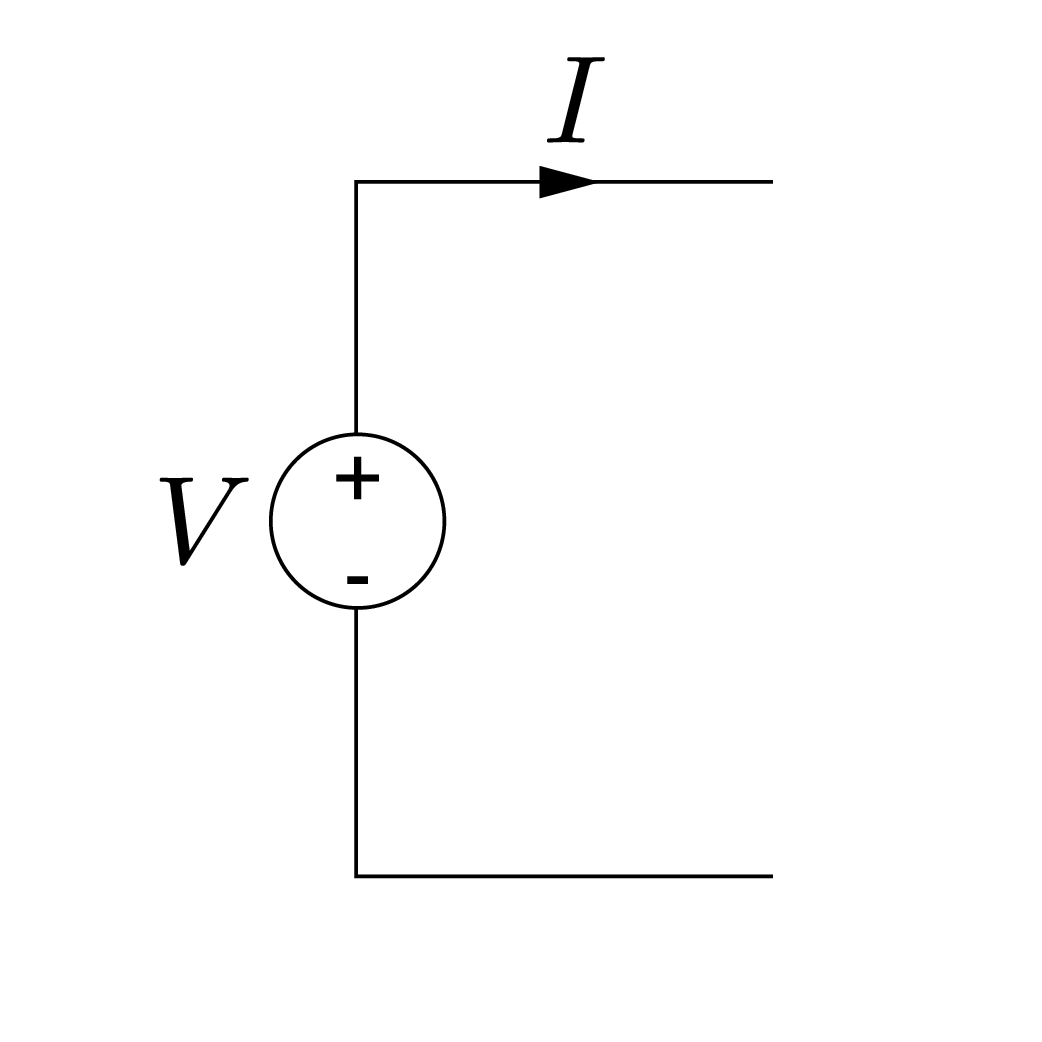} & \includegraphics[width=\figwidth]{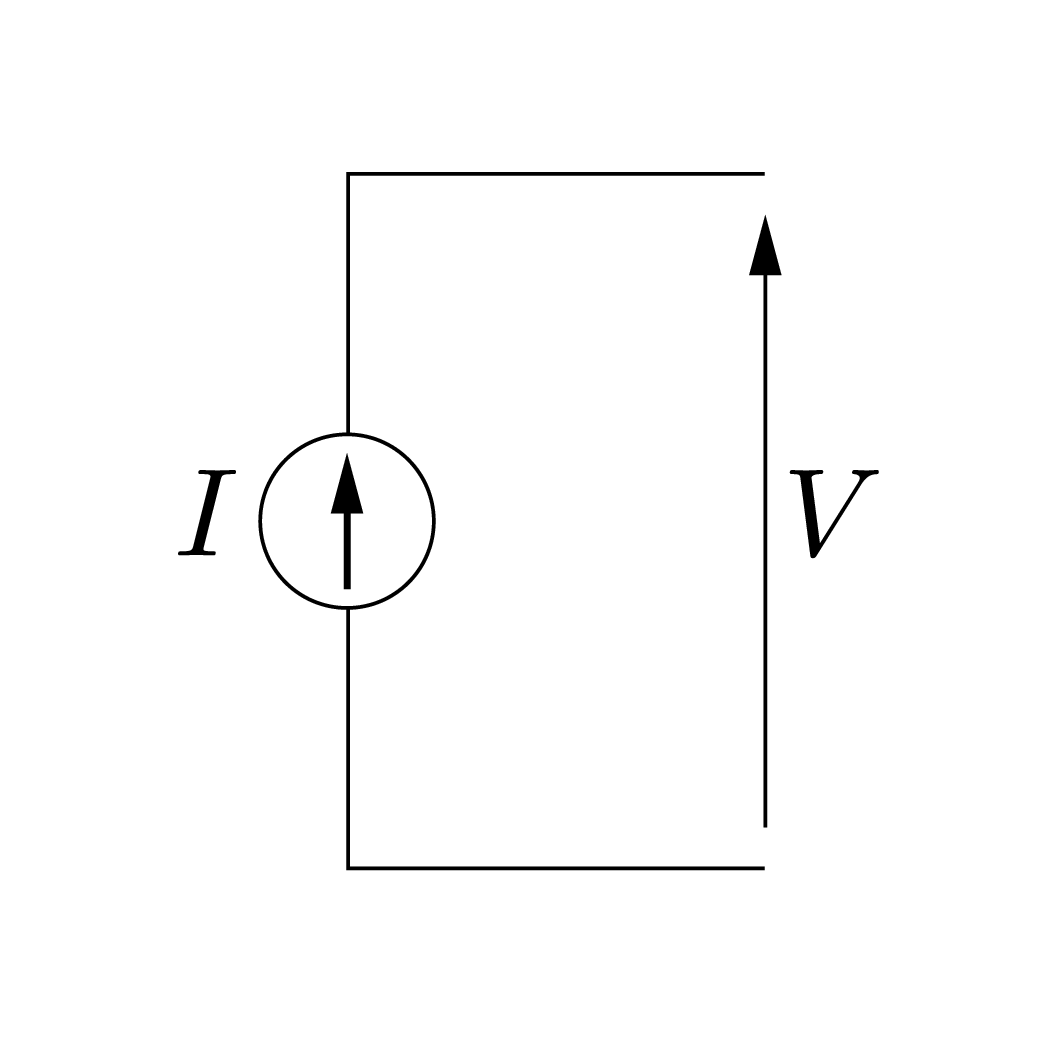}\\
(a) & (b) & (c) & (d) \\
\includegraphics[width=\figwidth]{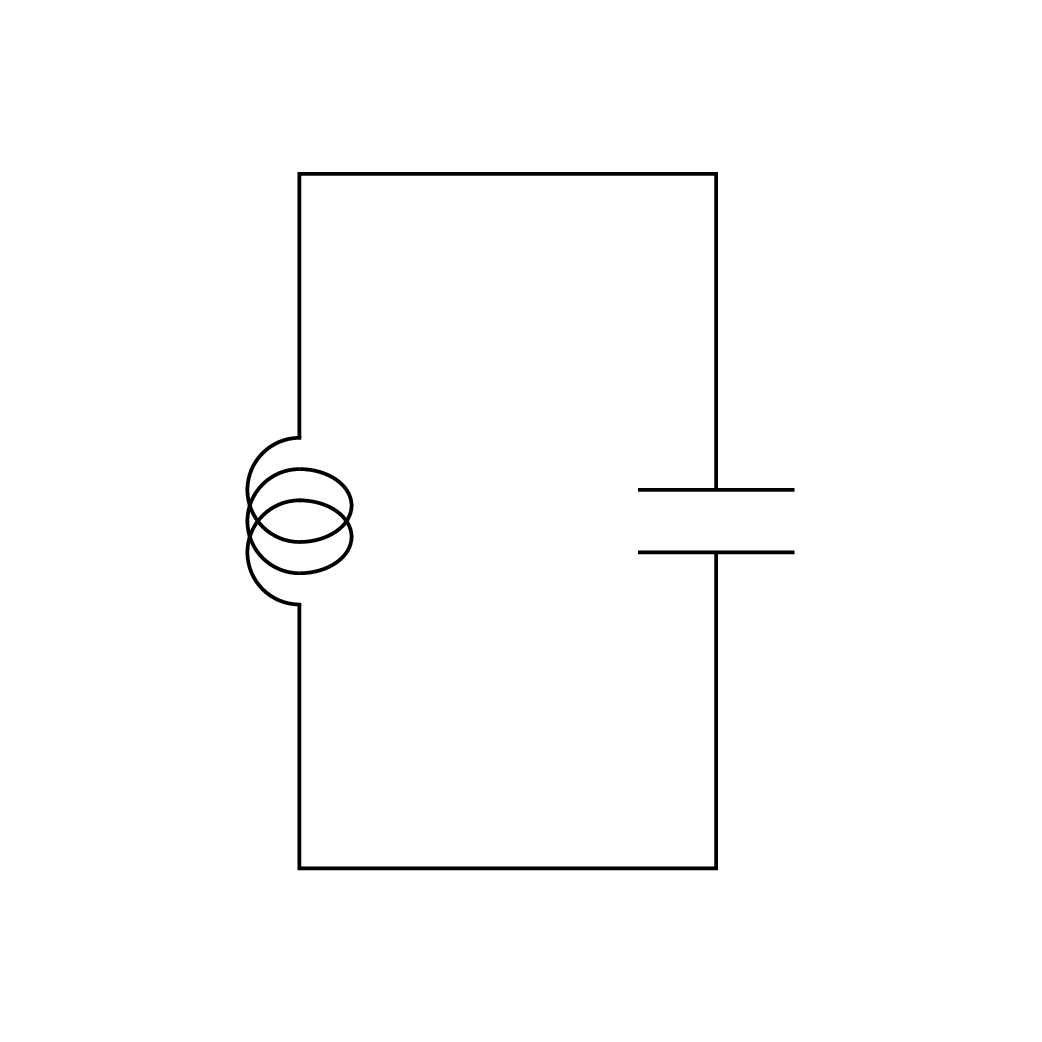} & \includegraphics[width=\figwidth]{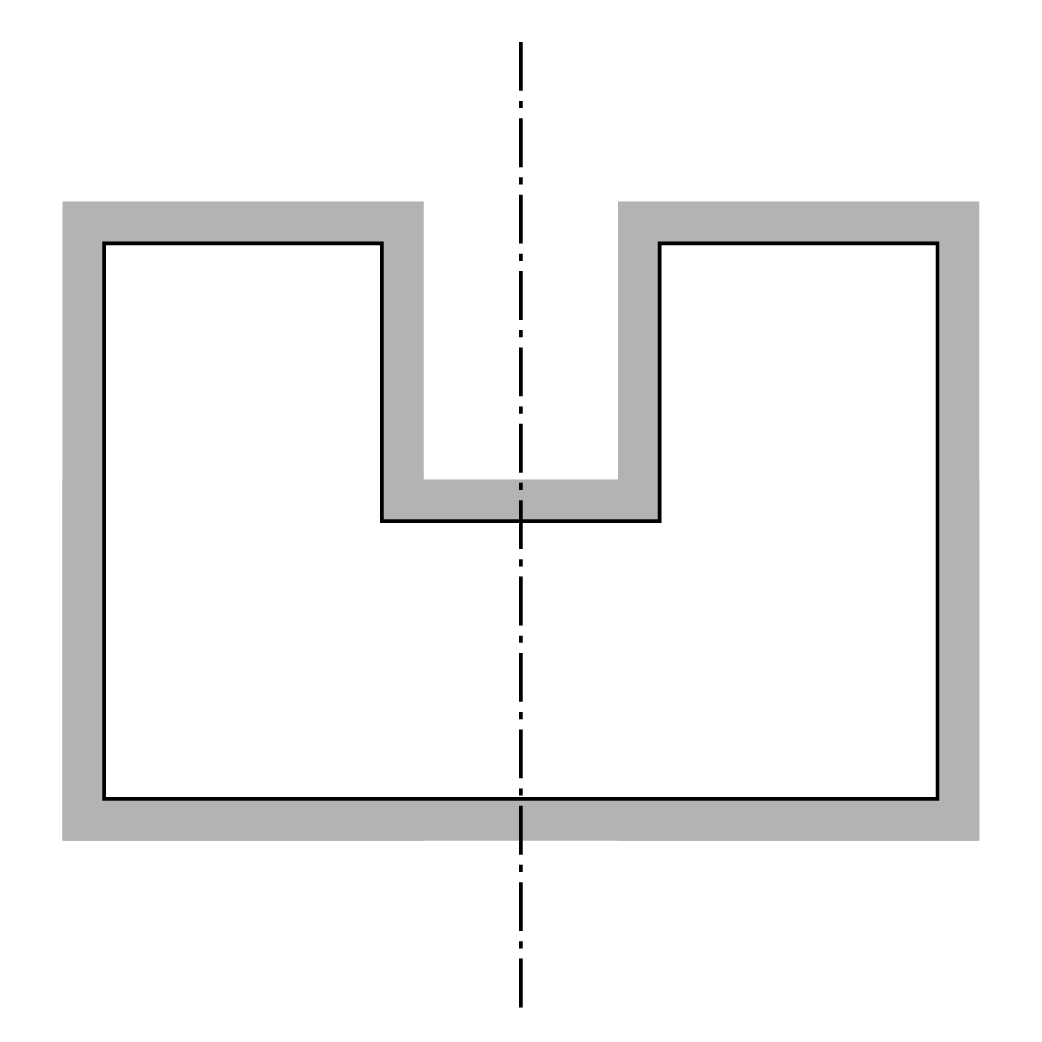} &  \includegraphics[width=\figwidth]{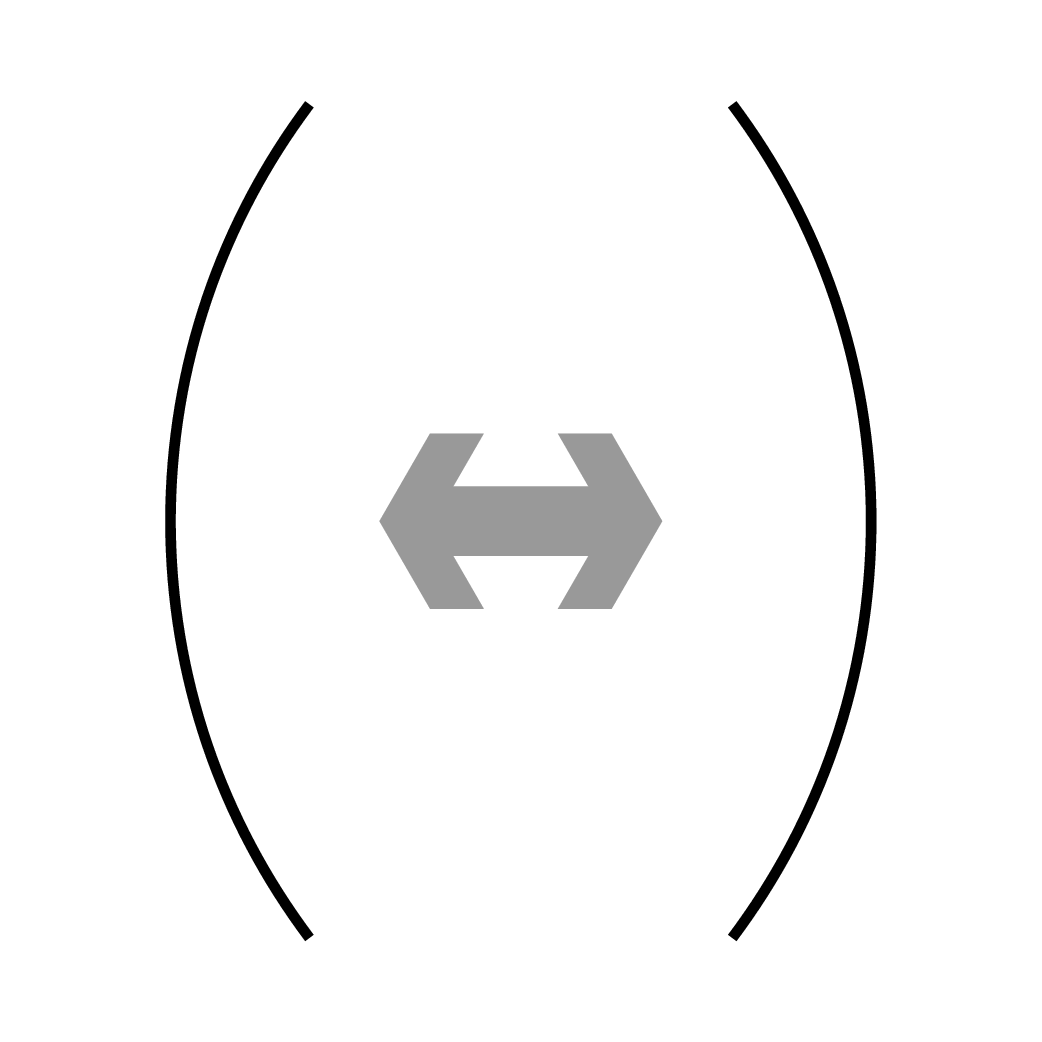} & \includegraphics[width=\figwidth]{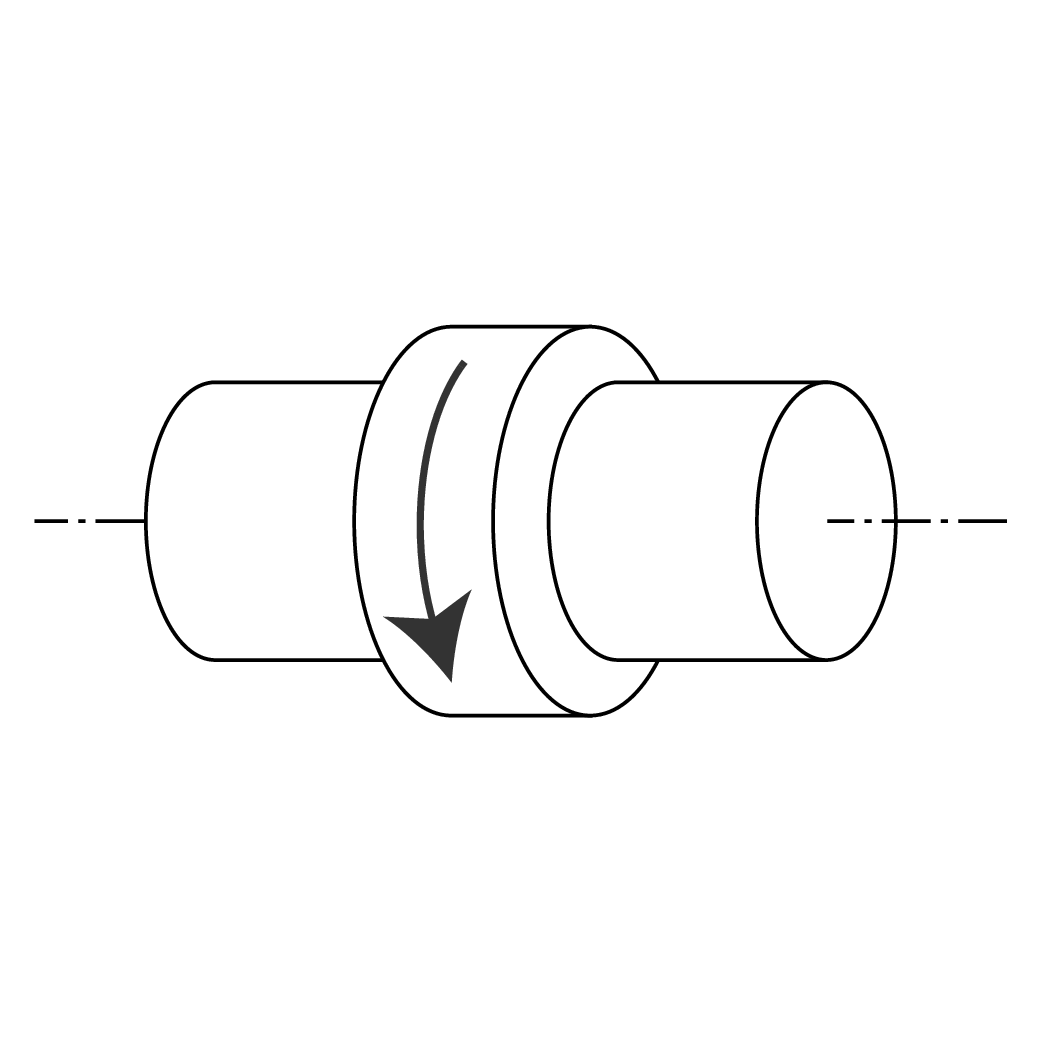}\\
(e) & (f) & (g) & (h) \\
\includegraphics[width=\figwidth]{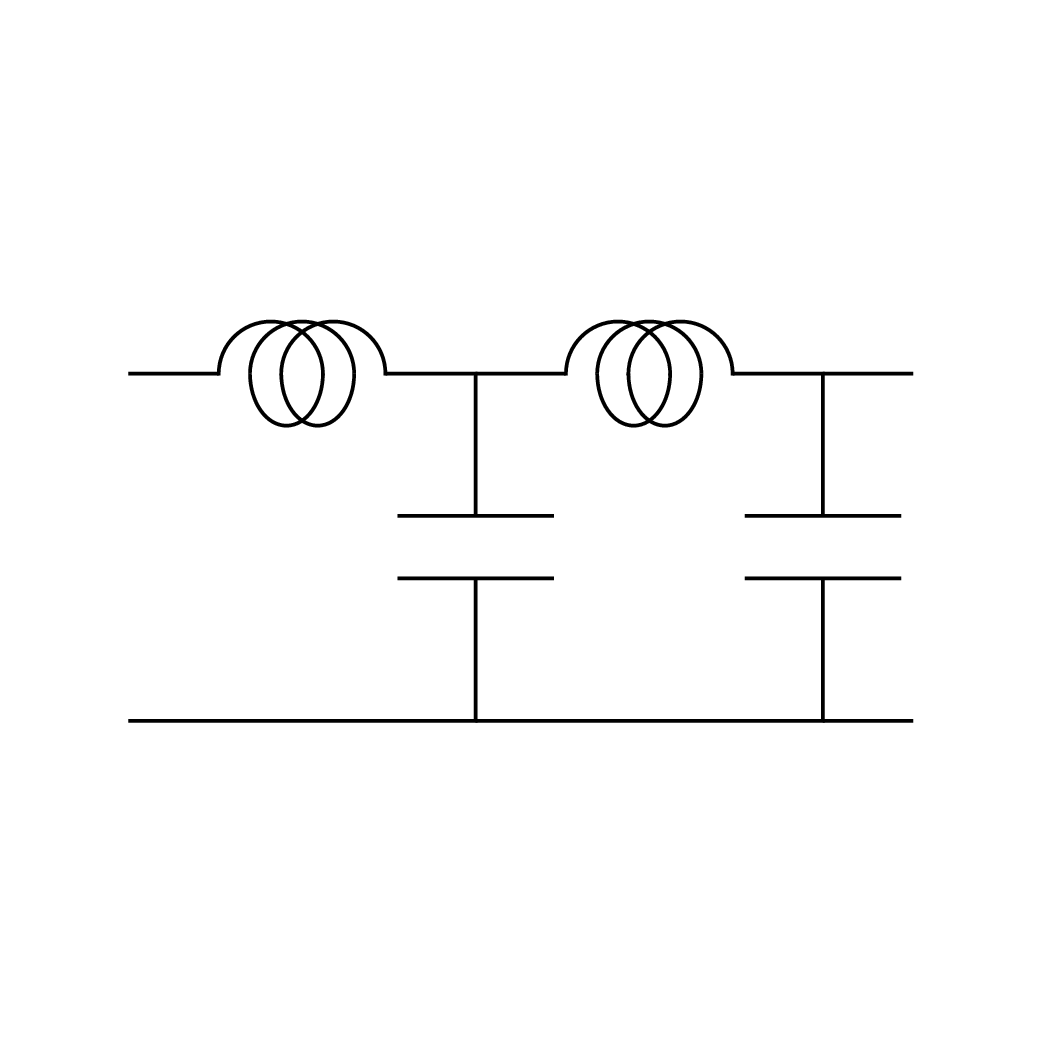} & \includegraphics[width=\figwidth]{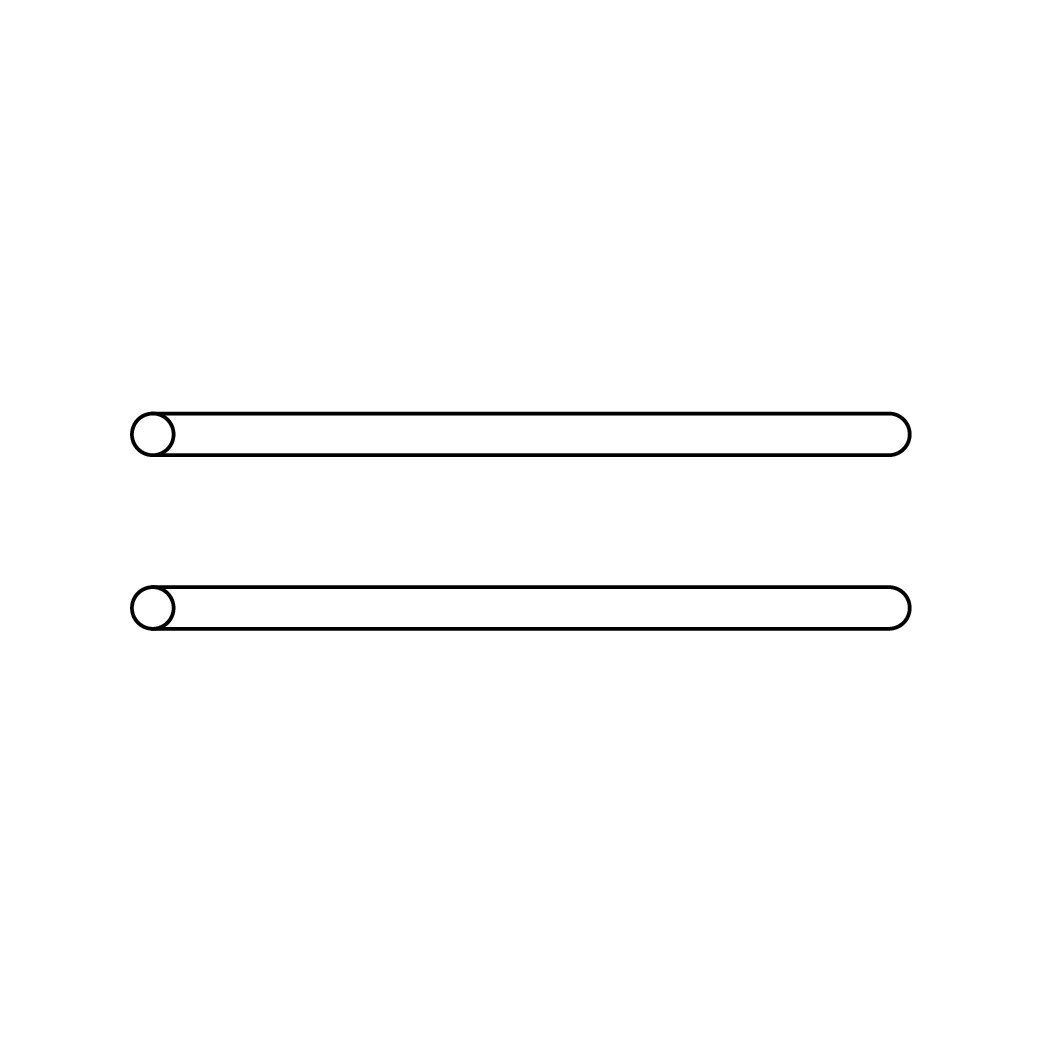} &  \includegraphics[width=\figwidth]{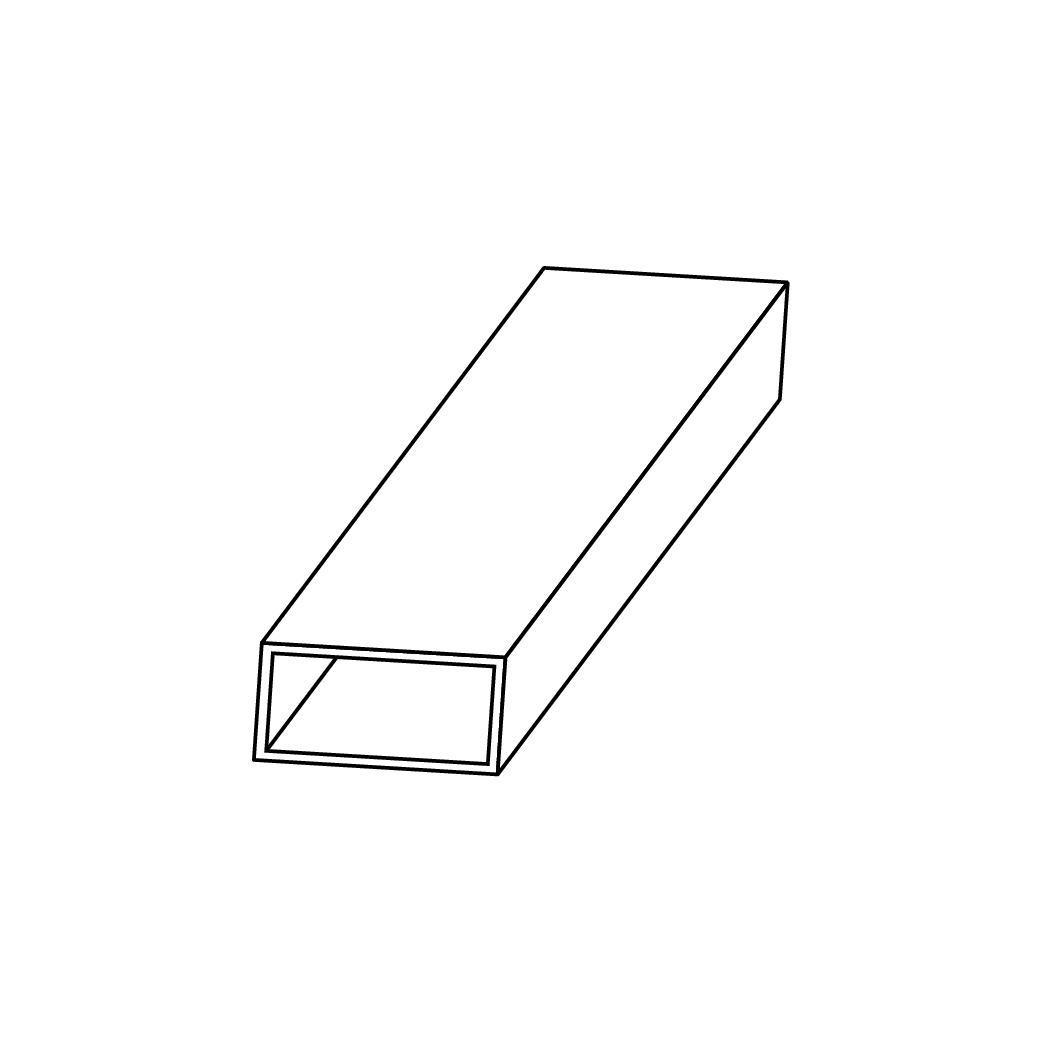} & \includegraphics[width=\figwidth]{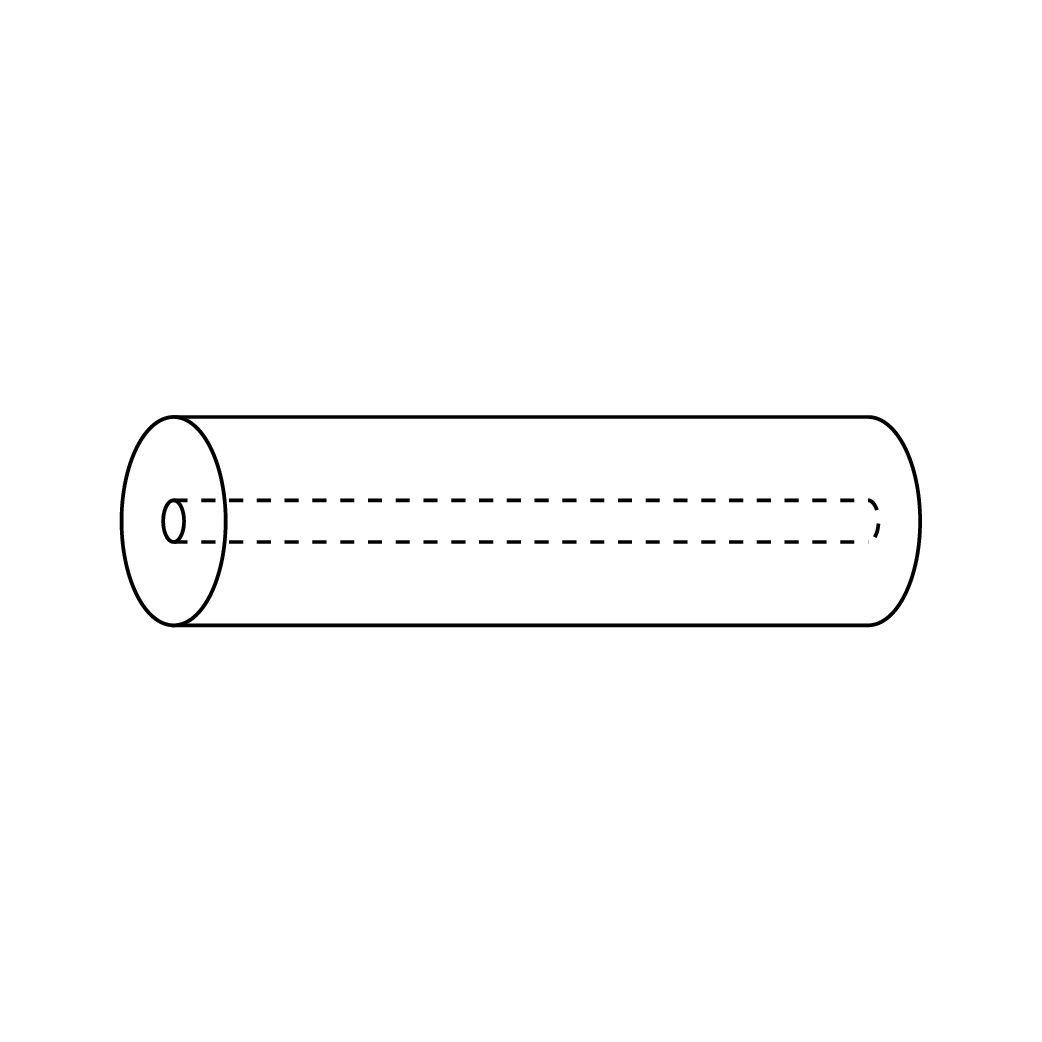}\\
(i) & (j) & (k) & (l) \\
\includegraphics[width=\figwidth]{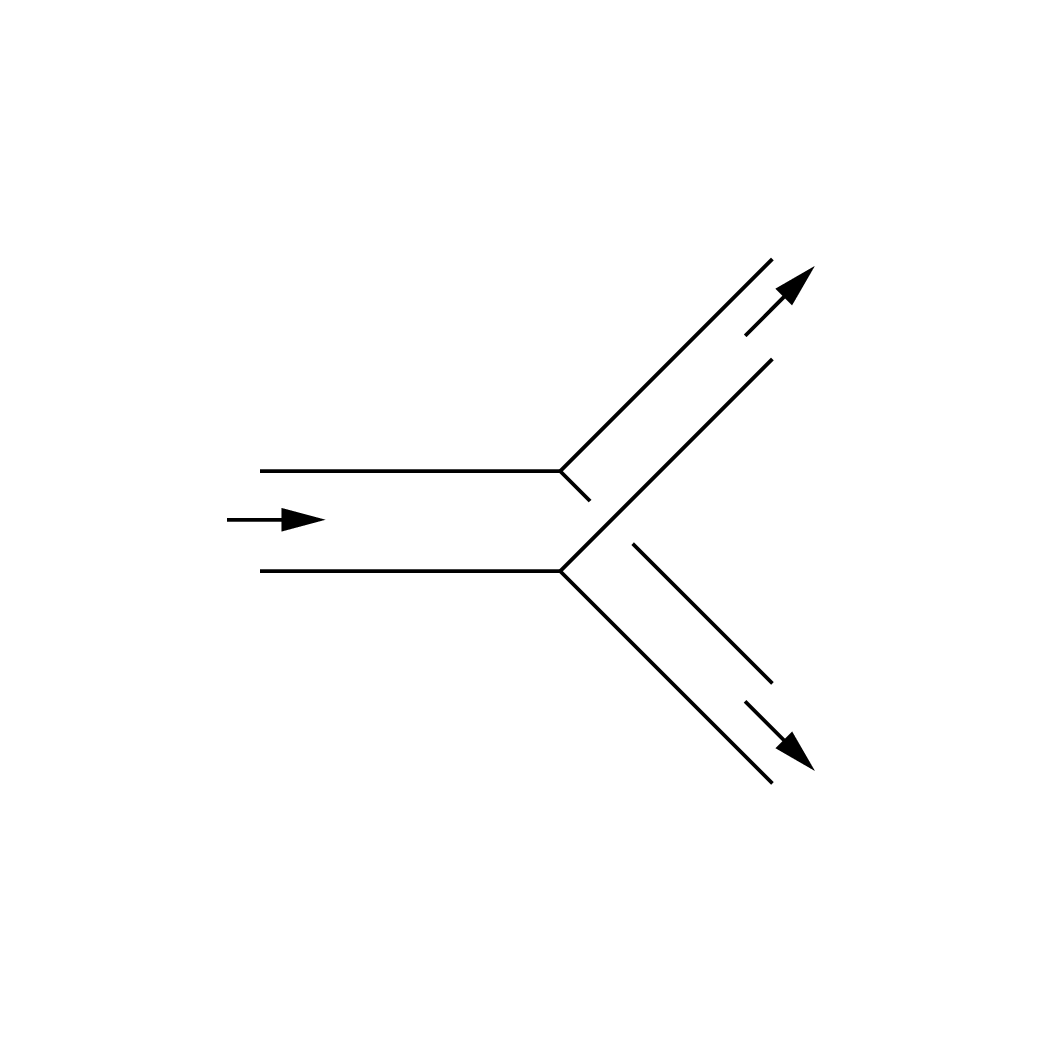} & \includegraphics[width=\figwidth]{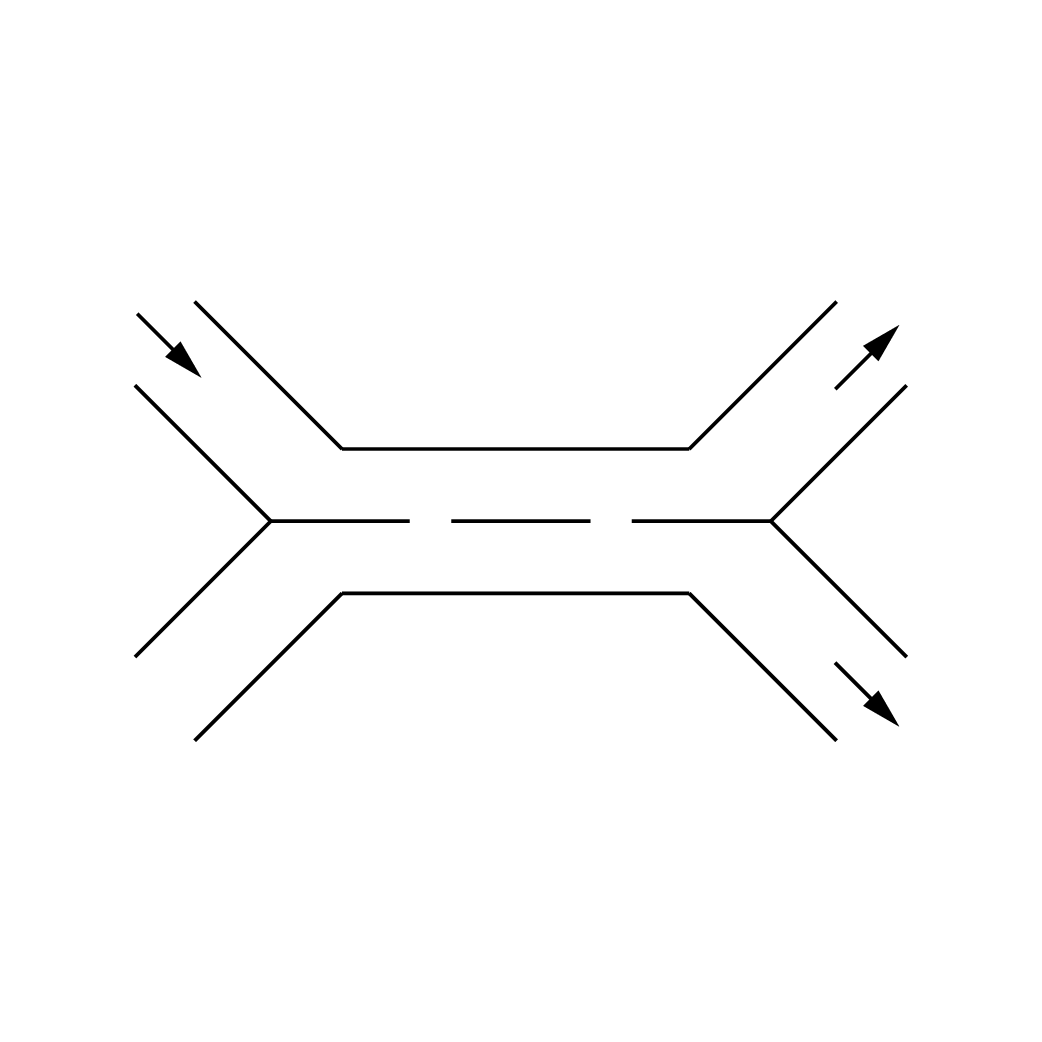} &  \includegraphics[width=\figwidth]{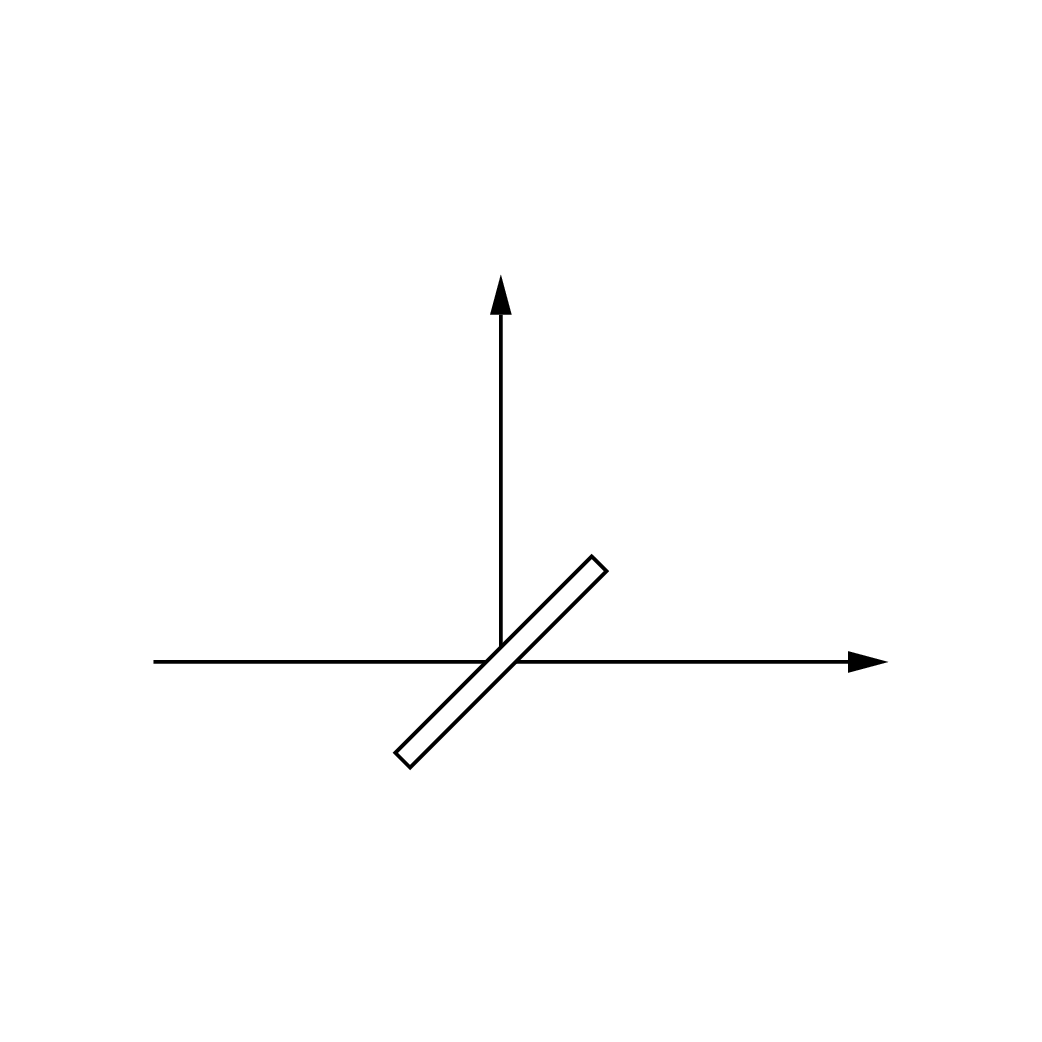} & \includegraphics[width=\figwidth]{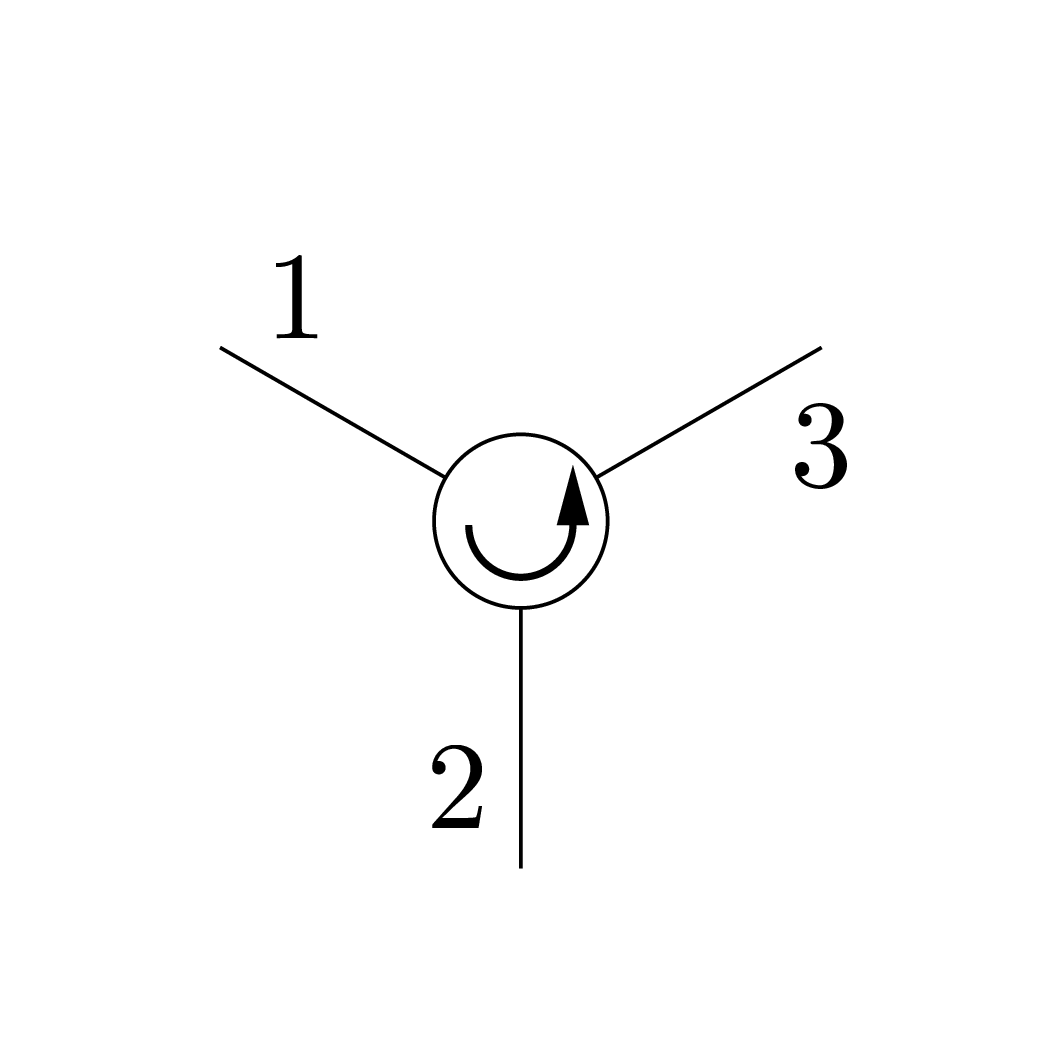}\\
(m) & (n) & (o) & (p) \\
\end{tabular}
\caption{a) There is a linear relation between the potential $V$ and the current $I$ at some frequency $\om$ (generalized Ohm law $I=Y(\om)V$).  b) Illustrates the Kirchhoff law: $I_1+I_2+I_3=0$. c) Represents a potential source with $V$ independent of $I$. d) Represents a current source with $I$ independent of $V$. e) Inductance-capacitance resonating circuit. f) Cavity employed, e.g., in reflex klystrons. g) Fabry-Pérot-type optical resonator with two curved mirrors facing each other. h) Whispering-gallery mode resonator. i) Low-pass filter, j) Parallel conductors, k) Waveguide, l) Optical fiber. Waves may be split in various ways: m) A transmission line is connected to two transmission lines whose characteristic conductances sum up to the original line characteristic conductance. n) Directional coupler. The two holes are spaced a quarter of a wavelength apart. o) The beam splitter is an optical equivalent of the directional coupler. p) The circulator is ideally a loss-less non-reciprocal device.}
\label{circuitfig}
\end{figure} 

\subsection{Capacitances}\label{capacitance}

A capacitance $C=\epsilon_o A/d_C$ may consist of two parallel perfectly-conducting plates of area $A$ separated by a distance $d_C\ll \sqrt A$. The constant $\epsilon_o$ in this formula is called the free-space permittivity. Its exact numerical value is given at the end of Section \ref{history}. There are two wires connected respectively to the upper and lower plates, so that electrical charges may be introduced or removed.  If an electrical charge $q$ is displaced from the (say, lower) plate to the upper plate a potential $v$ appears between the two plates given by $v=q/C$. The energy stored in the capacitance is $E_C=Cv^2/2=q^2/(2C)$, a result obtained by considering elementary charges $dq$ being displaced from the lower to the upper plate of the initially-uncharged capacitance until a final charge $q$ is reached. If $q$ is a function of time and $C$ is kept constant, we have $v(t)=q(t)/C$. We may set $q=-Ne$, where the number $N$ of electrons is supposed to be so large that $q$ varies almost continuously. As before, $e$ denotes the absolute value of the electron charge. 

Let now $v(t)$ be of the sinusoidal form given above. Because the current $i(t)$ represents a flow of electrical charges into one plate or flowing out from the other plate, $i(t)$ is the time-derivative of the electrical charge: $i(t)=dq(t)/dt$. The relation between the complex current $I$ and the complex potential $V$, as defined above, thus reads $I=-\ii C \om V$. The admittance is in the present case $Y(\om)=-\ii C\om$. If we set $Y=G+\ii B$, we have therefore for an ideal capacitance $G=0$ and $B=- C\om$. The stored energy averaged over a period $2\pi/\om$, reads $\ave{E_C}=C\abs{V}^2/2$. In following paragraphs, we will consider a light-emitting device driven by a very large capacitance (instead of, say, a battery) with a very large initial charge $q$ such that the potential $U=q/C$ across the capacitance has the desired value, for example 1 volt. If the light-emitter operation duration is denoted by $T$, the capacitance supplies a current $i$ during that time, and thus loose a charge $\De q=iT$. Because $q$ is very large we have $\De q\ll q$ provided the experiment does not last too long. As a consequence the potential $U$ across the capacitance does not vary appreciably. We realize in that manner a constant-potential source, that is a source whose potential does not depend appreciably on the delivered current.

A constant-potential source at optical frequency $\om$ may be realized in a similar manner. Again the capacitance $C$ and the initial charge $q$ are supposed to be arbitrarily large, but we now allow the spacing $d_C$ between the capacitance plates to fluctuate\footnote{Practically-minded readers may object that mechanical motion may not be feasible at high frequencies. Let us recall here that the numerical value of $\om$ is arbitrary. What we call "optical" frequency $\om/2\pi$ may be as low as 1Hz provided that the other frequencies considered be much lower, e.g., 1 mHz.} at the optical frequency $\om$. This spacing variation entails a fluctuation of the capacitance, and thus of the potential across the capacitance since the charge is nearly constant as was discussed above. The potential across the capacitance may be written as $U+v(t)$. The important point is that the optical potential $v(t)$ as well as the static potential $U$ are independent of the current delivered. That is, if atoms are present between the two capacitance plates, processes occurring in the atomic collection have no influence on the field. We have just described an essential component of our circuit-theory schematic. In contradistinction, the potential across a resonating inductance-capacitance circuit modeling a single-mode cavity does depend on atomic processes. For that resonator configuration the assumption that the optical field is nearly constant holds only in the large potential (or large photon number) limit. 

\subsection{Inductances}\label{inductance}

An inductance $L$ may be constructed from a cylinder of area $A$ and height $d_L\gg \sqrt A$, split along its hight, so that an electrical current may flow along the cylinder perimeter. In that case $L=\mu A/d_L$, where $\mu$ denotes the permeability. One may assume that the cylinder contains electrons that have magnetic moments. Just above the Curie temperature, $\mu$ much exceeds the free-space permeability, so that the latter may be set equal to zero, as was discussed earlier. For an inductance $L$, the magnetic flux (or magnetic charge) is $\phi(t)=Li(t)$, and the potential across the inductance is $v(t)=d\phi/dt$. It follows that for a constant $L$, $v(t)=Ldi(t)/dt$, or, using the complex notation $V=-\ii L\om I$. Thus $Y(\om)\equiv I/V=\ii/  \p L\om  \q$. The energy stored in an inductance with a current $i$ flowing through it is $E_L=Li^2/2$. For a sinusoidal current represented by the complex number $I$, the time-averaged energy is $\ave{E_L}=L\abs{I}^2/2$. If a large inductance supports a large magnetic flux, the current flowing through the inductance is nearly independent of the potential across the inductance. In that manner, we may realize constant-current sources, either static or oscillating at optical frequencies through a change of $L$ (e.g., by changing the coil length if a coil instead of a simple cylinder is employed). 

The linear relationships outlined in previous paragraphs are sometimes referred to as the "generalized Ohm laws". Let us recall that there is a well-known duality between potentials and currents and between  electrical charges (expressed in coulombs) and magnetic fluxes (expressed in webers), so that expressions obtained for capacitances may be translated into expressions relating to inductances. 

\subsection{Energy and power}\label{energy}

As an application of the above energy formulas, let us consider a circuit consisting of an inductance $L$ and a capacitance $C$ connected in parallel. Since the system is isolated the total admittance must vanish and we obtain the resonance formula $LC\om_o^2=1$, where $\om_o$ denotes the resonant frequency. The sum $E$ of the energy $E_L(t)$ located in the inductance and the energy $E_C(t)$ located in the capacitance does not vary in the course of time. This is twice the time-average energy stored in the capacitance (or inductance). Using above formulas we find that the rms (root-mean-square) field across the capacitance is 
\begin{align}
\label{field}
\mathcal{E}=\sqrt{\frac{E}{\eps_o \mathcal{V}}},
\end{align}
where $\mathcal{V}\equiv Ad_C $ denotes the capacitance volume. We later show that when a resonator such as the one presently considered is in a cold environment it eventually reaches a state corresponding to an energy $\hbar\om_o/2$, where $\hbar$ denotes the Planck constant (divided by $2\pi$). According to the above formula, the so-called "vacuum (rms) field" reads $\mathcal{E}_{vacuum}=\sqrt{\frac{\hbar\om_o/2}{\eps_o \mathcal{V}}}$. The two oppositely-charged capacitance plates attract one another with an average force $F=d\p \hbar\om_o/2\q /d(d_c)=\hbar\om_o/\p 4d_c\q$.

If two sub-systems are connected to one another by two perfectly conducting wires with a potential $v(t)$ across them and a current $i(t)$ flowing into one of them (the current $-i(t)$ flowing in the other one), the power flowing from one sub-system to the other at some instant $t$ is equal to $v(t)i(t)$. For sinusoidal time-variations, the power averaged over an oscillation period reads $P=\Re\{V I^\star\}$. 

Finally, let us recall that at a node, that is, at the junction between perfectly conducting wires, the sum of the currents entering into the node vanishes as a consequence of the fact that the electric charge is a conserved quantity. For three wires traversed by currents $i_1(t),~i_2(t),~i_3(t)$, for example, we have at any instant $i_1(t)+i_2(t)+i_3(t)=0$. It follows that the complex currents sum up to zero, that is $I_1+I_2+I_3=0$. Both the real and the imaginary parts of the sum vanish. Such relations are sometimes called "generalized Kirchhoff laws". The above discussion suffices to treat circuits consisting of conductances, capacitances and inductances arbitrarily connected to one another. Some circuits require a more complicated description involving for example (non-reciprocal) gyrators. These latter
components are useful to separate reflected and incident waves. 

\subsection{The tuned circuit}\label{tuned}

For the sake of illustration and later use, let us generalize the resonator previously considered by introducing in parallel with the capacitance $C$ and the inductance $L$ a conductance $G$. The relation between a complex current source $\C$ at frequency $\om$, supposed to be independent of frequency, and the potential $V$ across the circuit reads
\begin{align}
\label{lossreson}
V(\om )=\frac{\C}{Y(\om )}=\frac{\C}{G-\ii \p C\om-1/L\om\q}.
\end{align}

The power dissipated in the conductance $G$ at frequency $\om$ reads
\begin{align}
\label{power'}
P(\om)=G\abs{V(\om )}^2\approx \frac{G \abs{\C}^2}{G^2+4C^2\p \om-\om_o\q^2}
\end{align}
in the small-loss approximation. Thus $P(\om)$ drops by a factor of 2 from its peak value when $2C\p \om_±-\om_o\q=±G$. The full-width at half power (FWHP) $\de\om$ of the resonance that is, the difference of (angular) frequencies at which the dissipated power drops by a factor of two, is
\begin{align}
\label{width}
\de \om= \om_+-\om_- =\frac{G}{C}\equiv \frac{1}{\tau_p},
\end{align}
where $\tau_p=C/G$ is sometimes called the "photon lifetime". If the resonator is left alone in a cold environment ($T_m$=0K) its classical energy decays according to an $\exp(-t/\tau_p)$ law. For a Fabry-Pérot resonator with mirrors of small power transmissions $T_1,~T_2$, respectively, and spacing $L$, we have
\begin{align}
\label{fp}
\frac{1}{\tau_p}=\frac{T_1+T_2}{2L/v},
\end{align}
where $v$ denotes the group velocity and $2L/v$ is the round-trip time.

The energy contained in the resonator is twice the average energy contained in the capacitance whose expression was given earlier. We then obtain in the small-loss approximation
\begin{align}
\label{en'}
E(\om)=C\abs{V(\om )}^2=  \frac{C\abs{\C}^2}{G^2+4C^2\p \om-\om_o\q^2}\approx \frac{\tau_p \abs{\C}^2/G}{1+x^2},
\end{align}
where $x\equiv2\tau_p \p \om-\om_o\q$. 

\subsection{Derivative of an admittance with respect to frequency}\label{derivative}

For late use, note the expression of the derivative with respect to $\om$ of the admittance of a linear circuit, submitted to a voltage $V$
\begin{align}
\label{dispers}
\ii V^2\frac{dY(\om)}{d\om}=-\ii I^2\frac{dZ(\om)}{d\om}=\sum_{k}{C_k V_k^2-L_k I_k^2},
\end{align}
where the sum is over all the circuit capacitances and inductances. $V_k$ denotes the (complex) voltage across the capacitance $C_k$ and $I_k$ the (complex) current flowing through the inductance $L_k$. The circuit resistances or conductances do not enter in the sum. This relation is readily verified for an inductance in series with a resistance and a capacitance in parallel with a conductance. Thus the relation holds for any combination of elements, connected in series and in parallel.

\subsection{Matrix formulation}\label{matrix}

For an arbitrary circuit, the task is to "extract", figuratively speaking, the (positive or negative) conductances from the given circuit, each conductance being connected to the conservative circuit that remains after extraction of the conductances. If $N$ (positive or negative) conductances are involved, the circuit becomes an $N$-port conservative device. For an $N$-port circuit, we define the vectors $\boldsymbol{V}\equiv [V_1, V_2,...V_N]^t$ and $\boldsymbol{I}\equiv [I_1, I_2,...I_N]^t$ where the upper $t$ denotes transposition. The linear relation is written in matrix form $\boldsymbol{I}=\boldsymbol{Y}(\om)\boldsymbol{V}$, where $\boldsymbol{Y}(\om)$ is called the circuit admittance matrix. For a conservative circuit the total entering power $\Re\{ \boldsymbol{V}^{t\star} \boldsymbol{I} \}=0$. Since this relation must hold for any source this implies that $\boldsymbol{Y}^{t\star}+\boldsymbol{Y}=\boldsymbol{0}$.

It is convenient to view the connections between the conservative circuit and the conductances as ideal transmission lines of small length and characteristic conductances $G_c$. Supposing that $G_c=1$, the potential $V$ across one of the transmission lines and the current $I$ flowing through the (say, upper) wire, are combined into an ingoing wave whose amplitude is defined as $a=V+I$ and an outgoing wave defined as $b=V-I$. Since under our assumptions the circuit elements are linear, there is a linear relationship between the $a$-waves and the $b$-waves. The relation between $\boldsymbol{b}$ and $\boldsymbol{a}$, defined like $\boldsymbol{I}$ and $\boldsymbol{V}$ above, may be written in matrix form as $\boldsymbol{b}=\boldsymbol{Sa}$, where the $\boldsymbol{S}$ matrix is called the circuit "scattering matrix". Because the circuit is conservative, the outgoing power equals the ingoing power. It follows that the $\boldsymbol{S}$-matrix is unitary, i.e., $\boldsymbol{S}^{t\star} \boldsymbol{S}=\boldsymbol{1}$. We need not assume that the circuit is reciprocal, however, that is, the $\boldsymbol{S}$-matrix needs not be symmetrical.

\subsection{Various circuits}\label{various}

We have represented a number of important conservative (loss-less, gain-less) components in either their circuit form, their microwave form, or their optical form in Fig.~\ref{circuitfig}. The origin of the differences is that, as one goes to shorter wavelength (higher frequencies) some circuit elements become too small to be fabricated. It should also be noted that metals, such as copper, that are excellent electrical conductors up to microwave wavelengths, do not behave as electrical conductors any more at optical wavelengths because of electron inertia. On the other hand, while it is difficult to find very low-loss dielectrics at microwave frequencies, extremely low-loss glasses exist at optical frequencies. Fig.~\ref{circuitfig} represents four resonating circuits, that one may call "0-dimensional" devices. Namely, the inductance-capacitance circuit employed up to about 100 MHz, the cavity employed in reflex klystrons and masers for example, the Fabry-Perot resonator consisting of two mirrors facing each other, and the whispering-gallery-mode dielectric resonator, first demonstrated in the microwave range and now-a-days employed in the optical range. Resonators are primarily characterized by their resonant frequency $\om_o$. Small losses may be characterized by the so-called "photon life-time" $\tau_p$ defined earlier. When the resonator size is large compared with wavelength many resonating modes may be present. In most applications it is desirable that only one of them be loss-less, or nearly so (see, e.g. \cite{Arnaud1976}).

Figure~\ref{circuitfig} represents four one-dimensional devices called "transmission lines". The circuit form is a periodic sequence of series inductances and parallel capacitances. The microwave form consists of two parallel conductors characterized by a characteristic conductance $G_c$, with waves propagating at the speed of light. Above 1GHz one would rather use waveguides. The optical form is the now-a-day well-known \emph{optical fiber}. A glass fiber (core) in vacuum may guide optical waves by the mechanism of total reflexion. In order to increase the core size without having spurious modes propagating, the core is usually immersed into a lower-refractive-index glass.

Other useful devices are shown in Fig.~\ref{circuitfig}. The power carried by a transmission line of characteristic conductance $G_c$ may be split into two parts simply by connecting it to two transmission lines whose characteristic conductances sum up to $G_c$. This a three-port reciprocal conservative device. Alternatively, when two transmission lines are put side by side and coupled at two locations separated by a quarter of a wavelength, some of the power incident on a transmission line is transmitted into the other one. This device is called a directional coupler. This 4-port device may be reduced to a 3-port device by putting a matched load at the end of one of the transmission lines. The optical form of a directional coupler is called a beam-splitter, which may simply consist of a flat piece of glass. An important non-reciprocal 3-port device is the \emph{circulator}, which exists in microwave and optical versions. It is intrinsically loss-less: a wave entering into port 1 entirely exits from port 2, a wave entering into port 2 entirely exits from port 3, and a wave entering into port 3 entirely exits from port 1. Such a device is convenient to separate reflected waves from incident waves without introducing losses.

\newpage

\section{Electron motion and induced current}\label{electronmotion}

We first consider a single electron located between two parallel conducting plates pierced with two holes. A static potential source $U$ is applied to external plates, as shown in Fig.\ref{klystron} in (a). An alternating potential source $v(t)$ at frequency $\om$ are applied between the inner plates. These potentials are supposed to be independent of the current that the electron motion may induce. The sign convention is the one given in Fig.\ref{circuitfig} in (c). In both the Classical and Quantum Theories, the induced current $i(t)$ is, to within a constant, equal to the electron momentum $p(t)$, although the interpretations of $i(t)$ and $p(t)$ differ. The power $v(t)i(t)$ supplied by the alternating potential source, once averaged over a period, is denoted $P(t)$. 

The over-all effect of the electron motion is to transfer energy from the static source to the alternating source (stimulated emission) or the converse (stimulated absorption). Spontaneous emission in the usual sense does not occur. What may occur is that the electron is emitted or captured by one of the plates (Classical view-point) or tunnels into the plates (Quantum view-point). In the Classical treatment, one first evaluate 1) the electron motion under the static field, 2) the perturbation caused by alternating field, and 3) the electron momentum and the induced current. The same steps are taken in the Quantum treatment. Namely, we consider the stationary states of the electron submitted to the static field, the perturbation of those states due to the alternating field, and finally evaluate the induced current from the electron momentum.  

\subsection{Classical Equations of Electron Motion}\label{classicalelectronmotion}

The equations of motion of an electron of charge $-e$ and mass $m$ are first established for the case of a static (time-independent) potential. As an example consider an anode at zero potential and an electron emitted from a cathode at potential $-U$ in vacuum, and look for the electron motion and the induced current. If $i(t)$ denotes the current delivered by the potential source, the power $Ui(t)$ must be equal at any instant to the power delivered to the electron, which is the product of the velocity $p(t)/m$, where $p$ denotes the electron momentum, and the force $eU/d$ exerted upon it, where $d$ denotes the electrode spacing. Since the potential $U$ drops out from this equation, the current is
\begin{align}
\label{currentb}
i(t)=\frac{e}{md}~p(t).
\end{align}
Solving the equations of motion, we find that the diode current $i(t)=\p e^2U/md^2\q t$ increases linearly with time and drops to zero when the electron reaches the anode. Thus, each electron freed from the cathode entails a triangularly-shaped current pulse. If $i(t)$ is integrated over time from $t=0$ to $t=\tau$ we obtain the absolute value of the electron charge $e$. We will neglect the pulse duration (or transit time) $\tau$, so that triangularly-shaped current pulses are approximated by $e\de(t)$-functions. The above theory is applicable only when few electrons are emitted so that the initial electron velocities and space-charge effects are neglected. 

The expression for the current was established for the case where the potential generates a constant electric field. This is not the case for example if the anode and the cathode are coaxial cylinders of radii $r_a$ and $r_c$ with $r_c<r_a$. The electron is submitted to a force $\mathcal{F}(r)$ inversely proportional to $r$, assuming radial motion. The induced current $i(t)$ may nonetheless be obtained from the $p(t)$ trajectory by the same argument as above.

\begin{figure}
\centering
\begin{tabular}{cc}
\includegraphics[scale=0.35]{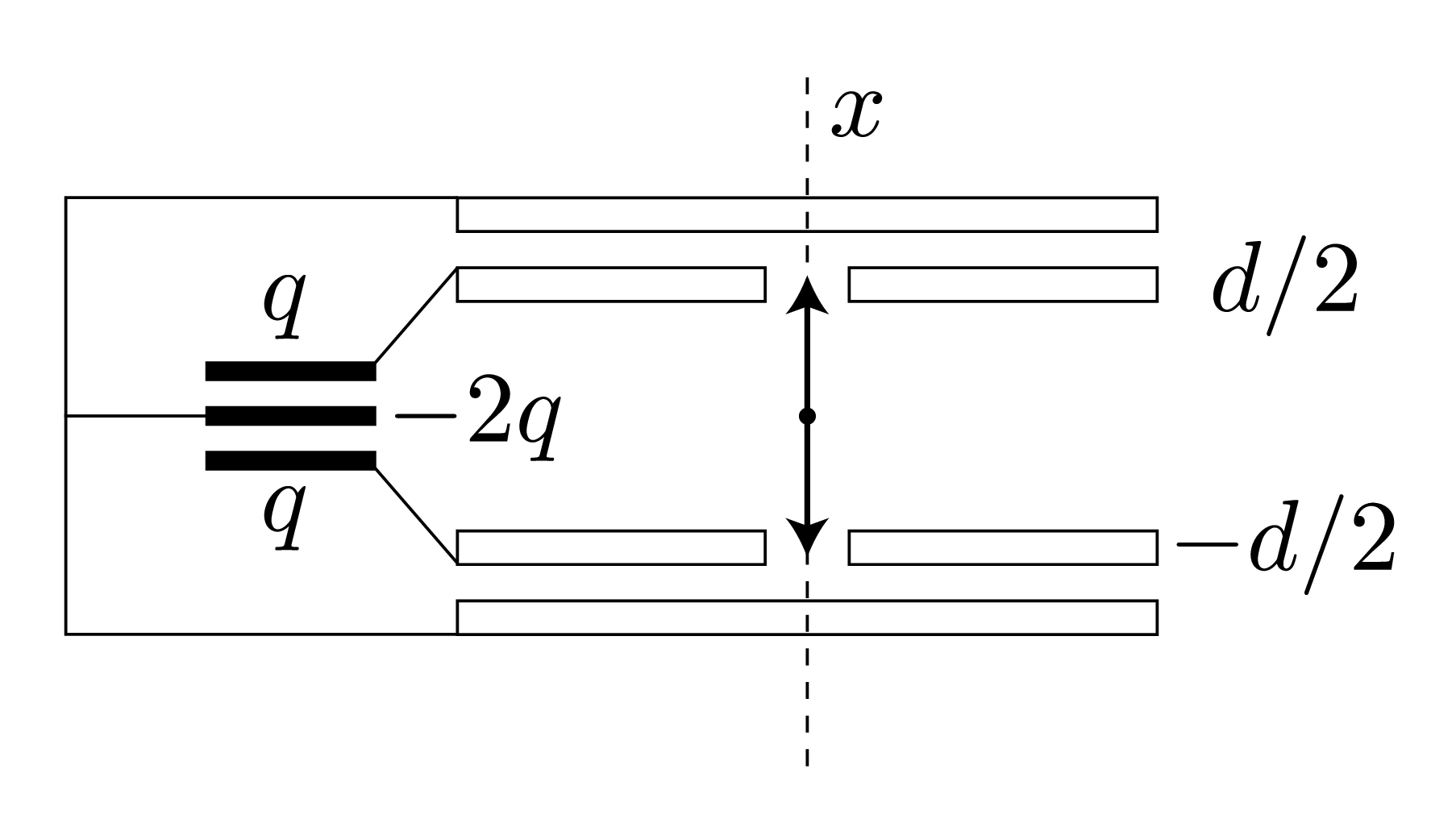}   &  \includegraphics[scale=0.35]{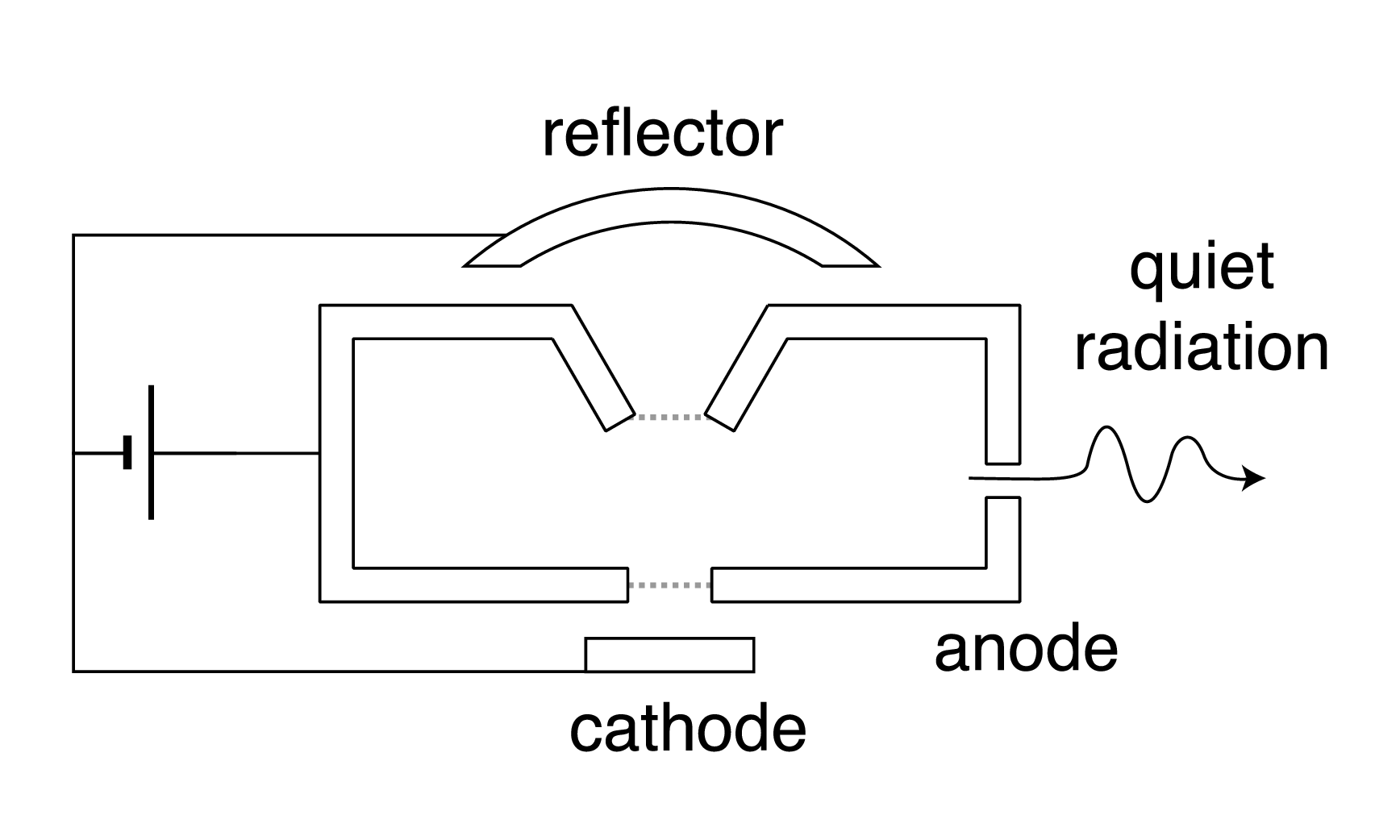}  \\
(a) & (b) \\
\includegraphics[scale=0.35]{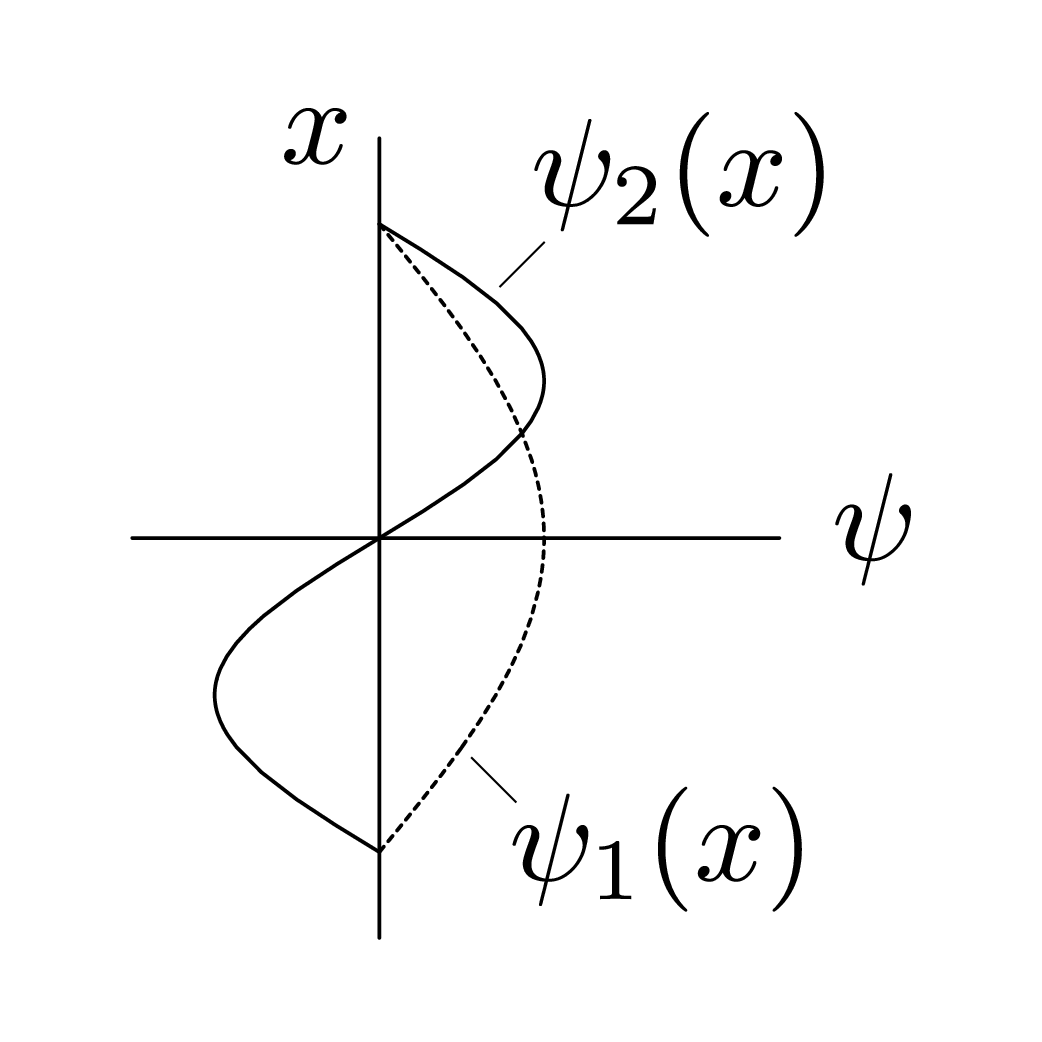}   &  \includegraphics[scale=0.35]{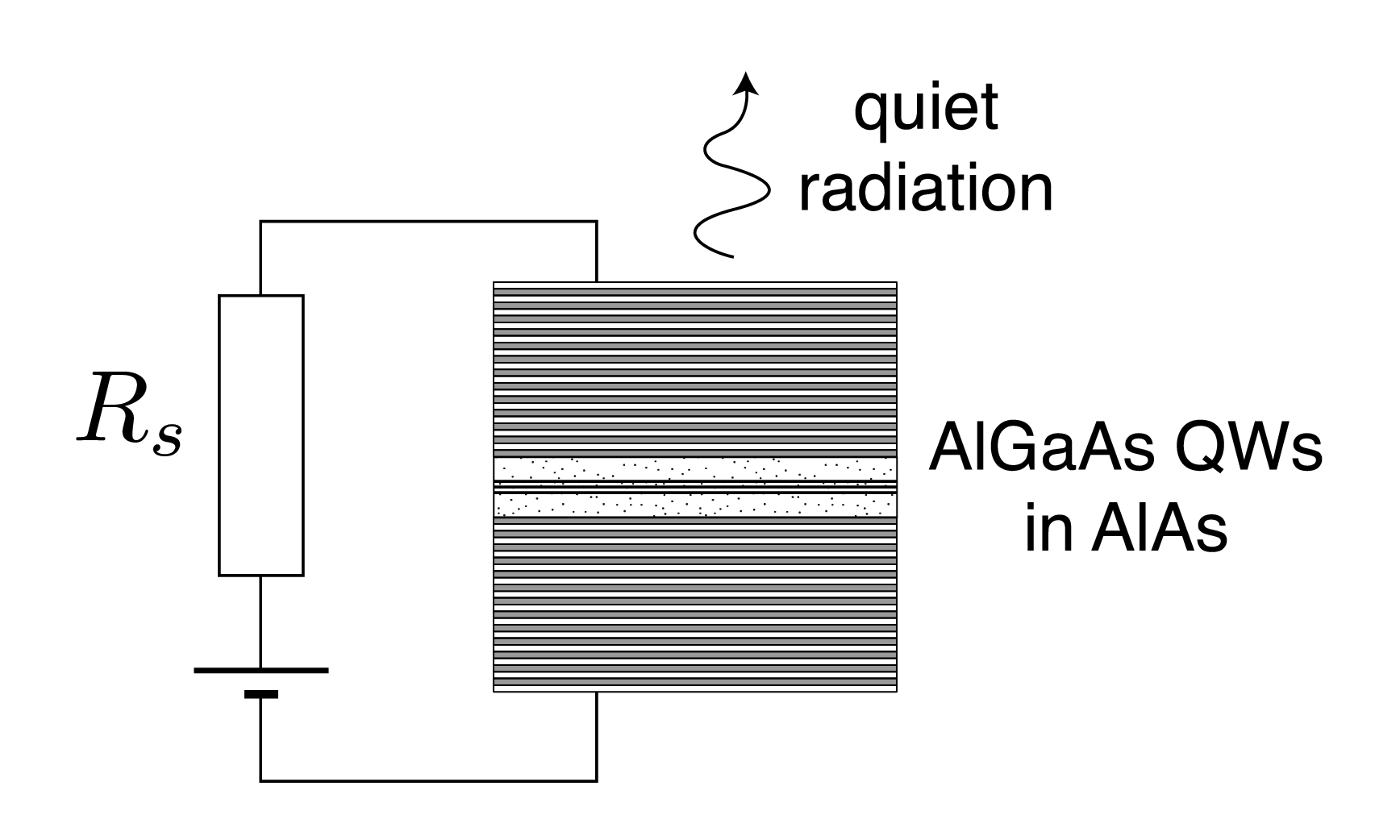}  \\
(c) & (d)
\end{tabular}
\caption{a) Illustrates the potential created by two anodes (inner electrodes) and two cathodes located just outside the anodes. The potential is generated by a large, charged, capacitance, shown on the left. According to the Classical Picture the space-time electron trajectory is almost  a zig-zag path, with slight incursions of the electron between the anodes and the cathodes. b) Represents a reflex klystron, which is similar to the previous schematic, but with a resonator added to it. The current is regulated by a space-charge limited cathode. c) Pictures the wave-functions of the ground state and first excited state of a square potential well. d) Represents a "surface-emitting" laser diode. The current is regulated by a large cold resistance $R_s$.}
\label{klystron}
\end{figure}

As a second example, consider a one-dimensional square-well, whose potential is equal to 0 for $\abs{x}<d/2$ and infinite (or nearly so) beyond. This potential may be generated by parallel anodes at potential 0 and cathodes at potential $-U$, as shown in Fig.~\ref{klystron}. The electron space-time trajectories $x=x(t)$ consist of straight lines with slopes $dx(t)/dt=±p/m$, where $p^2/2m=E$ is the electron energy, which may be selected arbitrarily from 0 to $eU$ so that the electron is not captured by the cathodes. The electron is prevented from being captured by the anodes by a strong magnetic field in the $x$-direction. The quick electron incursions between anodes and cathodes are here neglected. We may consider in particular a lower electron energy $E_1$ and a higher electron energy $E_2$, corresponding to small and large slopes in the $x=x(t)$ diagram, respectively.  

The electron motion induces an electrical current $i(t)$ in the potential source, which is proportional to the electron momentum $p(t)$, as said above. In the case of a static potential source the induced current does not correspond to any power delivered or received by the source on the average, so that the electron motion may go on, in principle, for ever.

If now the static potential $U$ is supplemented by a sinusoidal potential $v(t)$ of small amplitude, whose frequency is resonant with the electron motion described above, the \emph{unperturbed} electron momentum $p(t)$ does cause the alternating potential source to receive or generate power, depending of the electron state. However, if we consider a large collection of unperturbed electrons, the power averages out again to zero. It follows that a net energy transfer may be obtained only if we take into account the fact that the alternating potential \emph{perturbs} the electronic motion. In the present classical picture this amounts to bringing all the electrons with the appropriate phase with the alternating potential, an effect called "bunching". This name originates from the fact that electrons initially spread out uniformly eventually are forced by the alternating field to form periodic "bunches". This desirable bunching effect is limited by the velocity spread of electrons originating from the hot cathode and the fact that electrons tend to repel each others (space-charge effect).

This is not however the end of the story. The electron, initially in the lower energy state, may gain enough energy to be captured by a cathode. Once in a cathode, the electron flows through the static potential source to the anode, delivering an energy $eU$ to that source, and is emitted by the anode back into the lower energy state. The net effect of these processes is that some power is being transferred from the alternating potential source to the static potential source, or the converse, the electrons playing an intermediate role.

In more general situations, the Classical Equations of Motion of electrons of charge $-e$, mass $m$, and potential energy $-eu(x,t)$ are best based on the Hamiltonian formulation. The particle total energy $E(t)$ is expressed as a function of position $x$, momentum $p$, and time $t$ according to the relation
\begin{align}
\label{ham}
H(x,p,t)-E(t)\equiv \frac{p^2}{2m}-eu(x,t)-E(t)=0,
\end{align}
where $p^2/(2m)$ represents the kinetic energy. The Hamiltonian equations read
\begin{align}
\label{eqmot}
\frac{dx(t)}{dt}&=\frac{\partial H(x,p,t)}{\partial p }=\frac{p(t)}{m}\\
\frac{dp(t)}{dt}&=-\frac{\partial H(x,p,t)}{\partial x}=e\frac{\partial u(x,t)}{\partial x}.
\end{align}
The first equation says that the particle momentum $p(t)=m dx(t)/dt$, and the second equation may be written, with the help of the first equation, in the Newtonian form $m~d^2x(t)/dt^2=e~\left(\partial u(x,t)/\partial x\right)_{x=x(t)}$. Going back to the first example in this section, let us consider two parallel plates located at $x=0$ and $x=d$, and at potentials $0$ and $-u(t)$, respectively. We have $u(x,t)=-u(t)x/d$, and thus the equation of motion is $md^2x(t)/dt^2=-eu(t)/d$. The electron decelerates if $u(t)>0$. The electron is repelled by the negatively-charged cathode.

As far as static conditions are concerned, an electron submitted to a static potential source $U$ is analogous to an electron submitted to the Coulomb potential created by positively-charged nuclei. The potentials, on the order of 1 volt (corresponding to potential energies of 1.6 10$^{-19}$ joules) are comparable in the two situations. In the case of atoms, however, the Bohr radius, which is roughly equivalent to our distance $d$, is on the order of 0.05 nanometers while, in the case of two conducting plates, the distance can hardly be less than 100 nanometers for practical reasons. As a consequence there exist in the two-plate model many states whose energy is comprised between the lower-state energy $E_1\approx 0$ and the upper-state energy $E_2\approx eU$. In both cases the conductance (ratio of the induced current to the applied potential) is initially equal to zero and grows in time linearly until the electron is somehow absorbed or leaves the interaction region. 

\subsection{Quantum Equations of Motion}\label{quantum}

Before entering into the mathematical details, it is important to understand the significance of our schematics, and how these schematics may describe actual devices. The differences between schematics and real maser or laser devices, which we will point out below, are considered to be of minor importance as far as concepts are concerned.

The configuration that we have in mind is again the one shown in Fig. \ref{klystron} in (a), but with the electron motion quantized as in (c). In this picture, as was discussed earlier, the electron is submitted to a static potential source generated by a charged capacitance of arbitrarily large value. This potential is applied between inner electrodes and outer electrodes. The electron is constrained to move along the $x$-axis with the help of a magnetic field (not shown on the figure). Classically, the electron performs a zig-zag $x(t)$ path. From the time-independent Schrödinger-equation view point, the electron may reside only in a lower state 1 and an upper state 2. In real masers or lasers the potential is generated by the static potential of fixed, positively-charged point-like nuclei (because of their large mass, plates or nuclei recoils may be neglected). However, from our view-point, the two configurations differ only in the form of the wave-functions and the value of the transition element later on denoted by $x_{12}$. 

The static potential is supplemented by an alternating potential source, at a frequency on the order of 10 GHz for klystrons and 300 THz for lasers, generated in the picture of Fig. \ref{klystron} in (a) on the left by a sinusoidal motion of the inner capacitance plate. In a real klystron, the alternating potential is generated by an inductance-capacitance circuit, as shown in Fig. \ref{klystron} in (b), In general, this oscillator, resonant with the electron alternating motion, cannot be considered as a \emph{source}, because the potential depends on the induced current. It is only in the limit where the tuned-circuit capacitance would be extremely large and the inductance extremely small (so that the resonating frequency remains the same), that this tuned circuit could be considered as an alternating potential source. Indeed, in the limit considered, for a given alternating field, the tuned circuit energy is extremely large and little affected by the electron motion. In Quantum Optics, this limiting situation is described by saying that "the number of photons in the cavity is supposed to be extremely large, so that a classical treatment of the field is adequate". Let us emphasize that the configuration treated in the present section is only one idealized component of a complete laser device. We need to characterize this component accurately (in terms of conductances and event processes) before going on.

What is missing in the schematics of Fig. \ref{klystron} in (a) is the absorber of radiation. In that schematics this absorber could be realized by adding on the right a triple-plate capacitance, as already shown on the left. Similarly, in Fig. \ref{klystron} in (b) the wavy line, symbolizing the escape of radiation, could be replaced by a potential configuration similar to the one shown on the left, but with a slightly different static potential, so that power flows from the potential source on the left to the potential source on the right. The electron motion and the alternating field may be viewed as playing an intermediate role. If this is the case, one may wonder why complicated devices are needed to merely transfer energy from one capacitance (or battery) to another. The answer of course is that in the microwave or optical forms, energy may be carried over large distances with little (absorption or diffraction) loss. High-frequency electromagnetic waves also serve as sensors, e.g., in the radar. Transmission lines are not shown in Fig. \ref{klystron}.

The Quantum Equations of Motion of an electron of charge $-e$ and mass $m$ are first established for a static (i.e., time-independent) potential source. As an example we consider a one-dimensional square-well, whose potential is equal to 0 for $\abs{x}<d/2$ and infinite (or nearly so) beyond. This potential may be generated by parallel anodes at potential 0 and cathodes at potential $-U$, as shown in Fig.~\ref{klystron}. We solve the time-independent Schrödinger equation and obtain in particular a state 1 with lower energy $E_1$ and a state 2 with higher energy $E_2$. As we shall see, these two states correspond to wave-functions $\psi_1(x)=\cos(x)$ and $\psi_2(x)=\sin(2x)$, respectively, leaving aside constants. In the case of a static potential there is no energy exchange between the potential source and the electron when the electron is initially in a stationary state, so that the electron remains in the stationary state, in principle, for ever. There are no energy exchange either if we perform a time averaging when the electron is in a superposition of stationary states. This situation may be compared to the one discussed classically above. 

Let now the static potential source $U$ be supplemented by a sinusoidal potential source $v(t)$ of small amplitude, whose frequency is (in some sense to be defined later) resonant with the electron motion described above. A net energy transfer may be obtained only if we take into account the fact that the alternating potential \emph{perturbs} the electronic motion. In the Classical picture this amounts to bringing all the electrons with the appropriate phase, an effect called "bunching", as said previously. In the Quantum picture (time-dependent Schrödinger equation), the electron wave function $\psi(x,t)$ is the weighted sum of the unperturbed states defined above, with time-dependent weights. The theory leads to (Rabi) oscillations between the two states. Initially, the induced current is equal to zero and grows in proportion to time, but the conductance vanishes on the average. A non-zero positive conductance is obtained if the electron initially in the lower state remains in the interaction region for a finite time $\tau$.  We may then evaluate the average conductance "seen" by the optical potential source. A model based on the direct coupling between \emph{bands of states}, assuming that an equilibrium is quickly reached within each band separately, would be more realistic for semi-conductors.

The quantum treatment is based on the Schrödinger equation
\begin{align}
\label{sch}
[H(x,p,t)-E]\psi(x,t)=0,\quad E=\ii \hbar\partial/ \partial t,\quad p=-\ii \hbar\partial/ \partial x,
\end{align}
where the sign "$\partial$" denotes partial derivation.  $\psi(x,t)$ is called the wave-function, whose initial value $\psi(x,0)$ is supposed to be known, and $H(x,p,t)=\frac{p^2}{2m}-eu(x,t)$ as in the Classical Equations of Motion, but $p$ and $E$ are now operators of derivation. It is easily shown that, provided $\psi(x,t)$ decreases sufficiently fast as $x\to±\infty$, the integral over all space of $\abs{\psi(x,t)}^2$ does not depend on time. It therefore remains equal to 1 if the initial value is 1, a result consistent with the Born interpretation of the wave function. Because of linearity the sum of two solutions of the Schrödinger equation is a solution of the Schrödinger equation (superposition state). The wave-functions add up, but not in general the probabilities.

Let us evaluate the time derivative of the average value of $x$. We have, using the above Schrödinger equation to obtain $\partial \psi(x,t)/\partial t$ and the mathematical relation in \eqref{kkk} with $p=\hbar k$
\begin{align}
\label{eh}
\frac{d\ave{x(t)}}{dt}&=\frac{d}{dt}\int_{-\infty}^{+\infty} dx~x~ \psi(x,t)\psi^\star(x,t)\nonumber\\
&=\int_{-\infty}^{+\infty} dx~x~ \left( \frac{\partial\psi(x,t)}{\partial t}     \psi^\star(x,t) + \psi(x,t)\frac{\partial\psi^\star(x,t)}{\partial t}         \right) =\frac{\ave{p(t)}}{m}.
\end{align}
This is the first Ehrenfest equation. Thus the classical relation $p=m\frac{dx}{dt}$ still holds provided $x$ and $p$ be replaced by their QM-averaged values.

\subsection{Static potentials}\label{static}

Let us suppose that $u(x,t)\equiv u(x)$ does not depend on time. In that case solutions of the above equation of the form $\psi(x,t)=\psi_n(x)\exp(-\ii\om_nt)$ may be found, where $n=1,2...$. The $\psi_n(x)$ are real functions of $x$ and $E_n\equiv \hbar\om_n$ that form a complete orthogonal set of functions. 

For $n=1,2$ the wave functions obey the differential equations
\begin{align}
\label{diff}
\frac{\hbar^2}{2m}\frac{d^2\psi_1(x)}{dx^2}+eu(x)\psi_1(x)+E_1\psi_1(x)=0\nonumber\\
\frac{\hbar^2}{2m}\frac{d^2\psi_2(x)}{dx^2}+eu(x)\psi_2(x)+E_2\psi_2(x)=0.
\end{align}
with the appropriate boundary conditions. They may be ortho-normalized such that
\begin{align}
\label{norm}
\int_{-\infty}^{+\infty}dx~ \psi_m(x) \psi_n(x)=\de_{mn},
\end{align}
where $\de_{mn}=1$ if $m=n$ and 0 otherwise. 

\subsection{Potential well}\label{free}

As an example consider an electron of mass $m$ moving along the $x$ axis be reflected by boundaries at $x=-d/2$ and $x=d/2$ where the wave-function is required to vanish, that is, $\psi(±d/2)=0$. The lowest-energy state $n=1$ and the first excited state $n=2$ are
\begin{align}
\label{wf}
\psi_{1}(x,t)&=\sqrt{2/d} \cos(\pi x/d)\exp(-\ii \om_1 t)\\
\psi_{2}(x,t)&=\sqrt{2/d} \sin (2\pi x/d)\exp(-\ii \om_2 t)
\end{align}
 Notice that $\psi_{1}(x)$ is even in $x$, while $\psi_{2}(x)$ is odd in $x$. Substituting these expressions in the Schrödinger equation \eqref{sch} with $u(x,t)=0$, we obtain that
\begin{align}
\label{news}
\frac{\hbar^2}{2m}\frac{d^2\psi_n(x)}{dx^2}+\hbar \om_n \psi_n(x)=0
\end{align}
provided
 \begin{align}\label{En}
E_{n}\equiv \hbar \om_n=\frac{\pi^2 \hbar^2}{2md^2}n^2\qquad n=1,2.
\end{align} 
We will see later on that optical fields at frequency $\om_o=\om_2-\om_1=\p 3\pi^2\hbar\q/\p2md^2\q$ may cause the system to evolve from state 1 to state 2 and back. Numerically, $\hbar\om_o\approx 1.12$ electron-volt if $d=1$ nano-meter.

For later use let us evaluate
 \begin{align}\label{x12}
x_{12}&\equiv \int_{-d/2}^{d/2}dx~x~\psi_{1}(x)\psi_{2}(x)\nonumber\\
&=\frac{2}{d}\int_{-d/2}^{d/2}dx~x~\cos(\pi x/d)\sin (2\pi x/d) \nonumber\\ &=\frac{16d}{9\pi^2},
\end{align} 
where we have used the mathematical relation
 \begin{align}\label{int}
\int_{-\pi/2}^{\pi/2}t\cos(t) \sin(2t)dt=\frac{8}{9}.
\end{align}
The parameter $x_{12}$ determines the strength of the 
atom-field coupling. It is convenient to define a dimensionless 
\emph{oscillator strength} 
\begin{align}\label{f}
f\equiv \frac{2m\omega_{o}}{\hbar }x_{12}^2=\frac{256}{27\pi^2}\approx 
0.96.
\end{align} 
The maximum value of $f$ is 1.

\subsection{Perturbed motion}

We next suppose that a potential source $v(t)=\sqrt2V\cos(\om_o t)$ is applied between the two anodes in Fig.~\ref{klystron}. Since the potential varies linearly with $x$ the electron is submitted to a space-independent optical field $\E(t)=\E_o\cos(\om_o t), \E_o=\sqrt2V/d$, where $\om_o\equiv \om_2-\om_1$ is the 1-2 transition frequency defined in the previous section. In that case \eqref{sch} reads 
\begin{align}
\label{solve}
H\psi\equiv\left(\frac{p^2}{2m}-e\E_o\cos(\om_o t)x-E\right)\psi(x,t)=0,\quad E=\ii \hbar\partial/ \partial t,\quad p=-\ii \hbar\partial/ \partial x,
\end{align}
remembering that for stationary states $\psi_n(x)$
\begin{align}
\label{solvebis}
\left(\frac{p^2}{2m}-\hbar\om_n\right)\psi_n(x)=0.
\end{align}

The wave function may be expressed as an infinite sum of $\psi_n(x)\exp(-\ii\om_n t)$ functions with slowly time-varying coefficients $C_n(t)$, that is
\begin{align}
\label{solveter}
\psi(x,t)=\sum_{n=1}^\infty C_n(t)\exp(-\ii \om_n t)\psi_n(x).
\end{align}
We first evaluate 
\begin{align}
\label{soveter}
\left(\frac{p^2}{2m}-e\E_o\cos(\om_o t)x\right)\psi(x,t)&=\sum_{n=1}^\infty C_n(t)\exp(-\ii \om_n t)\left(\hbar\om_n -e\E_o\cos(\om_o t)x\right)\psi_n(x)\nonumber\\
E\psi(x,t)&=\sum_{n=1}^\infty\exp(-\ii \om_n t)\left(\hbar\om_n  C_n(t) +\ii\hbar \frac{dC_n(t)}{dt}\right)\psi_n(x).
\end{align}
If we subtract the first expression from the second and substitute this expression into the Schrödinger equation, taking \eqref{solvebis} into account, we obtain 
\begin{align}
\label{solebis5}
0=\sum_{n=1}^\infty\exp(-\ii \om_n t)\left(\ii\hbar\frac{dC_n(t)}{dt}+e\E_o\cos(\om_o t)~x~C_n(t)\right)\psi_n(x).
\end{align}
If we multiply \eqref{solebis5} throughout by $\psi_m(x)$, integrate with respect to $x$, and take into account the ortho-normality of the $\psi_m(x)$ functions, we obtain an infinite number of exact ordinary differential equations that can be solved numerically. 

Considering only states 1 and 2, we set
\begin{align}
\label{solvter}
\psi(x,t)= C_1(t)\exp(-\ii \om_1 t)\psi_1(x)+C_2(t)\exp(-\ii \om_2 t)\psi_2(x).
\end{align}
Introducing the resonance condition $\om_o=\om_2-\om_1$, we obtain from \eqref{solebis5}
\begin{align}
\label{lveter}
0=\ii\hbar\frac{dC_1(t)}{dt}+\exp(-\ii \om_o t)\cos(\om_o t)\E_oex_{12}C_2(t),\nonumber\\
0=\ii\hbar\frac{dC_2(t)}{dt}+\exp(-\ii \om_o t)\cos(\om_o t)\E_oex_{12}C_1(t),
\end{align}
where $x_{12}$ is given in \eqref{x12}. Because the wave-functions $\psi_1(x),\psi_2(x)$ are real $x_{12}$ is real, and because of the symmetry of the wave-functions $x_{11}=x_{22}=0$.

The rotating-wave approximation consists of keeping only the slowly-varying terms \cite{Loudon1983}, that is, replacing $\exp(-\ii \om_o t)\cos(\om_o t)$ by 1/2. Thus, the complex coefficients $C_1(t),C_2(t)$ obey the differential equations
\begin{align}
\label{formbisbis}
 \frac{dC_1(t)}{dt}=\ii\frac{\Om_R}{2}C_2(t)\qquad
 \frac{dC_2(t)}{dt}=\ii\frac{\Om_R}{2}C_1(t)\qquad C_1(t)C_1^\star(t)+C_2(t)C_2^\star(t)=1, 
\end{align}
where $\Om_R\ll\om_o$ is the Rabi frequency given by
\begin{align}
\label{rabi}
\hbar\Om_R=\E_oex_{12}.
\end{align}
For the potential considered and the value obtained in \eqref{f}, the above relation reads
\begin{align}
\label{om}
\hbar\Om_R=\frac{16}{ 9\pi^2}e\sqrt 2V\approx 0.17 ~e\sqrt 2V.
\end{align}
The pair of first-order differential equations in \eqref{formbis} is easily solved. Assuming that the electron is initially ($t=0$) in the absorbing state, we have the initial condition $C_2(0)=0$. The wave function thus reads
\begin{align}
\label{form}
\psi(x,t)&= C_1(t)\exp(-\ii \om_1 t)\psi_1(x)+C_2(t)\exp(-\ii \om_2 t)\psi_2(x)\nonumber\\
&=\cos(\frac{\Om_R}{2} t)\psi_{1}(x)\exp(-\ii\om_1t)+\ii \sin(\frac{\Om_R}{2} t)\psi_{2}(x)\exp(-\ii\om_2t).
\end{align}
It follows that
\begin{align}
\label{fom}
\rho_{22}(t)&\equiv C_2(t)C_2^\star(t)=\sin(\frac{\Om_R}{2} t)^2=\frac{1-\cos(\Om_R t)}{2}\nonumber\\
\rho_{12}''(t)&\equiv\I \{ C_1(t)C_2^\star(t)\}=-\frac{\sin(\Om_R t)}{2}.
\end{align}

\subsection{Momentum probability law}
Let us now evaluate the momentum probability law $P(p,t)$. The wave functions in momentum space are defined as, see \eqref{planche} with the Planck constant restored and at $t=0$, 
\begin{align}
\label{ppp}
\psi_1(p)&=\frac{1}{\sqrt{2\pi \hbar}}\int_{-d/2}^{+d/2}dx\exp(-\ii \frac{px}{\hbar})\psi_1(x)\nonumber\\
\psi_2(p)&=\frac{1}{\sqrt{2\pi \hbar}}\int_{-d/2}^{+d/2}dx\exp(-\ii \frac{px}{\hbar})\psi_2(x),
\end{align}
where $\psi_1(x),\psi_2(x)$ are given in \eqref{wf}. Because $\psi_1(x)$ is real even and $\psi_2(x)$ is real odd, $\psi_1(p)$ is real and $\psi_2(p)$ is imaginary. We set $\psi_2(p)\equiv \ii \psi_2''(p)$. The explicit result is 
\begin{align}
\label{ppx}
\psi_1(p)&=-\frac{\sqrt{ \pi d }  } { 2 }  \frac{ \cos(p d/2\hbar) } { (pd/2\hbar)^2-(\pi/2)^2 }       \nonumber\\
\psi_2''(p)&=\sqrt{ \pi d }    \frac{ \sin(p d/2\hbar) } { (pd/2\hbar)^2-\pi^2 }
\end{align}
These expressions, however, will not be needed.

Since the Fourier transform (with respect to $x$) is a linear operation, we obtain from \eqref{form} the expressions
\begin{align}
\label{ex}
\psi(p,t)&=C_1(t)\psi_{1}(p)\exp(-\ii\om_1t)+C_2(t)\psi_{2}(p)\exp(-\ii\om_2t)\\
&=\cos(\frac{\Om_R}{2} t)\psi_{1}(p)\exp(-\ii\om_1t)- \sin(\frac{\Om_R}{2} t)\psi_{2}''(p)\exp(-\ii\om_2t),
\end{align}
It follows that
\begin{align}
\label{e}
P(p,t)&=\abs{\psi(p,t)}^2\nonumber\\
&=C_1(t)C_1^\star(t)\abs{\psi_{1}(p)}^2+C_2(t)C_2^\star(t)\abs{\psi_{2}(p)}^2\nonumber\\
&+C_1(t)C_2^\star(t)\psi_{1}(p)\psi_{2}^\star(p)\exp(\ii\om_o t)
+C_2(t)C_1^\star(t)\psi_{2}(p)\psi_{1}^\star(p)\exp(-\ii\om_o t)\nonumber\\
&=\cos^2(\frac{\Om_R}{2} t)\psi_{1}(p)^2+ \sin^2(\frac{\Om_R}{2} t)\psi_{2}''(p)^2-2\sin(\frac{\Om_R}{2} t)\cos(\frac{\Om_R}{2} t)\psi_{1}(p)\psi_{2}''(p)\cos(\om_ot),
\end{align}
where we have used the expressions in \eqref{form} of $C_1(t),C_2(t)$, and taken into account the fact that $\psi_{1}(p),\psi_{2}''(p)$ are real. Of course, the integral of $P(p,t)$ over all $p$-values is unity at any time.

Without electron-optical field coupling, that is, when $\Om_R=0$, we have $P(p,t)=\psi_{1}(p)^2$ and thus $\ave{p(t)}=0$ according to the relation below \eqref{pars}. Since the average current induced in the potential source is proportional to $\ave{p(t)}$, the average current vanishes. It follows that for stationary states there are no energy exchange between the (static) potential source and the electron on the average.

\subsection{Average induced current}

To evaluate the average momentum when $\Om_R>0$, we notice that the first two terms in \eqref{e} do not contribute. We thus obtain
\begin{align}
\label{mom}
\ave{p(t)}&\equiv \int_{-\infty}^{+\infty}dp~p~P(p,t)\nonumber\\ &= -\sin(\Om_R t)\cos(\om_o t)\int_{-\infty}^{+\infty}dp~p~\psi_{1}(p)\psi_{2}''(p)
=\ii\sin(\Om_R t)\cos(\om_o t)p_{12},
\end{align}
where, according to \eqref{kmm} with $e_1=2mE_1/\hbar^2,e_2=2mE_2/\hbar^2$, 
\begin{align}
\label{whe}
p_{12}&\equiv  \int_{-\infty}^{+\infty}dp~p~\psi_{1}(p)\psi_{2}^\star(p)\nonumber\\
&=-\ii \frac{m\p E_2-E_1\q}{\hbar}\int_{-d/2}^{+d/2}dx~x\psi_{1}(x)\psi_{2}(x)=-\ii m\om_o x_{12}
\end{align}
Thus $\ave{p(t)}=\sin(\Om_R t)\cos(\om_o t)m\om_o x_{12}$. The QM-averaged induced current is
\footnote{The expression $\ave{i(t)}=\frac{e}{md}~\ave{p(t)}=\frac{e}{d}\frac{d\ave{x(t)}}{dt}$ (using the first Ehrenfest equation) is often expressed in a different but equivalent form in terms of the electron dipole moment $ex$. Omitting for brevity the QM-averaging signs and time arguments, and considering $N$ electrons in a volume $\V\equiv Ad$, the current density $J_{opt}\equiv \frac{i}{A}=\frac{N}{\V}\frac{d(ex)}{dt}$. One may view the system as a medium of susceptibility $\chi$ with the polarization related to the optical field by $P_{opt}\equiv \frac{N}{\V}ex=\epsilon_o \chi \E$. The expression $J_{opt}=dP_{opt}/dt$ coincides with the previous one. The current density $J_{opt}$, the field $\E$, and the electric induction $D=\epsilon_o \E+P_{opt}$, are the quantities that enter into the Maxwell equations.  }.
 \begin{align}
\label{qqq}
\ave{i(t)}=\frac{e\ave{p(t)}}{md}=\om_o\frac{ex_{12}}{d}\sin(\Om_R t)\cos(\om_ot),
\end{align}
It follows that $\ave{i(t)}$ varies, like the optical potential $v(t)\equiv\sqrt 2 V \cos(\om_ot)$, essentially according to $\cos(\om_o t)$-law, but with a slowly-varying factor proportional to $\sin(\Om_R t)$. Since the electrons are not directly coupled to one another, the induced averaged current is proportional to the total number of electrons\footnote{An alternative model, more appropriate for semiconductors is that of nearly-resonant coupling between two narrow bands of states, assuming that an equilibrium is quickly reached within the two bands separately. For example at $T$=0K, electrons are supposed to decay instantly to the lowest available level of the band considered (only one single-spin state electron being allowed in each level, according to the Pauli principle). The lower band is called in that case the valence band and the upper band the conduction band, see the book by Landau and Lifchitz \cite[p. 420]{Landau1984}. These authors consider from the outset broadened levels and introduce early in their calculation of the electron-optical field coupling population ratios at thermal equilibrium. The end result is a relation between the variance of the induced current and the conductance (or susceptibility) of the material.}.

The QM-averaged energy $E(\tau)$ supplied by the optical potential source to the electron from time 0 (when the electron energy is equal to zero) to time $\tau$ is obtained by integrating the (QM averaged) power $P(t)=\ave{i(t)}v(t)$. Averaging over the fast optical variations, that is, replacing $\cos^2(\om_o t)$ by 1/2, we obtain
 \begin{align}
\label{qqk}
E(\tau)=\hbar\om_o \sin^2(\frac{\Om_R}{2}\tau)
\end{align}
where we have employed the expression in \eqref{rabi} of the Rabi frequency in terms of $x_{12}$. The expression in \eqref{qqk} shows that, if the time $\tau$ of interaction of the electron with the field is such that $\Om_R\tau=\pi, 3\pi...$, the energy supplied by the optical potential source up to that time to the electron is equal to the energy $eU\approx \hbar\om_o$. In general the electron evolves from the absorbing state at $t=0$ to a state superposition, receiving from the optical source an average energy $\hbar\om_o \sin^2(\frac{\Om_R}{2}\tau)$, and has some non-unity probability of being in the emitting state at the exit time $\tau$. If $\Om_R\tau\ll1$ this probability is very small, yet measurable.

The (QM) average conductance "seen" by the (deterministic) alternating potential source $v(t)=\sqrt2V \cos(\om_o t)$ is initially
 \begin{align}
\label{gb}
G(t)\equiv \frac{\ave{i(t)}}{v(t)}=\om_o  \frac{e^2}{\hbar }\p \frac{x_{12}}{d}\q^2~\frac{\sin(\Om_R t)}{\Om_R}\approx  \frac{e^2}{\hbar }\p \frac{x_{12}}{d}\q^2~\om_ot=\frac{e^2}{2md^2}t.  
\end{align}
In the last expression we have assumed an oscillator strength equal to 1 instead of the previously calculated value of 0.96. It is interesting that this expression does not involves $\hbar$. If some external mechanism interrupts the process after a time $\tau$ much smaller than the Rabi period, the time-averaged conductance is given by \eqref{gb} with $t$ replaced by $\tau/2$. Under such conditions the optical potential source "sees" a positive (QM and time)-average conductance. A similar discussion applies if the electron is initially in the upper state 2. The conductance then has the same absolute value as before, but is negative. 

\subsection{Electron energy.}

The electron energy reduces in the present situation to the kinetic energy $E=p^2/\p2m\q$. For the non-stationary state previously considered we first evaluate
 $\ave{p^2}$ from \eqref{quick}
 \begin{align}
\label{m}
\ave{p^2}&\equiv \int_{-\infty}^{+\infty}dp~p^2~P(p,t)\nonumber\\
&=\cos^2(\frac{\Om_R}{2} t) \int_{-\infty}^{+\infty}dp~p^2\psi_{1}(p)^2+ \sin^2(\frac{\Om_R}{2} t) \int_{\infty}^{+\infty}dp~p^2\psi_{2}''(p)^2\nonumber\\
&-2\sin(\frac{\Om_R}{2} t)\cos(\frac{\Om_R}{2} t) \sin(\om_ot)\int_{-\infty}^{+\infty}dp~p^2\psi_{1}(p)\psi_{2}''(p)\nonumber\\
\end{align}
where $\psi(p,t)$ is given in \eqref{ex}. 
Thus, using previous expressions
 \begin{align}
\label{mmm}
\ave{E}=\frac{\ave{p^2}}{2m}=E_1\cos^2(\frac{\Om_R}{2} t)+E_2\sin^2(\frac{\Om_R}{2} t).
\end{align}
If a measurement of the electron energy is performed at some time, the probability that the electron be found in state 1 of energy $E_1$ is $\cos^2(\frac{\Om_R}{2} t)$ and the probability that the electron be found in state 2 of energy $E_2$ is $\sin^2(\frac{\Om_R}{2} t)$.  

To summarize, we have evaluated in the present section the QM-averaged optical current $\ave{i(t)}$ induced by an electron residing between plates submitted to static and optical potential sources. The power $v(t)\ave{i(t)}$, when averaged over an optical period, is denoted $P(t)$. If the electron is initially in the lower state, the energy $E(t)$ supplied by the optical potential up to time $t$, namely the integral of $P(t)$ from $t=0$ to $t$ was found equal to the electron QM-average energy at time $t$.

If an electron initially in the lower state interacts with the field during a fixed time $\tau$, with $\Om_R\tau\ll1$, it has a small probability of being in the upper state at the exit time. If this happens, a detection event is recorded. Because the probability is small, the detection events form a Poisson point process, even if the electrons are injected regularly, say at times $t=0,\tau,2\tau,...$. On the other hand, we have shown that a constant (i.e., field -independent) positive conductance is obtained. In the next section, we show that such a Poisson distribution may be interpreted by supposing that a current source of spectral density $\hbar\om_o G$ is associated to a conductance $G$, in the condition previously considered that the electron remains most of the time in the absorbing state. This is the Nyquist-like formula. This optical noise-current itself is not measurable. But when the complete circuit equations are solved, we obtain accurate expressions for measurable photo-currents. 

In subsequent parts of this paper we suppose that $G$ depends on parameters such as the number of electrons. The relative variations of these parameters and of the conductances are small (e.g., $\De n/n\approx \De G/G\approx 0.01$), and therefore the spectral density of the Nyquist-like noise sources may be supposed to be unaffected. The dependence of $G$ on these parameters, however, entails drastic changes in the system behavior. In particular it explains the different behaviors of light-emitting diodes and above-threshold lasers. In the latter case, intensity fluctuations are very small compared with the mean intensities.

\subsection{C-state}\label{C}

Light waves are said to be in the C-state\footnote{C-state beams resemble the so-called "coherent" states of light employed in quantum optics. However, C-states are fundamentally states of propagating light while coherent states are primarily states of optical resonators. In the context of Quantum Optics, Glauber has shown in 1963 that a classical prescribed current (which we call a current source) radiates light in the so-called "coherent state". When coherent states are incident on a photo-detector the statistics of the photo-electrons is Poissonian. The results therefore are similar.} if they generate Poissonian photo-electrons irrespectively of the carrier phase\footnote{Concretely, the carrier phase may be changed by inserting on the optical beam (i.e., before detection) a second-order all-pass filter, which is a conservative (i.e., lossless, gainless) resonating device that changes the carrier phase without changing its amplitude. We suppose that this circuit bandwidth is very small compared with the Fourier frequencies of interest. If this is the case, the carrier phase may be changed arbitrarily from 0 to 2$\pi$ simply through a very small detuning. The fluctuations, on the other hand, are essentially unaffected by that all-pass filter. The description of second-order all-pass filters may be found in Circuit-Theory textbooks. }. We show that potential or current \emph{sources} radiate light in the C-state.
 
Let us first recall well-known observations. The current emitted by a cathode whose emission is temperature-limited consists of independently emitted electrons. Mathematically, this electronic emission process is referred to as a Poisson process. Let the emitted current be denoted by $\mathcal{J}\equiv J+\De J$, where $J$ denotes the time-averaged emitted current. The spectral density of the fluctuation $\De J$ is given by the formulas
\begin{align}
\label{7}
\spectral_{\De J}&=eJ,\\
\label{8}
\spectral_{\De D}&=D, 
\end{align}
where $e$ denotes the absolute value of an electron charge and double-sided spectral densities are employed. The letter "D" stands for "detection". The electronic rate is defined as $\D\equiv \mathcal{J}/e=D+\De D$, with $D=J/e$ and $\De D\equiv \De J/e$.
Relation \eqref{7} says that the average power dissipated in a 1$\Om$ resistance following a 1 Hz band-pass filter centered at any low frequency (white noise) is given by the shot-noise formula
\begin{align}
\label{9}
2\spectral_{\De J}=2eJ,
\end{align}
If a light source, for some reason, is supposed to emit photons independently each photon carrying an energy $\hbar \om_o $, where $\hbar$ denotes the Planck constant and $\om_o$ the light frequency, we are dealing again with a Poisson process. The spectral density of the light-power fluctuation $\De P$ is thus
\begin{align}
\label{10}
\spectral_{\De P}&=\hbar \om_o P,\\
\label{11}
\spectral_{\De Q}&=Q,
\end{align}
where the average photonic rate is $Q\equiv P/\hbar \om_o$ and the fluctuation is $\De Q\equiv \De P/\hbar \om_o$. If the optical beam is incident on an ideal photo-detector, the light is converted into an electron rate identical to the photon rate. Thus, the electron rate fluctuation spectral density $\spectral_{\De D}=D=Q$. However, the above discussion suggests that the introduction of photons is here superfluous.

In the present theory complex random current sources $C(t)\equiv C'(t)+\ii C''(t)$ are associated with positive conductances $G $ (Remember that $C(t)$ is the complex representation of a signal at the carrier optical frequency $\om$). The real and imaginary parts of $C' $ and $C''$ of $C$, respectively, are uncorrelated and have spectral density
\begin{align}
\label{12}
\spectral_{C'}=\spectral_{C''}=\hbar \om_o G.
\end{align}
Let us show that this formalism agrees with the previously-quoted shot-noise formulas. We have established in the previous section that when a large number of electrons are submitted to optical potential sources photo-detection events are Poisson distributed. We thus consider a potential source applied to a conductance $G$ endowed with its Nyquist-like noise source, and show that the photo-electrons are indeed Poisson distributed, irrespectively of the potential-source phase. The interest of this representation is that it dispenses us of considering real devices involving electrons. We substitute to it a simple circuit schematic.

Consider indeed a potential source $V$ applied to a conductance $G$. The current delivered by the source consists of two parts. First the current $GV$ flowing through the conductance, and secondly the noise current $C(t)$. It follows that the power delivered by the potential source (and received by the detector according to the law of average-energy conservation) reads
\begin{align}
\label{9m}
P(t)=\Re\{ V^{\star}\bigl( GV+C    \bigr)    \}=G\abs{V}^2+V'C'+V''C'',
\end{align}
The first term is the average power $P=G\abs{V}^2$. The second terms are fluctuating terms. The spectral density of the fluctuation reads (see Section \ref{random})
\begin{align}
\label{9mn}
\spectral_{\De P}=V'^2\spectral_{C'}+V''^2\spectral_{C''}=G\abs{V}^2 \hbar\om_o=P\hbar\om_o.
\end{align}
Thus, setting $D=P/\hbar\om_o$, $\De D=\De P/\hbar\om_o$, we recover the relation $\spectral_{\De D}=D$. The photo-current is Poissonian, irrespectively of the phase of $V$. A similar calculation can be made for current sources. Thus we have shown that potential and current sources radiate light in the C-state. To summarize, under the conditions outlined above, detection events are Poisson distributed and the conductance is constant. This Poisson behavior is accounted for by postulating Nyquist-like currents. 

In the present section we were mostly concerned with electrons that interact with the field during some small time $\tau$, and reside most of the time either in the absorbing state (positive conductance) or in the emitting state (negative conductance). In the subsequent section we consider electrons that interact permanently with the field, but may spontaneously transit from one state to the other at some known average rate $2\gamma$.

\newpage

\section{Generalized electron-field interaction}\label{general}

We consider in the present section an electron present all the time in the interaction region, with spontaneous transitions between the two levels considered above. Unlike the spontaneous events usually considered in the Quantum Optics literature, these spontaneous transitions do not involve an irreversible loss of energy. They merely convert the electron energy into an energy received (or delivered) by the static potential source. 

The waiting-time distribution $w(t)$ is defined as follows: Given that the electron is in the lower state at $t=0$ (implying that a transition event just occurred), $w(t)dt$ is the probability that the \emph{next} transition occurs between $t$ and $t+dt$. We first give an approximate form of the waiting-time distribution. Then we introduce generalized Rabi oscillations and evaluate exactly the waiting-time distribution and the power $P(t)$ supplied by the optical-potential source to the electron and to the static-potential source.  

\subsection{Waiting time probability for small decay rates}\label{jump}

We considered in the previous section an electron which may reside in either one of two states separated in energy by $\hbar\om_o\approx eU$. If an optical potential source $\sqrt2 V\cos(\om_o t)$ is added, the electron undergoes Rabi oscillations. To wit,  if the electron is initially ($t=0$) in the lower state, the probability that it be in the higher state at time $t$ is $\sin^2(\frac{\Om_R}{2} t)$, where $\Om_R$ denotes the Rabi frequency. The energy delivered by the optical potential source is equal to $\hbar\om_o$ when $t=\pi/\Om_R$. This is the energy that the electron would deliver to the static potential source if it were to tunnel through it at that time.

If an event occurred at $t=0$, the probability density of a downward transition is approximately the product of the probability $\sin^2(\frac{\Om_R}{2} t)$ that the electron be in the emitting state at time $t$ and some constant $2\gamma$. The transition events form an inhomogeneous Poisson process of rate $\lambda(t)=2\gamma\sin^2(\frac{\Om_R}{2} t)=\gamma\bigl(1-\cos(\Om_R t)\bigr)$. A change of the time-scale $\tau=\tau(t)$ with $d\tau(t)=\lambda(t)dt$ transforms this inhomogeneous Poisson process into a homogeneous Poisson process of density 1. For such a process it is known that the probability that an the next event occurs between $\tau$ and $\tau+d\tau$  is $\exp(-\tau)d\tau$, see Section \ref{stationary}. Let us set
 \begin{align}
\label{jbis}
f(t)&=\exp(-\tau(t))=\exp\left(-\int_0^t dt\lambda(t)\right)= \exp{\bigl(-\gamma\bigl(t-\frac{\sin{\Om_Rt}}{\Om_R}\bigr)\bigr)}\nonumber\\
\end{align}
If an event occurs at $t=0$, the probability $W(t)dt$ that the next event occurs in the interval $(t,t+dt)$ reads after integration
 \begin{align}
\label{jubis}
W(t)dt&=\exp(-\tau(t))d\tau(t)=-df(t)=f(t)\lambda(t)dt\nonumber\\
&=\gamma(1-\cos{\Om_R t)}\exp{\bigl(-\gamma\bigl(t-\frac{\sin{\Om_Rt}}{\Om_R}\bigr)\bigr)}dt.
\end{align}
From now on we select a time-scale such that $\Om_R=1$. Figure \ref{waitingtime} represents $W(t)$ for $\gamma=\Om_R/7=1/7$. The expression for $W(t)$ given above almost coincides with the one, $w(t)$, known to apply to resonant fluorescence \cite[see Figs. 8.4 and 8.5]{Carmichael1993}\cite{Carmichael1989} when $\gamma t\ll1$. The exact expression of the waiting-time probability density $w(t)$ is derived in Section \ref{$2a$}. Note that the probability that no transition event occurs before time $t$ is the integral of $W(t)$ from $t$ to $\infty$, that is, $f(t)$. If we knew the time at which the last event occurred before time $t=0$, the electron state at $t=0$ would be a pure state. If, however, we only know the probability density that this event occurred at time $t<0$, the electron state at $t=0$ must be described by a mixed-state density matrix. This is the situation described in the next sections.

\setlength{\figwidth}{0.6\textwidth}
\begin{figure}
\centering
\includegraphics[width=\figwidth]{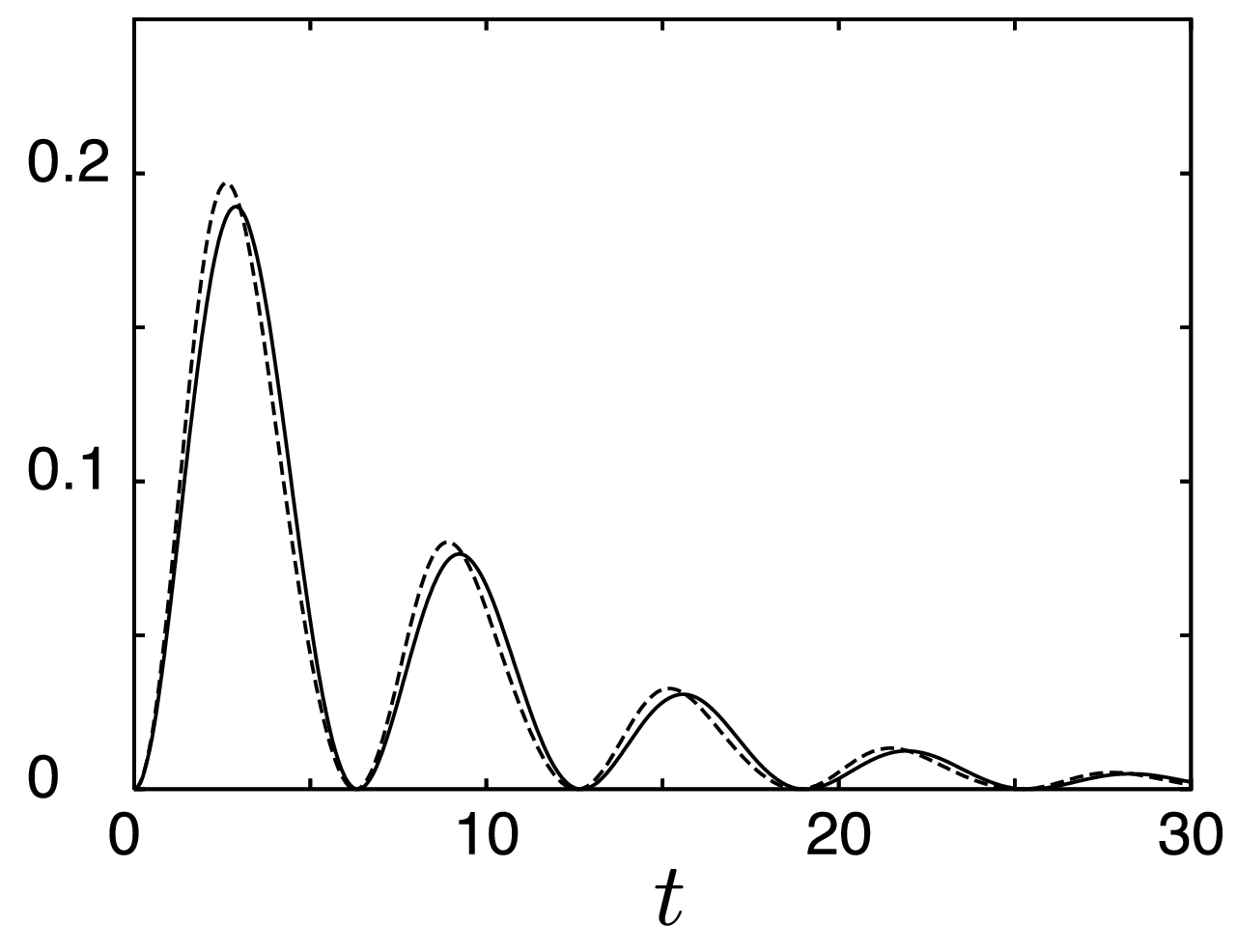}
\caption{The figure represents the waiting-time density $W(t)$ according to the approximate result in \eqref{jubis} (plain line) and  the waiting-time density $w(t)$ according to the exact result in \eqref{juterbis} (dashed line) for $\gamma=\Om_R/7$.}
\label{waitingtime} 
\end{figure}

In general, one should allow for the fact that an electron, having a probability $\sin^2(\frac{\Om_R}{2} t)$ of being in the upper state at time $t$ and therefore a probability $\cos^2(\frac{\Om_R}{2} t)$ of being in the lower state, has not only a probability density $2\gamma_2\sin^2(\frac{\Om_R}{2} t)$ of performing a transition to the lower state as we discussed above, but also a probability density $2\gamma_1\cos^2(\frac{\Om_R}{2} t)$ of performing a transition to the upper state. Previously, we assumed that $\gamma_1=0$ and $\gamma_2\equiv \gamma$, but this needs not be always the case.

\subsection{Density matrix}\label{dmat}

When there is some uncertainty concerning the initial form of the wave-function, it is convenient to employ the $2\times 2$ mixed-state density matrix whose mathematical properties were given in Section \ref{vector}. We first consider the Rabi equations as given earlier, and subsequently introduce the generalized Rabi equations.

The wave-functions  in space and momentum space are of the form given in \eqref{solvter} and \eqref{ex}
\begin{align}
\label{solvterbis}
\psi(x,t)&= C_1(t)\exp(-\ii \om_1 t)\psi_1(x)+C_2(t)\exp(-\ii \om_2 t)\psi_2(x)\nonumber\\
\psi(p,t)&= C_1(t)\exp(-\ii \om_1 t)\psi_1(p)+C_2(t)\exp(-\ii \om_2 t)\psi_2(p).
\end{align}
where the complex coefficients $C_1(t),C_2(t)$ obey the differential equations
\begin{align}
\label{formbis}
 \frac{dC_1(t)}{dt}=\ii\frac{\Om_R}{2}C_2(t)\qquad
 \frac{dC_2(t)}{dt}=\ii\frac{\Om_R}{2}C_1(t)\qquad C_1(t)C_1^\star(t)+C_2(t)C_2^\star(t)=1, 
\end{align}
and $\Om_R\ll\om_o$ is the Rabi frequency given by $\hbar\Om_R=ex_{12}\sqrt2V/d$. We assume that $C_1(0)C_2^\star(0)$ is imaginary by a proper selection of the phase of the wave-function. It then follows from the above differential equations that $C_1(t)C_2^\star(t)$ is imaginary at all times.

Proceeding as in the previous section, we evaluate
\begin{align}
\label{ehbis}
\ave{p(t)}=\int_{-\infty}^{+\infty} dp~p~ \psi(p,t)\psi^\star(p,t)=C_1(t)C_2^\star(t)\exp(\ii\om_o t)p_{12}+cc,
\end{align}
defining 
\begin{align}
\label{ehbi}
p_{ij}\equiv\int_{-\infty}^{+\infty} dp~p~ \psi_i(p)\psi_j^\star(p)\qquad i,j=1,2.
\end{align}
 and assuming as before that  $p_{11}=p_{22}=0$. For the situation presently considered $p_{12}=-\ii\hbar\om_o x_{12}=p_{21}^\star$. Because $C_1(t)C_2^\star(t)$ is imaginary, $C_1(t)C_2^\star(t)p_{12}$ is real, and we are left with a term proportional to $\cos(\om_ot)$, in phase with the driving field. We obtain the average induced current $\ave{i(t)}=\frac{ e\ave{p(t)} } {dm}$ as in \eqref{qqq}. The instantaneous power delivered by the optical potential source is $v(t)\ave{i(t)}$, where $v(t)=\sqrt2 V\cos(\om_o t)$ and $\hbar\Om_R=\frac{\sqrt2 V~ex_{12}}{d}$. Averaging over an optical period, that is, replacing $\cos^2(\om_o t)$ by 1/2, we may write the power in the form 
\begin{align}  \label{eis}
P(t)=\hbar\om_o \Om_R  \Im \{C_1(t)C_2^\star(t)\}\equiv \hbar\om_o \Om_R \rho''_{12}
\end{align}

Let us now define a $2\times2$ density matrix 
\begin{align}\label{s9}
 \boldsymbol{\rho}=  \left (   \begin{array}{ccc}
 \rho_{11}(t)&\rho_{12}(t)\\
\rho_{21}(t)&\rho_{22}(t) 
\end{array}\right)
\equiv\left (   \begin{array}{ccc}
 C_1 (t)C_1^\star(t)&C_1 (t)C_2^\star(t)\\
C_2(t)C_1^\star( t)&C_2 (t)C_2^\star(t) 
\end{array}\right)
\end{align}
The $\boldsymbol{\rho}(t)$-matrix is Hermitian, that is, equal to its complex conjugate transpose and $trace\{ \boldsymbol{\rho}^2(t) \} =trace\{ \boldsymbol{\rho}(t) \}= 1$. 

We readily obtain the differential equations obeyed by the elements of the $\rho(t)$-matrix from those in \eqref{formbis} as
\begin{align}
\label{rhobis}
\frac{d\rho_{11}(t)}{dt}&=\ii \frac{\Om_R}{2}(\rho_{12}(t)-\rho_{21}(t))\nonumber\\
\frac{d\rho_{22}(t)}{dt}&=\ii \frac{\Om_R}{2}(\rho_{21}(t)-\rho_{12}(t))\nonumber\\
\frac{d\rho_{12}(t)}{dt}&=\ii \frac{\Om_R}{2}(\rho_{11}(t)-\rho_{22}(t))\nonumber\\
\frac{d\rho_{21}(t)}{dt}&=\ii \frac{\Om_R}{2}(\rho_{22}(t)-\rho_{11}(t)).
\end{align}
The power delivered by the optical potential given in \eqref{eis} may be written in terms of the density matrix as
\begin{align}\label{obis}
P(t)=trace\{\boldsymbol{\rho}(t)\boldsymbol{P}  \}      
\end{align}
where
\begin{align}\label{z2}
 \boldsymbol{P}= \hbar\om_o\frac{\Om_R}{2} \left (   \begin{array}{ccc}
 0&\ii\\
-\ii&0 \end{array}\right) 
\end{align}

In view of the mathematical considerations in \eqref{vector}, it follows that the above expression for the power in \eqref{obis} holds also for mixed states, in which case $\boldsymbol{\rho}$ is the weighted sum of pure-state density matrices and is written as
\begin{align}\label{s9bis}
 \boldsymbol{\rho}=  \left (   \begin{array}{ccc}
 \rho_{11}(t)&\rho_{12}(t)\\
\rho_{21}(t)&\rho_{22}(t) 
\end{array}\right) 
\end{align}
without any reference to the $ C_1 (t), C_2 (t)$ coefficients anymore. In other words, the density-matrix concept has been introduced to generate an expression of the power delivered by the optical potential in the case where the wave-function is defined only statistically. We are now ready to generalize the Rabi equations.

\subsection{Generalized Rabi equations}\label{qp}

We have previously evaluated the conductance "seen" by an optical potential source under the assumption that the electron is submitted to the optical field during some known fixed time $\tau$. We now go back to the continuous configuration, and introduce instead a known average decay rate $2\gamma$. This leads to a generalized form of the Rabi equations. These equations involve a parameter $2a$. From the fact that the trace of the square of the density matrix may not exceed 1 we conclude that $2a$ may not be less than 1.  We justify the value $2a=1$ by considering that in the small-$\gamma$ limit the expression of the waiting-time distribution must agree with the expression obtained straightforwardly in a previous section. More generally we assert that the rate of decay of the coherence is as small as is allowed, additional possible sources of decoherence (collisions) being not present in our model. 

We again consider an electron with two states, submitted to a resonant field. In the absence of spontaneous transitions between states 1 and 2, the $2\times 2$-matrix $\rho(t)$ that describes the electron state obeys the differential equation given before in \eqref{rhobis}. If, however, an electron in the emitting state may decay spontaneously to the absorbing state at an average rate $2\gamma$, additional terms must be added on the right-hand-side of these differential equations. For generality, we consider both a spontaneous-transition rate $\gamma_2$ from 2 to 1, and a spontaneous-transition rate $\gamma_1$ from 1 to 2. The previously mentioned equations for $d\rho(t)/dt$ become
\begin{align}
\label{rho}
\frac{d\rho_{11}(t)}{dt}&= \frac{\ii\Om_R}{2}(\rho_{12}(t)-\rho_{21}(t))+2\gamma_2\rho_{22}(t)-2\gamma_1\rho_{11}(t)
\nonumber\\
\frac{d\rho_{22}(t)}{dt}&= \frac{\ii\Om_R}{2}(\rho_{21}(t)-\rho_{12}(t))+2\gamma_1\rho_{11}(t)-2\gamma_2\rho_{22}(t)\nonumber\\
\frac{d\rho_{12}(t)}{dt}&= \frac{\ii\Om_R}{2}(\rho_{11}(t)-\rho_{22}(t))-2a (\gamma_1+\gamma_2)\rho_{12}(t)\nonumber\\
\frac{d\rho_{21}(t)}{dt}&= \frac{\ii\Om_R}{2}(\rho_{22}(t)-\rho_{11}(t))- 2a(\gamma_1+\gamma_2)\rho_{21}(t)
\end{align}
These equations are unchanged under the interchange of the subscripts 1 and 2. If $\rho$ is Hermitian with trace 1 initially (that is, $\rho_{11},\rho_{22}$ are real, $\rho_{11}+\rho_{22}$=1, and $\rho_{21}^\star=\rho_{12}$), these conditions are preserved in the course of time. The trace condition expresses the fact that the electron resides in either one of the two states considered. The Hermitian condition expresses the fact that observables are real. In the low-field limit ($\Om_R=0$), the above equations describe accurately the spontaneous decay from level 2 to level 1, and the spontaneous promotion from level 1 to level 2.

The second equation in \eqref{rho} (or equivalently the first) may be written as
\begin{align}
\label{rhoh}
\ave{E(\tau)}=\hbar\om_o\Om_R\int_0^\tau dt \rho''_{12}(t)=\hbar\om_o\rho_{22}(\tau)+\hbar\om_o\int_0^\tau dt \bigl(2\gamma_2\rho_{22}(t)-2\gamma_1\rho_{11}(t)\bigr),
\end{align}
where we have assumed that the electron is initially in the absorbing state $\rho_{22}(0)=0$, and $\rho''_{12}(t)$ denotes the imaginary part of $\rho_{12}(t)$. $\ave{E(\tau)}$ represents the energy supplied by the optical potential source from $t=0$ to $t=\tau$, that is the time integral from 0 to $\tau$ of the power $P(t)$ given in \eqref{obis}. This energy split into two parts. First the energy delivered to the electron, i.e., the increment of $\hbar\om_o\rho_{22}(\tau)$. Second, the energy transferred to the static potential source through the decay rate $2\gamma_2\rho_{22}(t)$, minus a similar term for the upward spontaneous transition. Thus the first two equations in \eqref{rho} are almost obvious, as they express the law of average-energy conservation. From now on, for brevity, we set $\Om_R=1$.

As far of the third equation is concerned (or equivalently the fourth), let us note that if initially the trace of $\boldsymbol{\rho}^2$ equals 1 (pure state), the time derivative of the trace of $\boldsymbol{\rho}^2(t)$ may not be positive since $trace\{\boldsymbol{\rho}(t)\}≤1$. For $\gamma_1=0$ or $\gamma_2=0$, a short calculation shows that this implies that $2a≥1$, see Section \ref{vector} in which we set $\gamma=\gamma_1+\gamma_2, b\gamma=\gamma_1-\gamma_2$. Since for a single electron there are no decoherence due to elastic collisions, we may assume that the decoherence rate is as small as possible, implying that $2a(\gamma_1+\gamma_2)=\left(\sqrt{\gamma_1}-\sqrt{\gamma_2}   \right)^2$.

After a sufficiently long time a steady-state is reached, obtained by setting the time derivatives equal to 0. We have $\rho'_{12}(\infty)=0$ and
\begin{align}
\label{rhok}
\rho_{12}''(\infty)&=\frac{\gamma_2-\gamma_1}{1+4a\bigl(\gamma_1+\gamma_2\bigr)^2}=\frac{\gamma}{1+4a\gamma^2}=\frac{\gamma}{1+2\gamma^2}\nonumber\\
\rho_{22}(\infty)&=\frac{\frac{1}{2}+4a\gamma_1(\gamma_1+\gamma_2)}{1+4a\bigl(\gamma_1+\gamma_2\bigr)^2}=\frac{1}{2+8a\gamma^2}=\frac{1}{2+4\gamma^2}
\end{align}
For $\gamma_1=\gamma_2$ we have, as expected because of symmetry, $\rho_{12}''(\infty)=0,\rho_{22}(\infty)=\rho_{11}(\infty)=\frac{1}{2}$. In the second expressions above we have set $\gamma_1=0,\gamma_2=\gamma$ and in the third we have further set $2a=1$.

\subsection{Solution for arbitrary times}\label{solu}

Let us now solve the equations for arbitrary times. Using the fact that  $\rho_{21}=\rho_{12}^\star$, $\rho_{11}=1-\rho_{22}$, $\rho_{22}$ real, we only need the second and third equations. Taking the imaginary part of the third equation we have
\begin{align}
\label{rho4}
\frac{d\rho_{12}''(t)}{dt}= \frac{1}{2}-\rho_{22}(t)-2a(\gamma_1+\gamma_2)\rho_{12}''(t),
\end{align}
The second equation may be written as
\begin{align}
\label{rho5}
 \rho''_{12}(t)=\frac{d\rho_{22}(t)}{dt}-2\gamma_1+2(\gamma_1+\gamma_2)\rho_{22}(t).
\end{align}
Substituting $\rho_{12}''(t)$ from \eqref{rho5} into \eqref{rho4} we obtain for $\gamma_1=0,\gamma_2\equiv \gamma$ the second-order differential equation for $\rho_{22}(t)$
\begin{align}
\label{rho62}
\frac{d^2\rho_{22}(t)}{dt^2}+2(1+a)    \frac{d\rho_{22}(t)}{dt}+
\bigl( 1+4a\gamma^2  \bigr)\rho_{22}(t)= \frac{1}{2}.
\end{align}
Considering the function $y(t)\equiv \rho_{22}(t)-1/(2+8a\gamma^2)$, the constant term cancels out. Next, we try as usual solutions of the form $y(t)\equiv \exp(\lambda t)$ and find that $\lambda$ must be a solution of the second-degree equation
\begin{align}
\label{rho8}
\lambda^2+2(1+a)\gamma\lambda +1+4a\gamma^2=0,
\end{align}
whose solution is
\begin{align}
\label{rho8bis}
\lambda_\pm=-(1+a)\gamma\pm\kappa \qquad  \kappa\equiv \sqrt{(1-a)^2\gamma^2-1}.
\end{align}
With the initial condition $\rho_{22}(0)=0$ the solution is of the form
\begin{align}
\label{rho9}
\rho_{22}(t)=\frac{1}{2+8a\gamma^2}\{1+A\exp(\lambda_+ t)-(1+A)\exp(\lambda_- t)\},
\end{align}
Since the condition $\rho_{12}(0)=0$ implies that the first derivative of  $\rho_{22}(t)$ with respect to time vanishes at $t=0$, the expression of $\rho_{22}(t)$ is
\begin{align}
\label{rho10}
\rho_{22}(t)&=\frac{1}{2+8a\gamma^2}\{1+\frac{\lambda_-}{2\kappa}\exp(\lambda_+ t) -\frac{\lambda_+}{2\kappa}\exp(\lambda_- t)    \}
\end{align}
Note that this expression is always real. We interpret $G(t)\equiv2\gamma\rho_{22}(t)$ as the probability density of having an event at time $t$, not to be confused with the conductance $G(V)$, where $V$ denotes the applied potential. Since an event occurred at $t=0$, the normalized correlation reads $g(\tau)=G(\tau)/G(\infty)$, where $G(\infty)=2\gamma\rho_{22}(\infty)=\frac{\gamma}{1+4a\gamma^2}$ is the probability density of the process. The relative noise is obtained straightforwardly from the Fourier transform of $g(\tau)$ according to \eqref{g}.

\subsection{Correlation when 2a=1}\label{spe}

When $2a=1$, the expression of $G(t)\equiv2\gamma\rho_{22}(t)$ in \eqref{rho10} reads
 \begin{align}
\label{rhozz}
G(t)=\frac{\gamma}{1+2\gamma^2}\{1+\frac{\lambda_-}{2\kappa}\exp(\lambda_+ t) -\frac{\lambda_+}{2\kappa}\exp(\lambda_- t)    \} \quad \lambda_{±}=-\frac{3\gamma}{2}±\kappa\quad \kappa=\sqrt{\frac{\gamma^2}{4}-1}
\end{align}
In particular
 \begin{align}
\label{rho11un}
G(t)&\approx\frac{\bigl(1-\exp(-\gamma t)\bigr)^2}{2\gamma}\qquad\gamma\gg1\nonumber \\
G(t)&\approx\gamma\bigl(1-\exp(-\frac{3\gamma t}{2})\cos( t)\bigr)\qquad\gamma\ll 1\nonumber \\
G(t)&=2\frac{1-(1+3t)\exp(-3t)}{9}\qquad \gamma=2\nonumber\\
G(\infty)&=\frac{\gamma}{1+2\gamma^2}. 
\end{align}

\subsection{Average conductance and fluctuations when $2a=1$}\label{$cond$}

Let us rewrite \eqref{rhozz} with the Rabi frequency $\Om_R$ restored. We set $\gamma_o\equiv \gamma/\Om_R$.
 \begin{align}
\label{restx}
G(t)&=\frac{\gamma_o\Om_R}{1+2\gamma_o^2}\{1+\frac{\lambda_-}{2\kappa}\exp(\lambda_+\Om_R t) -\frac{\lambda_+}{2\kappa}\exp(\lambda_- \Om_R t)    \} \nonumber\\
 \lambda_{±}&=-\frac{3\gamma_o}{2}±\kappa\quad \kappa=\sqrt{\frac{\gamma_o^2}{4}-1}\nonumber\\
G(\infty)&=\frac{\gamma_o\Om_R}{1+2\gamma_o^2}
\end{align}
In the large $\gamma$ limit
 \begin{align}
\label{restxx}
G_o\equiv G(\infty)_{\gamma\to\infty}=\frac{\Om_R^2}{2\gamma}.
\end{align}
Since each event absorbs an energy $\hbar\om_o$ and $G_o$ (not to confused with a conductance) denotes the average number of events per unit time, the power delivered by the optical potential $V$ to the static potential $U$ is
 \begin{align}
\label{restxxx}
P\equiv GV^2=G_o \hbar\om_o=\frac{\Om_R^2 \hbar\om_o}{2\gamma}.
\end{align}
Because, according to \eqref{rabi}, $\Om_R^2=2\bigl( \frac{eVx_{12}}{\hbar d}  \bigr)^2$ is proportional to $V^2$, the conductance $G$ does not depend on the potential $V$, that is, the system is linear in that limit. The conductance has the same form as the one given in \eqref{gb} for an electron returning to the absorbing level after a fixed time $\tau$, if we set $\tau=2/\gamma$. For larger values of $V$ the conductance decreases, and tends to 0. This non-linearity may be interpreted as usual from the fact that the lower population decreases and the upper population increases as a consequence of stimulated absorption. Indeed, the conductance is for arbitrary $\gamma$-values equal to the previous low-field expression multiplied by $\rho_{11}-\rho_{22}=1-2\rho_{22}=2\gamma_o^2/(1+2\gamma_o^2)$.

Let us now consider the normalized events correlation $g(\tau)=G(t)/G(\infty)$. For large $\gamma$-values we obtain
 \begin{align}
\label{restxxxz}
g(\tau)&=\bigl( 1-\exp(-\gamma t)   \bigr)^2\to 1\qquad \gamma\to \infty\nonumber\\
 \N(0)&=2\int_0^\infty d\tau (g(\tau)-1)\to 0\qquad \gamma\to \infty.
\end{align}
That is, the relative noise $\N$ vanishes at zero Fourier frequencies, so that the point-process may be considered Poissonian. For arbitrary $\gamma$-values, the spectral density of the point process is viewed as the sum of two terms. One originating from the lower-state population $\rho_{11}$ (positive conductance), and one originating from the upper-state population $\rho_{22}$ (negative conductance). If they were independent, the corresponding spectral densities should add up, that is, be proportional to $\rho_{11}+\rho_{22}=1$. We have in fact from \eqref{rhozz}
 \begin{align}
\label{rtxxxz}
g(\tau)-1&=\frac{1}{2\kappa}\bigl(\lambda_-\exp(\lambda_+\Om_R t)-\lambda_+\exp(\lambda_-\Om_R t)        \bigr)\nonumber\\
 \lambda_{±}&=-\frac{3\gamma_o}{2}±\kappa\quad \kappa=\sqrt{\frac{\gamma_o^2}{4}-1},
\end{align}
with $g(\infty)=1,~g((0)=0$.
The relative noise at zero Fourier frequency is
\begin{align}
\label{txxxz}
\N(0)=2\int_0^\infty d\tau (g(\tau)-1)=\frac{1}{\kappa \Om_R}\bigl(\frac{\lambda_+}{\lambda_-}-\frac{\lambda_-}{\lambda_+}    \bigr)=-\frac{1}{\Om_R}\frac{6\gamma_o}{1+2\gamma_o^2}.
\end{align}
It follows that the fluctuations corresponding to the negative and positive conductances are correlated. This is to be expected because we are dealing with a single electron. In later applications, we consider instead two large collections of electrons, one in the emitting state and one in the absorbing state. In that case the fluctuations are independent.

\subsection{Waiting time probability}\label{qap}

The process we are presently considering is an ordinary renewal point process. For such a process it is straightforward to go from $G(t)$ to the waiting time probability density $w(t)$. The concept is that the probability density of an event at $t$ is the sum of the probabilities that this occur through 1 jump, 2 jumps,...and that k-jumps probability densities involve k-fold auto-convolutions of $w(t)$. The formula relates the Laplace transform $G(p)$ of $G(t)$ and the Laplace transform $w(p)$ of $w(t)$, see Section \ref{convolution}           
\begin{align}
\label{rhoml}
w(p)=\frac{1}{1+1/G(p)}.
\end{align}
It follows straightforwardly from the fact that $G(t)$ involves a constant term and that, consequently, $G(p)$ behaves as $1/p$ for small $p$, that
\begin{align}
\label{rhomo}
\frac{1}{\ave{\tau}}=G(\infty)\qquad \ave{\tau}\equiv\int_0^\infty dt~ t~w(t)=-\bigl(\frac{dw(p)}{dp}\bigr)_{p=0}
\end{align}
implying conservation of the average power.

To obtain $w(p)$ we must evaluate the Laplace transform of $G(t)$. Since the Laplace transform of $\exp(-\al t)$ is $1/(p+\al)$ and Laplace transforms are linear operations, the Laplace transform of $G(t)\equiv2\gamma\rho_{22}(t)$, where $ \rho_{22}(t)$ is given in \eqref{rho10}, is, with $\lambda_\pm=-(1+a)\gamma\pm\kappa, ~~ \kappa\equiv \sqrt{(1-a)^2\gamma^2-1}$ as before,
\begin{align}
\label{rho11}
G(p)&=\int_0^\infty dt \exp(-pt)2\gamma\rho_{22}(t)=\frac{\gamma}{1+4a\gamma^2}\bigl( \frac{1}{p}+\frac{\lambda_-/(2\kappa)}{p-\lambda_+}-\frac{\lambda_+/(2\kappa)}{p-\lambda_-}  \bigr)\nonumber\\
\frac{1}{G(p)}&=\frac{p(p-\lambda_+)(p-\lambda_-)}{\gamma}.
\end{align}

From \eqref{rhoml} and \eqref{rho11}, we obtain 
\begin{align}
\label{rhomp}
w(p)&=\frac{\gamma}{p(p-\lambda_+)(p-\lambda_- )+\gamma}\nonumber\\
&=\frac{\gamma}{p^3+2(1+a)\gamma p^2+(4a\gamma^2+1)p+\gamma}
\end{align}
We will first consider the case where $2a=1$ and subsequently give $w(t)$ for arbitrary $a$-values.

\subsection{Waiting time probability when $2a=1$}\label{$2a$}

Let us consider first the case $2a=1$. We have 
\begin{align}
\label{romp}
w(p)&=\frac{\gamma}{2\al^2}\bigl(\frac{1}{p+\gamma+\al} +\frac{1}{p+\gamma-\al}-2\frac{1}{p+\gamma}   \bigr)\qquad\al\equiv\sqrt{\gamma^2-1}
\end{align}
Taking the inverse Laplace transform of $w(p)$ in \eqref{romp} we get
\begin{align}
\label{juterbis}
w(t)=\frac{\gamma}{2\al^2}\{ \exp (-(\gamma-\al) t)+ \exp(-(\gamma+\al) t) -2\exp(-\gamma t)\} .
\end{align}
which is real non-negative for all $\gamma$-values and integrates to unity.  At small times we have $G(t)\approx w(t)\approx \gamma t^2/2$.
Special forms are
\begin{align}
\label{uter}
w(t)&=\frac{1}{2\gamma}\exp(-\frac{t}{2\gamma})\qquad t≠0, \quad  \gamma\gg1    \nonumber\\
&= \frac{1}{3} \{ \exp(-(2-\sqrt3)t)+\exp(-(2+\sqrt3)t)-2\exp(-2t) \}        \quad  \gamma=2 \nonumber\\
&=\gamma(1+\gamma^2)\bigl(1-\cos(t)\bigr) \exp(-\gamma  t)\quad  \gamma\ll1
\end{align}
In the small $\gamma$ limit, the expression in \eqref{uter} coincides with the expression obtained on intuitive grounds in \eqref{jubis}. Note that the three expressions in \eqref{uter} integrate to unity.

\subsection{Arbitrary $a$-parameter}\label{$gen$}

We now have to find the roots of $p^3+2(1+a)\gamma p^2+(4a\gamma^2+1)p+\gamma=0$ in \eqref{rhomp}. This is done by setting in \eqref{fou} and \eqref{three} $a_0=\gamma, a_1=4a\gamma^2+1, a_2=2(1+a)\gamma$. We then obtain $w(t;a)$ from the Heaviside theorem in \eqref{conv4}, and evaluate it numerically. We define the "distance" from $W(t)$ in \eqref{jubis} and $w(t;a)$ as defined above, in terms of the dimensionless parameter
\begin{align}\label{de}
\De(a)\equiv 10^6 \ave{\tau}\int_0^\infty (w(t;a)-W(t))^2,
\end{align}
where $\ave{\tau}$ denotes the average waiting time. We have obtained for $\gamma=0.001$ the following values

\begin{align}\label{value}
 \begin{array}{ccc}
2a&=     &     0.996,~ 0.998,~1,~1.04\\
\De(a)&= & 1.6 ,~0.066,~0.1,~3.2  
\end{array}
\end{align}
It follows from these numbers that the expression of $w(t)$ in the $\gamma\to 0$ limit for $2a≠1$ does not coincide with an almost obvious result. This is, it seems to us, a strong argument for setting $2a=1$ in the starting Rabi equations.

\subsection{ $\gamma_1=\gamma_2$ }\label{gamma}

The generalized Rabi equations in \eqref{rho} involve both a downward transition rate $2\gamma_2\rho_{22}$ and an upward transition rate $2\gamma_1\rho_{11}$. Plausible expressions for the two parameters are
\begin{align}\label{deltx}
\gamma_{1,2}=\varpi \exp  \left(   \pm \frac{eU-\hbar \om_o}{\hbar \varpi} \right)\qquad \varpi\ll\om_o.
\end{align}
We will not attempt to justify these expressions here. If $eU=\hbar \om_o$ strictly, $\gamma_1=\gamma_2\equiv \varpi, ~b=0$. The Rabi equations in vector form \eqref{ens2} become, setting $a=0, ~x(t)=0$
\begin{align}\label{deltax}
\frac{dy}{dt}&=-z,\nonumber\\
\frac{dz}{dt}&=y-2\varpi z.
\end{align}
In the large-time limit we have $y=z=0$, that is, there is no population inversion and the net conductance vanishes. Fluctuations however remain. In a small time interval $0,dt$ we may have zero event, a photo-absorption event (+) or a photo-emission event (--). As said above, the average numbers of (+) and of (-) are equal so that the average current vanishes. Solving the above equations and supposing that at time $t=0$ a photo-absorption event occurred ($\rho_{11}(0)=1\Longrightarrow z(0)=-1,y(0)=0$), we obtain the probability density for the sum of (+) and (-) events as $G(t)=\gamma z(t)$, which can easily be obtained explicitly.

If $\varpi\ll\Om_R\equiv1$, the electron performs many Rabi oscillations before a transition of any kind occurs. It follows that we can average the transition probabilities over a Rabi period, so that the populations are essentially equal to 1/2. In that limit, the electron population follows a telegraphic random motion between state 1 and state 2, with equal upward and downward transition probability densities. The spectral density of the photo-current at zero Fourier frequency is then the sum of that of the upward and downward processes. If the condition $\varpi\ll1$ is not fulfilled a more involved analysis would be required.

\newpage 

\section{Alternative approaches to Nyquist-like noise sources}\label{variousbis}

The present theory is based on two fundamental concepts. One is the introduction of Nyquist-like current sources, the other is the law of conservation of average energy. In the present section we consider alternative approaches to the link that exists between fluctuation and dissipation, usually referred to as the "fluctuation-dissipation theorem". Fluctuating rate sources are uncorrelated and their spectral densities are equal to the average rates.

Brown was the first to observe the random motion of small particles in viscous fluids (Brownian motion) and Johnson later on measured the electrical noise associated with conductances (Johnson noise). The first interpretation of Brownian motion was offered by Einstein and the first interpretation of electrical noise by Nyquist. The two phenomena are closely related. The Einstein and Nyquist interpretations rest on Classical Statistical Mechanics ($\kB T_m\gg\hbar\om$). For our purposes we need consider instead the situation where $\kB T_m\ll\hbar\om$. The latter "quantum" situation was treated by Callen and Welton in 1951, see for example \cite{Hanggi2005}. We first recall the Quantum Optics viewpoint. We consider next simpler, but admittedly partly heuristic, explanations. These arguments tend to prove that the spectral density of the Nyquist-like noise sources associated with two-level atoms exhibiting a peak conductance $G$ at frequency $\om=\om_o$ is equal to $\hbar\om_o \abs{G}$ when either $\rho_{11}\approx 1$ (positive conductance) or $\rho_{22}\approx 1$ (negative conductance). 

In the first argument we show that there is a (perhaps unique) way of generalizing to the quantum domain the classical expression $\ave{E}=\kB T_m$ for the average energy of an oscillator resonating at frequency $\om_o$ that avoids divergences. The expression obtained differs from the 1901 Planck formula by an additional energy $\hbar\om_o/2$ called the vacuum energy. This expression does not lead to "ultraviolet" divergences if only measurable quantities are being considered. 

In the second approach we consider an oscillator in a state of thermal equilibrium with a small negative conductance and a slightly larger positive conductance (in absolute value). The ratio of these two conductances is equal to the ratio of lower and higher-state populations, given by Classical Statistical Mechanics. 

The third approach is quite different from the previous one since we consider an \emph{isolated} oscillator containing two-level atoms in a state of \emph{highly-non-thermal} equilibrium. Statistical mechanics tells us that states of equal energy are equally likely to occur, as long as no information concerning these states is available. The noise terms (Langevin forces) must be such that the variance of the photon number derived from the above Statistical Mechanical law be obtained in the long-time limit.

\subsection{Quantum-optics approach}\label{q}

In Quantum Optics treatments, loss-less optical oscillators oscillating at frequency $\om_o$ are viewed as being akin to loss-less mechanical oscillators with quantized energy levels $E_m=\hbar\om_o\p m+\frac{1}{2}\q,~m=0,1...$. If we apply to such oscillators the Boltzmann result that the probability that a level of energy $E$ be occupied is proportional to $\exp(-\beta E),~\beta\equiv1/\kB T_m$ when the system is in contact with a bath at absolute temperature $T_m$ (subscript "$m$" unrelated to the integer $m$), the Planck law of black-body radiation results through a summation over $m$. We obtain
\begin{align}\label{pl}
\ave{E}= \frac{\sum_{m=0}^{\infty}E_m~ \exp(-\beta E_m)}{\sum_{m=0}^{\infty} \exp(-\beta E_m)}=\frac{\hbar \om_o}{2}\frac{\exp(\beta \hbar \om_o)+1}{\exp(\beta\hbar \om_o)-1}.
\end{align}

The fluctuation-dissipation theorem (FDT) applies to systems having a linear causal response, in thermal equilibrium. One first evaluates the symmetrized correlation $C_{i,q}(\tau)\equiv\ave{\De i(\tau)\De q(0)+\De q(0)\De i(\tau)}/2$, where the charge $q$ and the current $i=dq/dt$ are conjugate operators (as $x$ and $p/m=dx/dt$ in Quantum Mechanics). The end result of the calculation is that the spectral density of the current, that is the Fourier transform of the correlation $C_{i,i}(\tau)$ reads \cite{Hanggi2005}
\begin{align}
\label{ny}
\spectral_{i}(\om)\equiv\spectral_{i,i}(\om)=2\p\frac{\hbar\om}{2}+\frac{\hbar\om}{\exp(\hbar\om/\kB T)-1}\q \R \{Y(\om)\}.
\end{align}
where $\R \{Y(\om)\}\equiv G(\om)$ denotes the system conductance, a function of the frequency $\om$. In the following, operators are not employed, that is, $i(t)$ is viewed as a classical function of time.

\subsection{Heuristic approach}\label{heuristic}

A simple heuristic derivation of the black-body formula is suggested here. In the classical regime, single-mode oscillators at frequency $\om_o$ have a probability $\exp(-\beta E)$ of having an energy $E$, according to Classical Statistical Mechanics. Thus the average energy reads
\begin{align}\label{n4}
\ave{E}=\frac{ \int_0^\infty dE E\exp(-\beta E)} { \int_0^\infty dE \exp(-\beta E)}=\frac{1}{\beta}=\kB T_m,
\end{align}
independent of $\om_o$. The action $f$ of an oscillator is the ratio of its average energy and frequency, and accordingly $f(x)=1/x$, setting $x\equiv\beta E= \om/\kB T_m$. Thus $f(x)$ obeys the Riccati equation
\begin{align}
\label{ric}
\frac{df(x)}{dx}+f(x)^2=0.
\end{align}
However, as was noted at the end of the 19th century, the expression $\ave{E}=\kB T_m$ leads to infinite radiated heat since the number of electromagnetic modes is infinite in a cavity with perfectly reflecting walls. It apparently did not occur to scientists at the time that this difficulty is resolved simply by adding a constant $\p \hbar/2\q^2$ on the right-hand side of \eqref{ric}, that is, supposing
\begin{align}
\label{ric'}
\frac{df(x)}{dx}+f(x)^2=\p\frac{\hbar}{2}\q^2,
\end{align}
where $\hbar$ is a universal constant with the dimension of action. The solution of this modified equation reads
\begin{align}
\label{solter}
x=\int \frac{df}{\p\hbar/2\q^2-f^2}=\frac{1}{\hbar}\log(\frac{2f+1}{2f-1})+x_o,
\end{align}
where $x_o$ denotes an arbitrary constant. For $x_o$=0 we obtain that 
\begin{align}
\label{sol}
f(x)=\frac{\hbar}{2}\frac{\exp(\hbar x)+1}{\exp(\hbar x)-1}
\end{align}
which coincides with the Planck formula except for the vacuum energy mentionned above. The arbitrary constant $x_o$ on the right-hand-side of \eqref{solter} must vanish to obtain agreement with the classical result. It was noted by Einstein and Stern in 1913 that $f(x)-1/x\to0$ if $x\to0$, that is, the expansion of $f(x)$ is of the form
\begin{align}
\label{solbis}
f(x)=\frac{1}{x}+ax+...,
\end{align}
where $a$ is a constant, without an $x$-independent term. One may conjecture that there are no other differential equation but \eqref{ric'} which possesses only one solution satisfying the above classical limit.
To conclude, the classical expression of the average energy generalizes to 
\begin{align}
\label{entotbis}
\ave{E}=\frac{\hbar \om_o}{2}\frac{\exp(\beta \hbar \om_o)+1}{\exp(\beta\hbar \om_o)-1},
\end{align}
where $\beta\equiv1/\kB T_m$. If we next consider a cavity with perfectly-conducting walls, solutions of the Maxwell equation exist only for a series of real resonating frequencies $ \om_1, ~\om_2, ~\om_3, ~...$. Each of these modes of resonance is ascribed an average energy given by the above expression with $\om_o$ replaced by $ \om_1, ~\om_2, ~\om_3, ~...$. For a d-dimensional cavity the mode density $\rho(\om)$, where $\rho(\om)d\om$ denotes the number of modes whose frequencies are between $\om$ and $\om+d\om$, grows in proportion of $\om_o^{d-1}$. It follows that the total energy is apparently infinite. This is perhaps why, in his original work, Planck subtracted the vacuum energy $\hbar \om_o/2$ from the expression given in \eqref{entotbis}. The total energy in a cavity of volume $\V$ in thermal equilibrium at absolute temperature $T_m$ is then found to be finite and proportional to the fourth power of $T_m$. If the cavity is pierced with a small hole that does not perturb much the state of thermal equilibrium, the measured output-power spectral density is supposed to be proportional to the internal energy spectrum, that is, to the product of $\ave{E(\om)}-\hbar\om/2$ and the mode density $\rho(\om)$. This is the famous black-body spectrum measured around 1899, and obtained by Planck in 1900 on the basis of an "educated guess" of the entropy function. This formula agrees very well with measurements. In our model the power is collected by a small external conductance $g$ at $T_m=$0K, having its own noise source. A finite result is then obtained from the expression of $\ave{E(\om)}$ as given in \eqref{entotbis}.

\subsection{Johnson-Nyquist noise}\label{nyquist}

When a frequency-independent conductance $G$ is in equilibrium with a bath at absolute temperature $T_m$, there is associated with it a random current source $j(t)$ whose (double-sided) spectral density $\spectral_j$ equals $2\kB T_m G$, as discovered by Johnson and Nyquist in 1927, 1928. This expression holds in the so-called "classical regime", that is, at frequencies much smaller than $\kB T_m /\hbar$. In the present section currents are denoted by $j$ instead of $i$ because we are mostly concerned with low frequencies.

To justify the Nyquist expression in the classical regime, let us recall that, according to Statistical Mechanics, the average energy must equal $\kB T_m/2$ per degree of freedom. As a consequence, the energy stored in a capacitance $C$ with a conductance $G$ in parallel must equal $\kB T_m/2$. The modulus square $\abs{V/I}$ of the impedance of the circuit considered at frequency $\om$ is $1/\p G^2+C^2\om^2\q$. Since the energy stored in a capacitance $C$ submitted to an (root-mean-square) voltage $V$ is $C\abs{V}^2/2$, as we have seen in section \ref{circuit}, we must have
 \begin{align}
\label{ny2}
\int_{-\infty}^{+\infty}\frac{d\om}{2\pi}\frac{C \spectral_j}{2\p G^2+C^2\om^2\q}=\frac{\kB T_m }{2}
\end{align}
from which it follows that $\spectral_j=2\kB T_m G$. What we have given above is essentially the Nyquist argument.

The Nyquist result may be generalized to any (time-invariant, linear) circuit consisting of any number of conductances all of them being at the same temperature $T_m$, capacitances and inductances. If the conductance between any two terminals of the circuit is $G(\om)$, the circuit may be considered noiseless provided a current source $j(t)$ be applied to the two terminals with spectral density $\spectral_j(\om)=2\kB T_m G(\om)$. 

For later use note that when a white (i.e., with a frequency-independent spectrum) Nyquist current $i(t)$ is applied to an ideal narrow-band filter whose response is centered at $\om=±\om_o$, the filter output may be written in the form  
 \begin{align}
\label{current}
i(t)=\sqrt{2}C'(t)\cos(\om_o t)+\sqrt{2}C''(t)\sin(\om_ot)\equiv \Re\{ \sqrt{2}C(t)\exp(-\ii\om_o t)   \}
\end{align}
where the real random functions of time $C'(t)$ and $C''(t)$ vary slowly, are uncorrelated, and have (double-sided) spectral density $\spectral_{C'}=\spectral_{C''}=2\kB T_m G$.

\subsection{Thermal equilibrium approach}\label{thermal}

In the present section we generalize the previous Johnson-Nyquist expression to arbitrary frequencies. Consider an inductance-capacitance ($L-C$) oscillator resonating at frequency $\om_o$ with a small positive conductance $G_a$ and a negative conductance $-G_e$  (subscript "$e$" for "emitting") in parallel, with $G_a$ (subscript "$a$" for "absorbing") exceeding $G_e$ so that there is a small net loss. We set $G_a-G_e\equiv G>0$. The conductance $G_a$ is supposed to represent the absorption by two-level atoms in the lower state (energy $E_a$), while $G_e$ represents the emission from atoms in the upper state (energy $E_e$). A near-resonance condition $\hbar\om_o\approx E_e-E_a$ holds. According to the Schrödinger equation the conductances are proportional to the corresponding numbers of atoms $n_a$ and $n_e$, respectively. 

\begin{figure}
\setlength{\figwidth}{0.45\textwidth}
\centering
\begin{tabular}{cc}
\includegraphics[width=\figwidth]{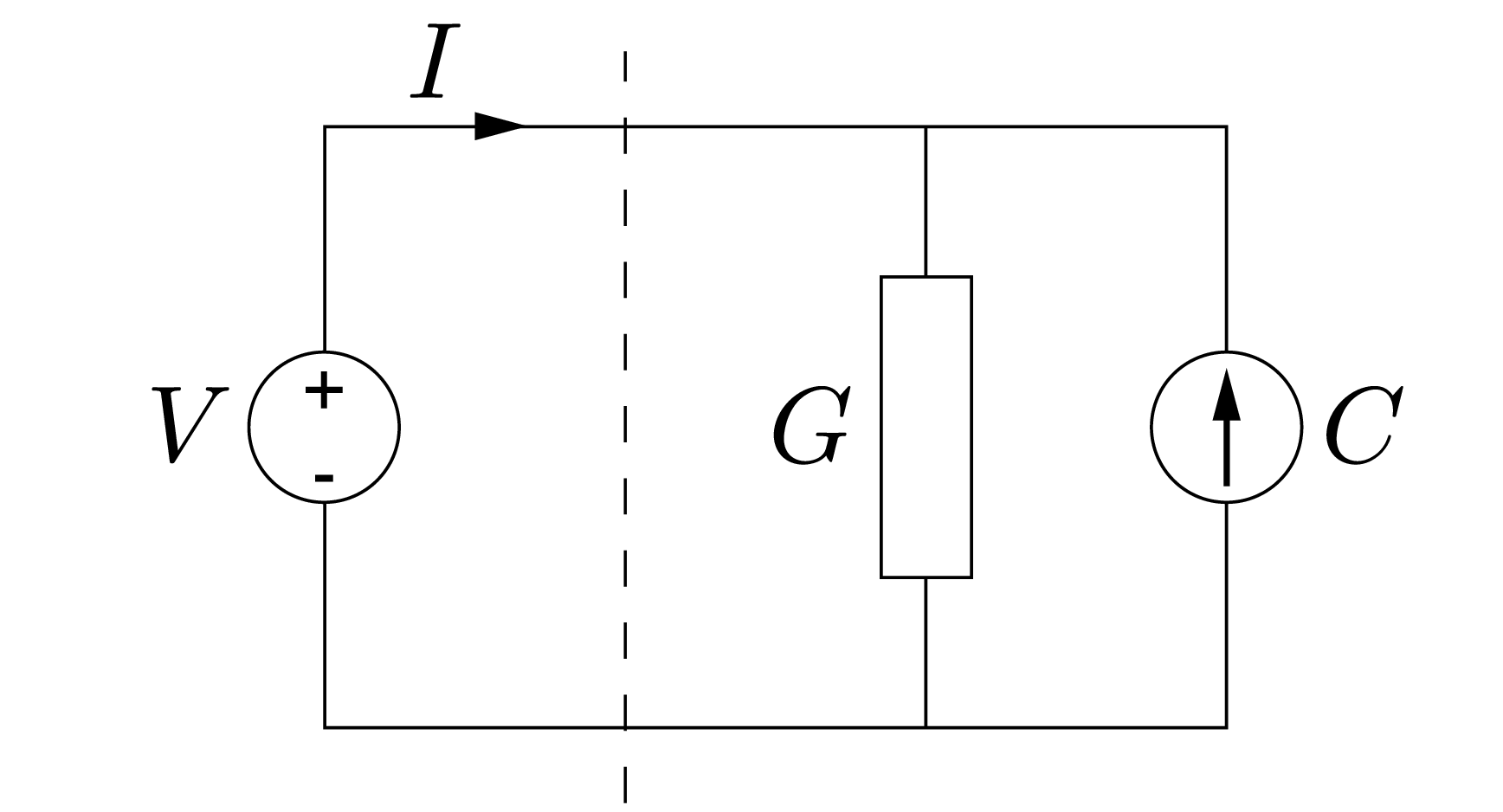} & \includegraphics[width=\figwidth]{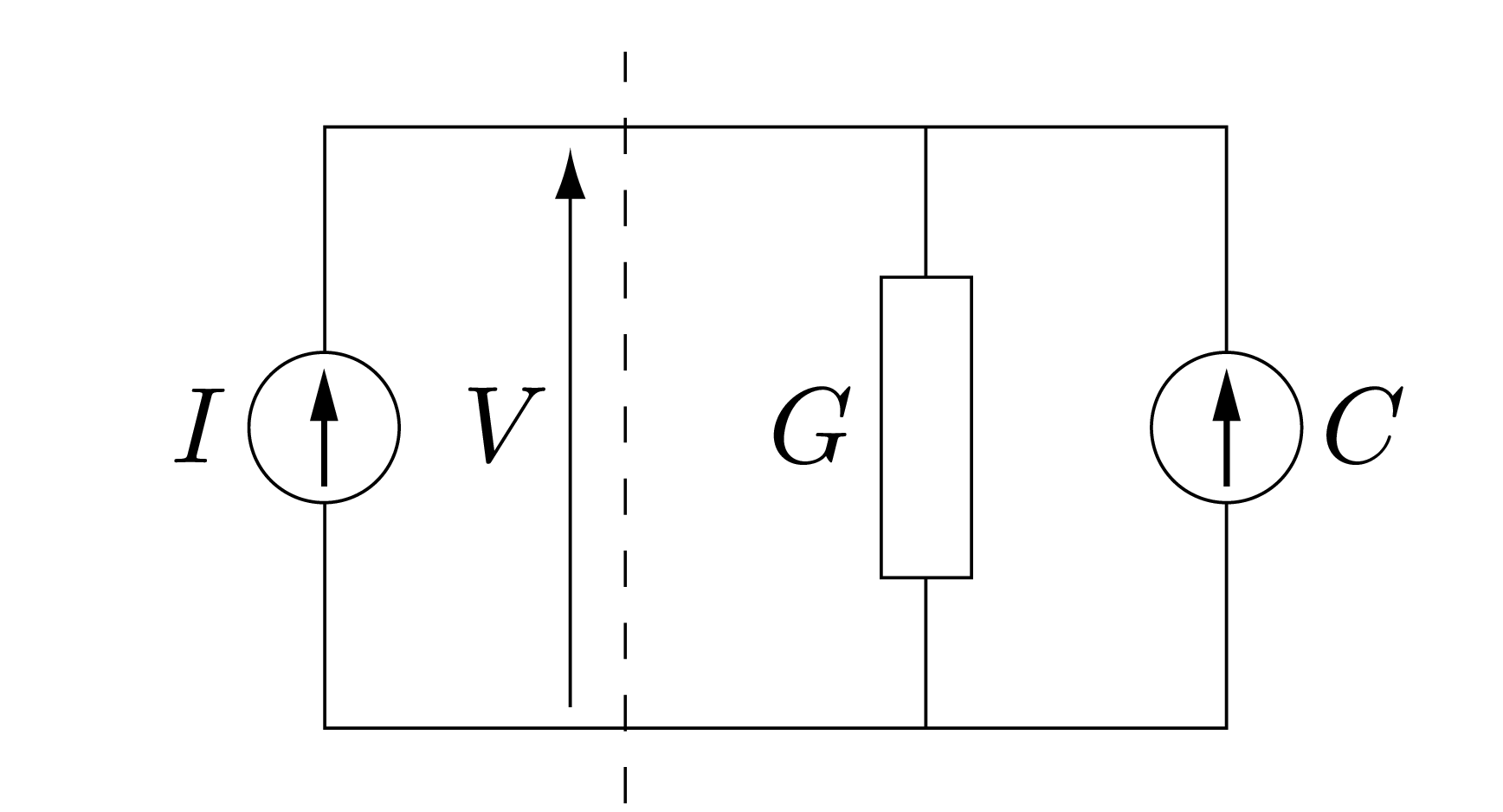} \\
(a) & (b) \\
\includegraphics[width=\figwidth]{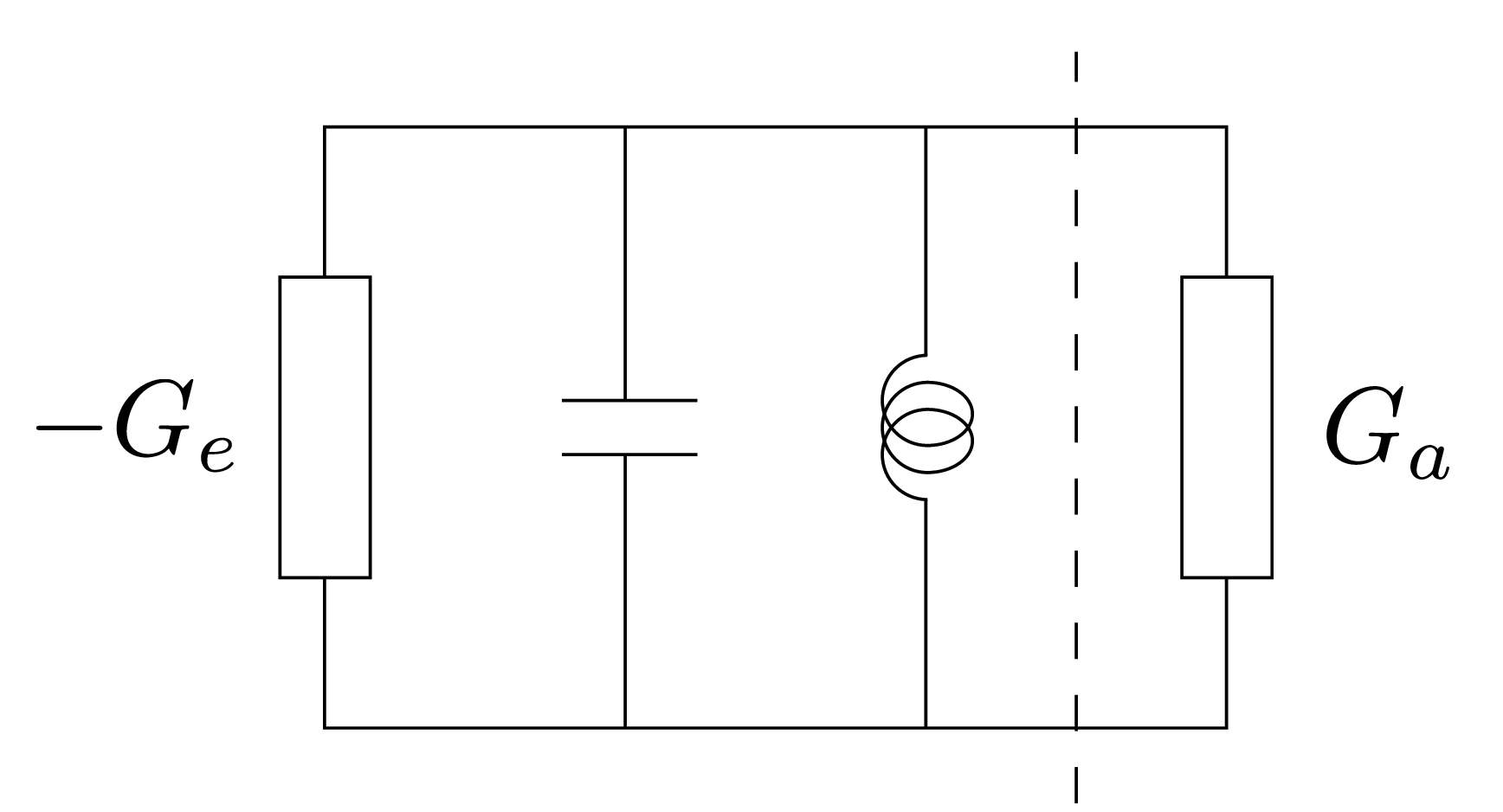} & \includegraphics[width=\figwidth]{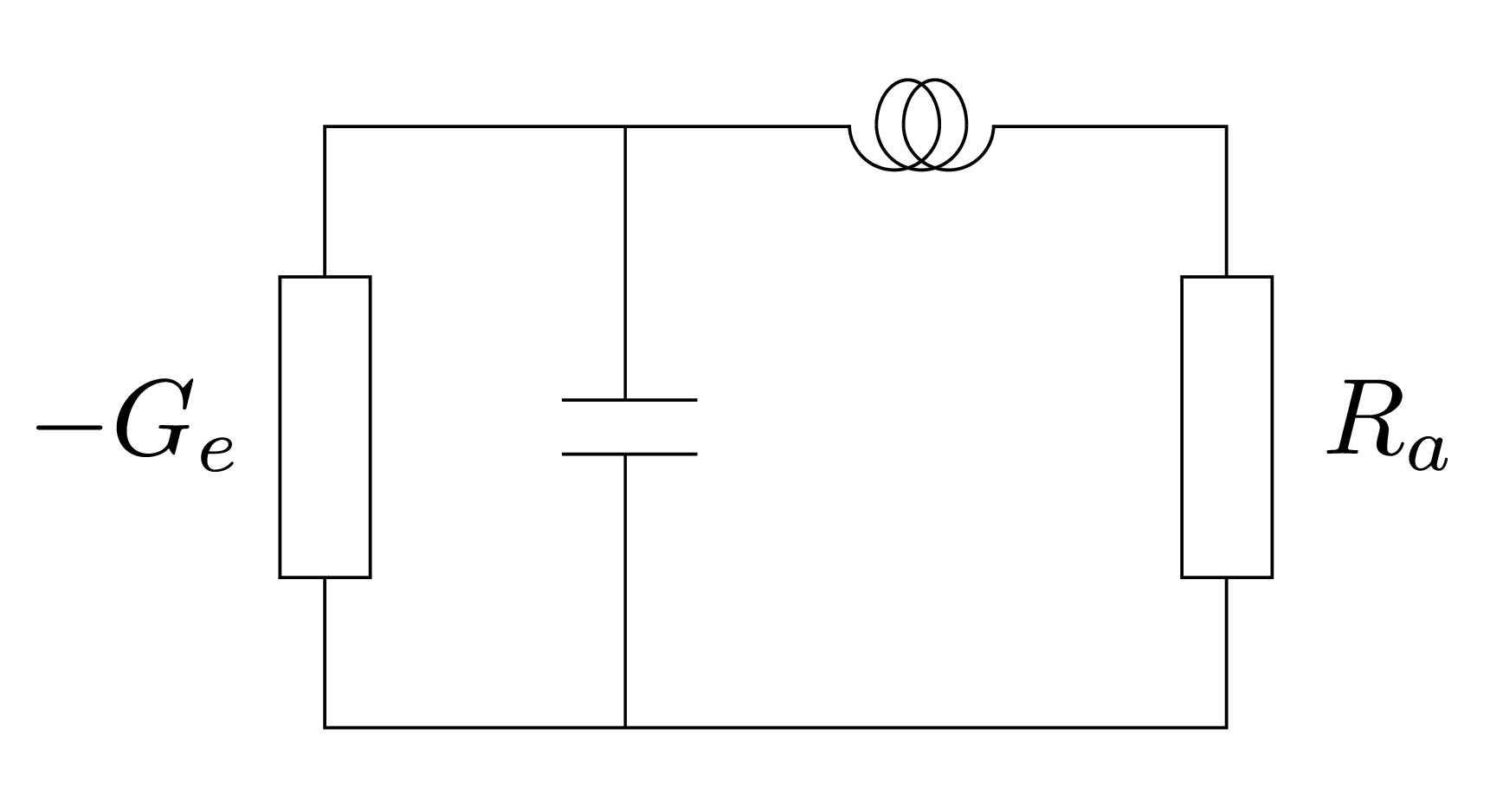} \\
(c) & (d) \\
\end{tabular}
\caption{a) represents a potential source $V$ applied to a conductance $G$. $C$ represents the Nyquist-like noise current source associated with $G$. We are mostly concerned with the power flowing from left to right through the dashed line, b) represents a conductance driven by a current source $I$, c) is a laser model with a negative conductance $-G_e$, a resonating circuit, and a positive conductance $G_a$, representing the detector of radiation, connected in parallel, d) represents a laser model that exhibits a Petermann-like linewidth-enhancement factor.}
\label{circuitlaser}
\end{figure}

Because the conductances considered are constant the circuit is linear. The potential $V$ across the circuit is therefore equal to $\C/Y(\om)$, where $\C$ denotes the driving current, assumed to be independent of frequency, and the resonating circuit admittance $Y(\om)=G+\ii B(\om)$ where $B(\om)$ represents the sum of the $L$ and $C$ susceptances. Referring to \eqref{en'} the oscillator energy is
\begin{align}
\label{ennmk}
E= \int_{-\infty}^{+\infty}\frac{d\om}{2\pi}\frac{C\abs{\C}^2 }{G^2+4C^2\p \om-\om_o\q^2}
= \frac{\abs{\C}^2 }{4G} \frac{1}{\pi} \int_{-\infty}^{+\infty}\frac{dx}{1+x^2}=\frac{\abs{\C}^2 }{4G}.
\end{align}

Let this oscillator be in thermal contact with a bath at absolute temperature $T_m$. We replace the deterministic current $\C$ by a complex random function of time\footnote{Note that in the present linear regime the regulation mechanism at work in above-threshold lasers does not occur, and the fluctuations of $V$ are comparable to average values. Supposing that the current source is gaussian distributed, this is also the case for the optical potential $V$ and optical current $I$. Power should in general be evaluated as the real part of $V^\star ( I+ C(t))$, but in the linear regime presently considered the term $C(t)$, much smaller than the fluctuations of $I$, may be neglected. The power $\Re\{V^\star  I\}$ is Rayleigh-distributed.}. Because the processes are stationary we expect that the statistical properties of the random source $C(t)\equiv C'(t)+\ii C''(t)$ are unaffected by an arbitrary phase change, that is, we require that $C(t) \exp(\ii \phi)$ has the same statistical property as $C(t)$ for any phase $\phi$. This entails that $C'(t)$ and $C''(t)$ are uncorrelated and have the same statistical density. We thus set $\spectral_C'=\spectral_C''\equiv \spectral$. We have seen that for $n$ independent atoms in some state both $G$ and $\spectral$ are proportional to $n$. We therefore expect that $\spectral=\al G$, where $\al$ is a constant to be determined. 

Because of the symmetry between stimulated absorption and stimulated emission implied by the Schrödinger equation the spectral densities have the same form for positive conductances $G_a$ and negative conductances $-G_e$, namely $\spectral_a=\al G_a$ and $\spectral_e=\al G_e$, with the same constant of proportionality $\al$. Note that $C'$ and $C''$ contribute equally and that double-sided spectral densities for $C'$, $C''$ in the Fourier $\Om$-domain are employed. If the conductances $G_a$ and $-G_e$ are connected in parallel the total conductance is $G=G_a-G_e$, as said above, and the total spectral density is $2\al \p G_a+G_e\q$. The average resonator energy follows from \eqref{ennmk}
\begin{align}
\label{emk}
\ave{E}=\frac{2\al (G_a+G_e) }{4G}=\frac{\al}{2}\frac{G_a/G_e +1}{G_a/G_e -1}.
\end{align}

Classical Statistical Mechanics tells us that at thermal equilibrium $\frac{n_a}{n_e}=\exp(\frac{E_e-E_a}{\kB T_m})$. It follows that
\begin{align}
\label{ennm}
\frac{G_a}{G_e}=\frac{n_a}{n_e}=\exp\p\frac{E_e-E_a}{\kB T_m}\q=\exp\p\frac{\hbar \om_o}{\kB T_m}\q
\end{align}
Classical Statistical Mechanics also tells us that in the classical limit the average energy equals $\kB T_m/2$ per degree of freedom, and thus $\ave{E}=\kB T_m$ for the oscillator considered when $\kB T_m\gg \hbar\om_o$. According to \eqref{emk} this is the case if and only if
\begin{align}
\label{em}
\al=\hbar \om_o
\end{align}

At low-temperatures, $\kB T_m\ll\hbar \om_o$, the resonator energy is therefore $\hbar \om_o/2$. Since a resonator may exchange energy resonantly with atoms only in units of $\hbar \om_o$, we conclude that the energy of a single-mode loss-less resonator may be written as
\begin{align}
\label{mb}
E_T=\p m+1/2\q\hbar \om_o,
\end{align}
where $m$ is an integer that one may call "number of photons in the resonator". However, we view linear loss-less resonators of any kind (mechanical, optical, or other) as abstractions. Such oscillators acquire physical meaning only when they are coupled to sources of energy. What need to be quantized are those energy-coupling mechanisms.

\subsection{Isolated cavity approach}\label{ratebis}

A single-mode optical cavity resonating at angular frequency $\om_o$
may be modeled as an inductance-capacitance ($L,C$) circuit with
$LC\om_o^2=1$.  The active atoms, located between the capacitor
plates, interact with a spatially uniform optical field through their
electric dipole moment. The
2-level atoms (with the lower level labeled "$a$" and the upper level
labeled "$e$") are resonant with the
field. This means that the atomic levels $a$ and $e$ are separated in
energy by $\hbar\om_o$. The concept of temperature nowhere enters in the present section.

\paragraph{(Highly non-thermal) equilibrium}

Consider $N$ identical two-level atoms.  For each atom, the zero of
energy is taken at the lower level and the unit of energy at the upper
level (typically, 1 eV).  The atoms are
supposed to be at any time in either the upper or lower state.  The
number of atoms that are in the upper state is denoted by $n$, and the
number of atoms in the lower level is therefore $N-n$.  With the convention $\hbar \om_o=1$, the atomic energy is equal to $n$.  Its maximum value
$N$ occurs when all the atoms are in the upper state.  There is
population inversion when the atomic energy $n>N/2$. The atoms are supposed to reach a state of equilibrium before other parameters have changed significantly. The strength of the atom-atom
coupling, however, needs not be specified.

The statistical weight $W(n)$ of the atomic collection is the number
of distinguishable configurations corresponding to some total energy
$n$.  For two atoms ($N=2$), for example, $W(0)=W(2)=1$ because there
is only one possible configuration when both atoms are in the lower
state $(n=0)$, or when both are in the upper state $(n=2)$.  But
$W(1)=2$ because the energy $n=1$ obtains with \emph{either one} of
the two (distinguishable) atoms in the upper state.  For $N$ identical
atoms, the statistical weight (number of ways of picking up $n$ atoms
out of $N$) is 
\begin{align}
    W(n)=\frac{N!}{n!(N-n)!}.
    \label{Wofn}
\end{align}
Note that $W(0)=W(N)=1$ and that $W(n)$ reaches its maximum value at
$n=N/2$ (supposing $N$ even), with $W(N/2)$ approximately
equal to $2^{N}\sqrt{2/\pi N}$. Note further that
\begin{align}\label{Z}
    Z\equiv \sum_{n=0}^{N}W(n)=2^{N}.
\end{align}

Consider next an isolated single-mode optical cavity containing $N$
resonant two-level atoms, and suppose that initially all of these atoms are in the emitting (upper) state. One observes that the number of atoms in the emitting state may vary from 0 to $N$, and thus one presumes that the missing energy is stored in the optical resonator. The atoms indeed perform jumps from one state to
another in response to the optical field so that the number of atoms
in the upper state is some function $n(t)$ of time.  If $m(t)$ denote the
number of light quanta at time $t$, the sum $n(t)+m(t)$ is a conserved
quantity (essentially the total atom+field energy).  Thus, $m$ jumps
to $m-1$ when an atom in the lower state gets promoted to the upper
state, and to $m+1$ in the opposite situation.  If $N$ atoms in their
upper state are introduced at $t=0$ into the empty cavity ($m=0$),
part of the atomic energy gets converted into field energy as a result
of the atom-field coupling and eventually an equilibrium situation is
reached.  

The basic principle of Statistical Mechanics asserts that in isolated systems
all states of equal energy are equally likely.  Accordingly, the
probability $pr(m)$ that some $m$ value occurs at equilibrium is
proportional to $W(N-m)$, where $W(n)$ is the statistical weight of
the atomic system.  As an example, consider two (distinguishable)
atoms ($N$=2).  A microstate of the isolated (matter$+$field) system
is specified by telling whether the first and second atoms are in
their upper (1) or lower (0) states and the value of $m$.  Since the
total energy is $N=2$, the complete collection of microstates (first
atom state, second atom state, field energy), is: (1,1,0), (1,0,1),
(0,1,1) and (0,0,2).  Since these four microstates are equally likely,
the probability that $m=0$ is proportional to 1, the probability that
$m=1$ is proportional to 2, and the probability that $m=2$ is
proportional to 1.  This is in agreement with the fact stated earlier
that $pr(m)$ is proportional to $W(N-m)$.  After normalization, we
obtain for example that pr(0)=1/4.
 
The normalized probability reads in general
\begin{align}\label{P}
    pr(m)=\frac{W(N-m)}{Z}=\frac{N!}{2^{N}m!(N-m)!}
\end{align}
The moments of $m$ are defined as usual as
\begin{align}\label{mr}
     \ave{m^{r}}\equiv \sum_{m=0}^{N}m^{r} pr(m)
\end{align}
where brakets denote averagings.  It is easily shown that $\ave{m}=N/2$ and $\mathrm{var}(m) \equiv
{\ave{m^{2}}-\ave{m}^{2}} = N/4$.  Thus the number $m$ of
light quanta in the cavity fluctuates, but the statistics of $m$ is
sub-Poissonian, with a variance less than the mean. For example, one may readily deduce from that principle that, if an isolated single-mode cavity initially contains no photons but $N$ two-level resonating atoms in the upper state, the system evolves to an equilibrium state in which the variance of the number of photons in the cavity is half the average number of photons, that is, the photon statistics is sub-poissonian. More generally, for atoms with $B$ evenly-spaced levels, the variance of the photon number is $(B+1)/6$ times the average photon number, a result that coincides with the previous one if we set $B=2$.

The expression of $pr(m)$ just obtained has physical and
practical implications.  Suppose indeed that the equilibrium cavity
field is allowed to escape into free space, thereby generating an
optical pulse containing $m$ quanta.  It may happen, however, that no
pulse is emitted when one is expected, causing a counting error.  From
the expression in (\ref{P}) and the fact that $\ave{m}=N/2$, the
probability that no quanta be emitted is seen to be
$pr(0)=4^{-\ave{m}}$.  For example, if the average number of light
quanta $\ave{m}$ is equal to $20$, the communication system
suffers from one counting error (no pulse received when one is
expected) on the average over approximately $10^{12}$ pulses.  Light
pulses of equal energy with Poissonian statistics are inferior to the
light presently considered in that one counting error is recorded on
the average over $\exp(\ave{m})=\exp(20)\approx 0.5~10^9$ pulses.

\paragraph{Time evolution of the number of light quanta in isolated
cavities}

Let us now evaluate the probability $pr(m,t)$ that the number of light
quanta be $m$ at time $t$.  Note that here $m$ and $t$ represent two
independent variables.  A particular realization of the process was
denoted earlier $m(t)$.  It is hoped that this simplified notation
will not cause confusion.
 
Let $R_e(m)dt$ denote the probability that, given that the number of
light quanta is $m$ at time $t$, this number jumps to $m+1$ during the
infinitesimal time interval [$t, t+dt$], and let $R_a(m)dt$ denote the
probability that $m$ jumps to $m-1$ during that same time interval
(the letters "$e$" and "$a$" stand respectively for "emission"
and "absorption"). The probability $pr(m,t)$ obeys the relation
\begin{align}\label{FP}
    pr(m,t+dt) =  pr(m+1,t)R_a(m+1)dt+pr(m-1,t)R_e(m-1)dt \nonumber\\+pr(m,t)[1-R_a(m)dt-R_e(m)dt].
 \end{align}
Indeed, the probability of having $m$ quanta at time $t+dt$ is the sum
of the probabilities that this occurs via states $m+1$, $m-1$ or $m$
at time $t$.  All other possible states are two or more jumps away
from $m$ and thus contribute negligibly in the small $dt$ limit.  After a sufficiently long time, one expects
$pr(m,t)$ to be independent of time, that is $pr(m,t+dt)=pr(m,t)\equiv
{pr(m)}$.  It is easy to see that the "detailed balancing" relation 
\begin{align}\label{ball}
    pr(m+1)R_a(m+1)=pr(m)R_e(m)
\end{align}
holds true because $m$ cannot go
negative. When the expression of
$pr(m)$ obtained in \eqref{P} is introduced in \eqref{ball}, one finds that
\begin{align}\label{EA}
    \frac{R_e(m)}{R_a(m+1)}=\frac{pr(m+1)}{pr(m)}=\frac{N-m}{m+1}.
\end{align}
$R_e$ must be proportional to the number $n=N-m$ atoms in the upper state while $R_a$ must be proportional to the number $N-n=m$ of atoms in the lower state. We therefore set $R_e(m)=\p N-m\q f(m), R_a(m)=mg(m)$, where $f(m)$ and $g(m)$ are two functions to be determined. Substituting in \eqref{EA} we find that
\begin{align}\label{Ei}
    f(m)=g(m+1).
\end{align}
Because we assume that atoms emit or absorb a single light
quantum at a time ("1-photon" process) the two functions $f(m)$ and $g(m)$ must be of the linear form $f(m)=am+b$ and $g(m)=cm+d$, where $a,b,c,d$ are constants. But $R_a$ is required to vanish for $m=0$ since, otherwise, $m$ could go negative, and thus $d=0$. Substituting into \eqref{Ei}, we find the relation $am+b=c(m+1)$ which must hold for any $m$-value. Therefore, $a=b=c$. Setting for brevity $a=b=c=1$ amounts to fixing up a time scale. Then $R_e(n,m)=n\p m+1\q, R_a(n,m)=\p N-n\q m$. We note here a lack of symmetry between the rate of stimulated emission (proportional to $m+1$) and the rate of stimulated absorption (proportional to $m$). Since according to the Schrödinger equation the two processes should be similar, we are led to define the field energy as $m+1/2$, and to express $R_e$ and $R_a$ in terms of the field energy at jump time, $E_{jump~ time}$, defined as the arithmetic average of the field energy just before and just after the jump. If we do so, we finally obtain, setting $n=n_e$ and $N-n=n_a$
\begin{align}\label{Ei'}
R_e(n_e,m)&=n_e~E_{jump~ time}\\
R_a(n_a,m)&=n_a ~E_{jump~ time},
\end{align}    
and the symmetry is indeed restored.

Let us now restrict our attention to the steady-state regime and large
values of $N$.  Since $m$ is large, it may be
viewed as a continuous function of time with a well-defined
time-derivative.  Because the standard deviation $\sqrt{N/4}$ of $m$
is much smaller than the average value, the so-called "weak-noise
approximation" is permissible. Within that
approximation, the average value of any smooth function $f(n,m)$ may
be taken as approximately equal to $f(\ave{n},\ave{m})$.
  
The evolution in time of a particular realization $m(t)$ of the
process obeys the classical Langevin equation
\begin{align}\label{L1}
    \frac{dm}{dt}= \mathfrak{E}-\mathfrak {A},
\end{align}
where  
\begin{align}\label{aux}
   {\mathfrak{E}} \equiv R_e(m)+e(t) \qquad {\mathfrak{A}} \equiv R_a(m)+a(t).
\end{align}    
In these expressions, $e(t)$ and $a(t)$ represent uncorrelated
white-noise processes whose spectral densities are set equal to $ \beta R_e(\ave{m})$ and $\beta R_a(\ave{m})$, respectively, where $\beta$ is a constant to be determined. 

Let us show that the variance of $m$ obtained from the above Langevin equation coincides with the result obtained directly from
Statistical Mechanics only if $\beta=1$.  Without the noise sources, the evolution of
$m$ in (\ref{L1}) would be deterministic, with a time-derivative
equals to the drift term $R_e(m)-R_a(m)$.  If the expressions (\ref{Ei})
are employed, the Langevin equation (\ref{L1}) reads
\begin{align}
    \frac{dm}{dt} & =  Nm-2m^2+e-a \nonumber\\
	\spectral_{e-a} & =  \beta \p R_e+R_a\q=\beta N\ave{m}=\beta N^{2}/2,
	\label{L3}
\end{align}
where the approximation $N\gg1$ has been made.
  
Let $m(t)$ be expressed as the sum of its average value $\ave{m}$ plus
a small deviation $\Delta m(t)$, and $Nm-2m^{2}$ in (\ref{L3}) be
expanded to first order.  A Fourier transformation of $\Delta m(t)$
with respect to time amounts to replacing $d/dt$ by $j\Omega$.
The Langevin equation now reads
\begin{align}\label{L4}
    j\Omega \Delta m= -N \Delta m+e-a\qquad \spectral_{e-a}=\beta N^{2}/2,
\end{align}
where $m$ has been replaced by its average value $N/2$. 

Since the spectral density of some random function of time $z(t)=ax(t)$, where $a\equiv{a'+ja''}$
is a complex number and $x(t)$ a stationary process, reads
$\spectral_{z}(\Omega)=|a|^2 \spectral_{x}(\Omega)$, see Section \ref{random}, one finds from (\ref{L4}) that the spectral density of the $\Delta m(t)$ process is
\begin{align}\label{spb}
    \spectral_{\Delta m}(\Omega)=\frac{\beta N^{2}/2}{N^{2}+\Omega ^{2}}.
\end{align} 
The variance of $m$ is the integral of $\spectral_{\Delta m}(\Omega)$ over
frequency ($\Omega/2\pi$) from minus to plus infinity, that is
var$(m)=\beta N/4$. There is agreement with the previous result derived from the basic Satistical Mechanics rule only if $\beta=1$. It follows that the spectral density of fluctuation rates such as $r(t)$ must be equal to the average rates, say $R$.

\subsection{Conclusion of the present section}\label{conc}

We have offered independent (but, admittedly, partly heuristic) methods of showing that the spectral density of the real part of the fundamental current source associated with a conductance $G$ at frequency $\om_o$ is equal to $\hbar\om_o\abs{G}$, or, equivalently, that the spectral density of the real part of the fluctuation-rate equals the absolute value of the average rate, $\abs{R}$, when only one of the two states is populated. 

In general the admittance $Y(\om,n,R,T_m)$ of a circuit depends on frequency $\om$, on the number $n$ of atoms or electrons in the conduction band or upper state, on the emitted rate $R$ of electromagnetic radiation, and possibly on temperature $T_m$. In the above discussion we have implicitly assumed that $Y$ does not depend on $R$, that is, no spectral-hole burning occurs. In laser diodes, the number $n$ of electrons in the conduction band is a monotonic function of the static electric potential $U$ applied, provided some unessential internal resistance be neglected. It follows that $Y$ does not depend on $n$ if the applied static potential $U$ is held constant. There are other circumstances where $n$ is approximately constant. There may exist an equilibrium between the electron injection rate and the spontaneous decay rate. As long as the departure from this equilibrium condition remains small, we may assume that $n$ is approximately constant. If this is the case, the admittance $Y$ is a function of $\om$ only, and the system is linear. We have assumed that the optical frequency $\om$ remains close to some average oscillation frequency $\om_o$.

\section{Conclusion}\label{conclusion}

We have shown that the photo-electron spectrum originating from a detector submitted to non-fluctuating-pump laser light may be understood in semi-classical terms. We did not employ the Quantum Optics concept that the optical field should be treated as an operator acting on the state of the light, nor the concept that light consists of point particles called "photons". The expression "photon rate" was used only as another name for electromagnetic power divided by $\hbar \om_o$, where $\om_o$ denotes the average laser frequency. Photon rates are written in the form $R+\De R(t)+r(t)$, where $R$ denotes the average rate, $\De R(t)$ is proportional to the fundamental noise sources in the linearized regime, and the spectral density of the fundamental noise source $r(t)$ is equal to $R$. This conclusion has been reached from different approaches, essentially requiring agreement with Classical Statistical Mechanics formulas. We restricted ourselves to the non-relativistic ($c\to\infty$), stationary, linear or linearized, approximations. Results obtained from the present theory for some particular configurations were listed in the first version of this paper.

Any laser-detector system is viewed as a set of capacitances, inductances, and conductances connected to one-another, with driving static potentials or currents. Conductances were modeled by a single electron located between two conducting plates, see Fig. \ref{klystron} in (a). The electron is submitted to a static potential source $U_e$, which determines the states, 1 (lower), and 2 (upper) the electron may reside in, with an energy difference $E_2-E_1=\hbar \om_o\approx eU$, where $U$ is slightly smaller than $U_e$. The applied (complex) potential source $V$ "sees" a conductance $G$, which under some conditions does not depend on $V$. The laser-detector ensemble is similar to the klystron represented in Fig. \ref{klystron} in (b). To emphasize the symmetry between stimulated emission and stimulated absorption, the absorber of radiation schematically shown on the right of that figure should be replaced by a device similar to the emitter of radiation, but with a static potential source $U_a$ slightly smaller that $U$. 

The optical-potential \emph{source} $V$ could be materialized by a tuned circuit with a very large capacitance and a very small inductance resonating at the frequency $\om_o$. Then $V$ is almost independent of the induced current. In a real device, the capacitance of the tuned circuit is not arbitrarily large, that is, the "photon lifetime" is finite. The optical potential $V$ across the tuned circuit  then depends on the induced current and it is necessary to evaluate the complete circuit response to the noise sources, whose statistics is known. The case of "quiet" lasers differs from the set-up just described in that the static potential source $U_e$ is being replaced by a static current source $J_e$. This current source could be materialized by a very large inductance with a current $J_e$ flowing through it. In the very large inductance limit the delivered current is nearly constant, provided the experiment does not last too long. Here, by "quiet laser" we mean that the power collected by the static potential $U_a$ does not fluctuate much in the course of time. More precisely, the spectral density of the collected power vanishes at small Fourier frequencies. This is also what happens for the "grand-mother pendulum", an existing purely classical mechanical system, described at the beginning of this paper introduction.  

The over-all effect of the laser action is to slowly transfer energy from one static source to the other. Whether the energy flow occurs from left to right, or the converse, in the configurations just described, depends on the small difference existing between $U_e$ and $U_a$. There is therefore a slight irreversible loss of energy, perhaps carried away from the structure by acoustical waves. This loss may be as small as one wishes, as long as one does not insist on having a fast power transfer. This is also what happens in Carnot heat engines, which are reversible in the small-power transfer limit. The main features of the (admittedly highly idealized) configurations described in this paper may be understood with a single electron in the emitting part and a single electron in the absorbing part, as said above. In reality, the two elements involve a large number of electrons. We generally neglect the direct electron-electron Coulomb interaction (space-charge effects), but need take into account the Pauli exclusion principle and the material temperature (Fermi-Dirac statistics).

In Part II we intend to consider applications to the linear regime. In that regime, the circuit considered involves only constant (positive and negative) conductances $G$, besides capacitances and inductances. That is, the conductances are independent of the potential applied to them and do not depend on any other parameter. It suffices in principle to evaluate the response of a linear system to current sources $C(t)\equiv C'(t)+\ii C''(t)$ of known spectral density. We suppose that the response spectrum is sharply peaked near some frequency $\om_o$, so that the spectral density density of the current sources may be written as $\spectral_{C'}=\spectral_{C''}=\hbar\om_o \abs{G},~\spectral_{C'C''}=0$. The linear regime is usually applicable when laser-diodes are driven by constant electrical potentials, or when the oscillating or amplified signals are small. We also consider the noise properties of linear attenuators and amplifiers.

In Part III of this paper we intend to consider the linearized regime, and in particular the relative noise of idealized laser diodes at high power levels. At such high powers, the time derivative of the number $n$ of electrons in the conduction band may be neglected (i.e., $dn/dt=0$). It follows from that assumption that the net rate of photons entering into the cavity is equal to the pumping rate. But the detected rate is different because photons may be stored for a while in the cavity before exiting. In that limiting situation no relaxation-oscillation appear. Spontaneous emission and gain compression are neglected. The more general theory will be given later on. Reference to a circuit simulating the lasers Fourier-frequency response is cited. We consider the effect of gain compression, and electrical feedback. Frequency noise is considered.

Part IV will discuss the principles of Quantum Statistical Mechanics, beginning with the fundamental concepts introduced by Carnot in 1824.

Part V would provide an evaluation of parameters such as the optical gain, that enter in laser-diode theories, essentially on the basis of the Fermi-Dirac distribution (grand-canonical ensemble). 

Part VI would discuss the properties of classical beams and optical fibers.

Part VII would consider random distortions of optical waveguides.

\newpage

\bibliographystyle{IEEEtran}
\bibliography{circuittheory}

\end{document}